\documentclass[usenatbib]{mnras}
\RequirePackage{rotating}
\usepackage{dcolumn}
\usepackage{bm}
\usepackage{lscape}
\usepackage{natbib}
\usepackage{graphicx}
\graphicspath{ {kids_ims} {model_ims} {post_thes_ims} {new_ims_26sep} {im_30Oct} {13Nov} }
\usepackage{amsmath}
\usepackage{hyperref}
\usepackage{float}
\usepackage{siunitx}
\usepackage{subcaption}
\usepackage{braket}

\newcommand{\zphot}{z_\mathrm{phot}}
\newcommand{\zspec}{z_\mathrm{spec}}

\begin{document}

\twocolumn

\title[GAMA DR4]{Galaxy And Mass Assembly (GAMA): Data Release 4 and the $\mathbf{\emph{z}<0.1}$ total and $\mathbf{\emph{z}<0.08}$ morphological galaxy stellar mass functions}
\author[Driver et al.] 
{Simon P.\,Driver$^1$, 
Sabine Bellstedt$^1$, 
Aaron S.\,G.\,Robotham$^1$, 
Ivan K.\,Baldry$^2$, \newauthor
Luke J.\,Davies$^1$, 
Jochen Liske$^3$, 
Danail Obreschkow$^1$,  
Edward N.\,Taylor$^4$, \newauthor 
Angus H.\,Wright$^5$, 
Mehmet Alpaslan$^{6}$, 
Steven P.\,Bamford$^{7}$, 
Amanda E.\,Bauer$^{8}$, \newauthor
Joss Bland-Hawthorn$^9$,    
Maciej Bilicki$^{10}$,     
Mat\'ias Bravo$^{1}$,
Sarah Brough$^{11}$, \newauthor  
Sarah Casura$^3$, 
Michelle E.\,Cluver$^{4,12}$,  
Matthew Colless$^{13}$,  
Christopher J.\,Conselice$^{14}$, \newauthor  
Scott M.\,Croom$^{9}$, 
Jelte de Jong$^{15,16}$, 
Franceso D'Eugenio$^{17,18}$,
Roberto De\,Propris$^{19}$, \newauthor
Burak Dogruel$^{4}$,
Michael J.\,Drinkwater$^{20}$, 
Andrej Dvornik$^{5}$,  
Daniel J.\,Farrow$^{21}$,  \newauthor
Carlos S.\,Frenk$^{22}$,    
Benjamin Giblin$^{23}$, 
Alister W.\,Graham$^4$, 
Meiert W.\,Grootes$^{24}$, \newauthor
Madusha L.P.\,Gunawardhana$^{15}$,   
Abdolhosein Hashemizadeh$^1$, 
Boris H{\"a}u{\ss}ler$^{25}$, \newauthor        
Catherine Heymans$^{5,23}$, 
Hendrik Hildebrandt$^{5}$, 
Benne W.\,Holwerda$^{26}$,  \newauthor   
Andrew M.\,Hopkins$^{27}$, 
Tom H.\,Jarrett$^{28}$, 
D.\,Heath\,Jones$^{29}$,  
Lee S.\,Kelvin$^{30}$, \newauthor
Soheil Koushan$^{1}$, 
Konrad Kuijken$^{15}$,  
Maritza A.\,Lara-L\'opez$^{31}$,   
Rebecca Lange$^{32}$, \newauthor 
\'Angel R.\,L\'opez-S\'anchez$^{27}$,   
Jon Loveday$^{33}$,  
Smriti Mahajan$^{34}$, 
Martin Meyer$^{1}$, \newauthor
Amanda J.\,Moffett$^{35}$, 
Nicola R.\,Napolitano$^{36}$, 
Peder Norberg$^{22}$, 
Matt S.\,Owers$^{37,38}$, \newauthor 
Mario Radovich$^{39}$, 
Mojtaba Raouf$^{15}$,
John A.\,Peacock$^{23}$, 
Steven Phillipps$^{40}$, \newauthor   
Kevin A.\,Pimbblet$^{41}$, 
Cristina Popescu$^{42}$, 
Khaled Said$^{20}$, 
Anne E.\,Sansom$^{42}$, \newauthor   
Mark Seibert$^{43}$,  
Will J.\,Sutherland$^{44}$,  
Jessica E.\,Thorne$^1$, 
Richard J.\,Tuffs$^{45}$, \newauthor
Ryan Turner$^{1,4}$, 
Arjen van\,der\,Wel$^{18}$,
Eelco van\,Kampen$^{25}$,  
Steve M.\,Wilkins$^{33}$
\\
\\
The authors' affiliations are shown after the references.}


\pubyear{2021} \volume{000}


\maketitle
\label{firstpage}

\vspace{-2.0cm}

\begin{abstract}
In Galaxy And Mass Assembly Data Release 4 (GAMA DR4), we make available our full spectroscopic redshift sample. This includes 248\,682 galaxy spectra, and, in combination with earlier surveys, results in 330\,542 redshifts across five sky regions covering $\sim 250$\,deg$^2$. 
The redshift density, 
is the highest available over such a sustained area, has exceptionally high completeness (95 per cent to $r_{\rm KiDS} = 19.65$\,mag), and is well suited for the study of galaxy mergers, galaxy groups, and the low redshift ($z<0.25$) galaxy population. 
DR4 includes 32 value-added tables or Data Management Units (DMUs) that provide a number of measured and derived data products including GALEX, ESO KiDS, ESO VIKING, WISE and Herschel Space Observatory imaging. Within this release, we provide visual morphologies for 15\,330 galaxies to $z<0.08$, photometric redshift estimates for all 18 million objects to $r_{\rm KiDS} \sim 25$ mag, and stellar velocity dispersions for 111\,830 galaxies. We conclude by deriving the total galaxy stellar mass function (GSMF) and its sub-division by morphological class (elliptical, compact-bulge and disc, diffuse-bulge and disc, and disc only). This extends our previous measurement of the total GSMF down to $10^{6.75}$\,M$_{\odot} h_{70}^{-2}$ and we find a total stellar mass density of $\rho_* = (2.97 \pm 0.04) \times 10^{8}$\,M$_{\odot} h_{70}$ Mpc$^{-3}$ or $\Omega_*=(2.17 \pm 0.03) \times 10^{-3} h_{70}^{-1}$. We conclude that at $z<0.1$, the Universe has converted $4.9 \pm 0.1$ per cent of the baryonic mass implied by Big Bang Nucleosynthesis into stars that are gravitationally bound within the galaxy population.
\end{abstract}

\begin{keywords}
surveys,galaxies:distances and redshift,galaxies:fundamental parameters,galaxies:luminosity function, mass function,cosmological parameters, catalogues
\end{keywords}

\section{Introduction}
Spectroscopic surveys of galaxies are one of the mainstays of observational extragalactic astronomy. These {\it redshift surveys} started in the 1980s with the Harvard Center for Astrophysics survey led by John Huchra and Margaret Geller \citep{huchra83,delapp86,geller1989}. This continued with numerous shallow and medium-deep surveys conducted through the 1980s and 1990s, operating mainly on the new 4-metre class telescopes, e.g., {\it Stromlo-APM Redshift Survey}, \citep{loveday92}; {\it Durham-UKST Redshift Survey}, \citep{dur96}; {\it Las Campanas Redshift Survey}, \citep{LCRS2}; {\it ESO slice Project}, \citep{ESP97}; {\it Southern Sky Redshift Survey}, \citep{dacosta98}; {\it Canadian Network for Cosmology}, \citep{CNOC2} and many more. 

In the period leading into the millennium, the subject underwent an `industrial revolution' through the advent of wide-area multiplexed fibre-fed systems, as used by the {\it 2-degree-Field Galaxy Redshift Survey} (2dFGRS) on the Anglo-Australian Telescope at Siding Spring Observatory in Australia \citep{colless01}, and the {\it Sloan Digital Sky Survey} at Apache Point Observatory in New Mexico, USA \citep{york00}. These two surveys provided $\sim 250,000$ and over 1 million redshifts respectively.

Both the 2dFGRS and SDSS surveys based their input catalogues on flux limited samples with  minimal pre-selection other than stringent star-galaxy classification criteria, see for example the SDSS selection described in \cite{straus02}. Both surveys strove to pursue complete flux-limited samples with relatively high spectroscopic completeness ($>80$ per cent).

The SDSS survey, in particular, not only advanced the field through the provision of redshifts, but through the release of moderate signal-to-noise spectra (S/N\,$\sim 5-30$), and a dedicated CCD based imaging survey conducted in multiple $ugriz$ bandpasses across $\sim 8\,000$\,deg$^2$ of the Northern and Equatorial sky \citep{stoughton02}. SDSS has continued since this time, diversifying into more focused and niche sub-areas (i.e., SDSS II, \citeauthor{sdssII} \citeyear{sdssII}; SDSS III, \citeauthor{sdssIII} \citeyear{sdssIII}; SDSS IV, \citeauthor{sdssIV} \citeyear{sdssIV}; SDSS V, commencing in 2021, see \citeauthor{sdss5} \citeyear{sdss5} and also \url{http://www.sdss5.org}).

Today SDSS remains the preeminent low-$z$ spectroscopic survey, responsible for transforming our understanding of the nearby extragalactic sky. While part of the capacity to transform came from the scale, scope and quality of the data, this was magnified by the manner in which the data were made available. As of today there have been 16 SDSS Public data releases \citep{SDSS16}, as well as the efforts of the many who provided derived data products in an Open Source fashion suitable for immediate science e.g., \cite{kauff03,brinchmann2004,tremonti2004,blanton2005,simard2011} and many more.

Since the 2dFGRS and SDSS, and post millennium, there has been a bifurcation in the design and implementation of redshift surveys. One branch has pursued complex target selections, usually colour and/or photometric-redshift based, to maximise survey efficiency for constraining cosmological parameters. This essentially trades completeness for sky-coverage, e.g., the Australian-led {\it WiggleZ} \citep{wigglez} and the US-led {\it Baryon Oscillation Spectroscopic Survey} (an SDSS extension: \citeauthor{boss} \citeyear{boss}). While these surveys do remain useful for some galaxy population science (e.g., \citeauthor{bossgals} \citeyear{bossgals}), the more complex selection and sub-sampling does render some science cases unviable. Obvious examples include the study of merger rates via close dynamical pairs, group finding, and the low-mass end of the galaxy stellar mass function, all areas where high-completeness is paramount. 

The second branch in the bifurcation followed the path of conducting high-density high-completeness surveys often over more modest regions of sky, with the exception of the very local hemispheric surveys, and with a greater focus on complementary panchromatic data, e.g.,  {\it the Millennium Galaxy Catalogue} \citep{liske03}; {\it the 6-degree-Field Galaxy Survey} (6dFGS; \citeauthor{6dfgs} \citeyear{6dfgs}); {\it the Galaxy And Mass Assembly survey} (GAMA; \citeauthor{driver2011} \citeyear{driver2011}); and {\it the Smithsonian Hectospec Lensing Survey} \citep{geller2016}.  These surveys also built on the multi-wavelength direction started by SDSS, and in particular capitalised on the available UV (via GALEX) and near-infrared (via 2MASS, UKIRT, VISTA and WISE) data. Through collaboration with the Herschel-ATLAS survey \citep{hatlas}, the wavelength coverage of GAMA was extended into the far-IR and now spans 0.15-500$\mu$m \citep{driver2016}. These surveys, while optimised for galaxy population studies, are sub-optimal for cosmology due to their limited coverage \citep{blake2013}. However, we note the ability of the very local hemispheric 6dFGS survey to place significant constraints on the Hubble Constant via the detection and measurement of baryonic acoustic oscillations \citep{florian11}, and GAMA to assist in improving the cosmological constraints from the ESO KiDS weak-lensing survey (e.g., \citeauthor{kids2018} \citeyear{kids2018}; \citeauthor{kids2018b} \citeyear{kids2018b}; \citeauthor{kids2019} \citeyear{kids2019}).

With the advent of the 8-metre class facilities, spectroscopic surveys were extended out to higher redshift, e.g., {\it the VLT Very Deep Survey}, \citep{vvds}; the zCOSMOS survey, \citep{zcosmos}; {\it the Deep Extragalactic Evolutionary Probe 2}, \citep{deep2} and {\it the VIMOS Public Extragalactic Redshift Survey}, \citep{vipers}. Here completeness is also an issue, as on the whole these surveys are below 50 per cent redshift completeness (see \citeauthor{devils} \citeyear{devils} Figure 1). However, this is less by design and more imposed by either the difficulty of obtaining redshifts for very distant targets, or the logistical restrictions in using multi-slit devices. Recently the Deep Extragalactic Visible Legacy Survey (DEVILS) \citep{devils}, via stacked long-exposure integrations on the 4m class Anglo-Australian Telescope, is revisiting notable deep fields (COSMOS, XMMLSS, ECDFs), seeking to raise the spectroscopic completeness to $>90$ per cent, at intermediate magnitudes ($Y < 21.0$ mag) and depth ($z<1$). 

In the very near future, forthcoming multi-fibre systems on 4-metre (e.g., DESI, \citeauthor{desi} \citeyear{desi}; WEAVE, \citeauthor{weave} \citeyear{weave}; 4MOST, \citeauthor{4most} \citeyear{4most}), and 8-metre (MOONS, \citeauthor{moons} \citeyear{moons}; PFS, \citeauthor{pfs} \citeyear{pfs}) class facilities, will transform the existing low, intermediate, and high-redshift domains taking us from the million galaxy redshift scale and into the tens of millions. In the slightly longer-term the proposed and planned  12m Mauna Kea Spectroscopic Explorer (MSE), a dedicated optical/near-IR multiplexed spectroscopic facility \citep{MSE}, will extend this to the hundreds of millions, and is essentially capable of sampling the entire observable Universe at masses $>10^{9}$\,M$_{\odot} h_{70}^{-2}$ since $z \approx 5$. Also notable are the forthcoming European Space Agency Euclid \citep{euclid} and NASA SPHEREx \citep{spherex} missions that will survey very large samples within specific high or low redshift windows at low wavelength resolution with grism or linear variable filters respectively.

In parallel to the progression of spectroscopic survey campaigns, has been the rise of broad-band photometric redshift techniques (see for example the comparison of methods reported in \citeauthor{abdalla11} \citeyear{abdalla11}), and the narrow-band filter surveys that define the middle ground, e.g., COMBO17, \citep{combo17}; COSMOS, \citep{cosmos15}; ALHAMBRA, \citep{alhambra}; J-PAS \citep{benitez2015}; PAUS \citep{eriksen2019}; and OTELO \citep{otelo}. For many purposes, photometric redshifts are sufficient, but once again for merger, group, and very low redshift ($z<0.1$) science, the traditional photometric surveys struggle with velocity resolutions typically at $\sim 10\,000$\,km\,s$^{-1}$ (broad-band) to $\sim 1\,000$\,km\,s$^{-1}$ (narrow-band) compared to the typical galaxy pairwise velocity of 200-600\,km\,s$^{-1}$ \citep{love18} and typical low mass group velocity dispersions of $< 500$\,km\,s$^{-1}$ \citep{robotham}.

The Galaxy And Mass Assembly Survey (GAMA; \citeauthor{driver2011} \citeyear{driver2011}), commenced in 2008 with the goal of building upon the legacy of the original 2dFGRS and SDSS surveys to produce a {\it highly complete} redshift survey with {\it maximal multi-wavelength data} (x-ray-to-radio via eROSITA, GALEX, VST, VISTA, WISE, Herschel, ASKAP and MWA). GAMA data thus far, have been used to explore merger rates \citep{depropris2014,robotham14,casteels14,davies15}, galaxy groups \citep{robotham,khosroshahi2017,raouf2019,taylor20,raouf2021}, the low-$z$ Universe \citep{madusha11,driver12, kelvin2012,loveday2012,lara-lopez2013, oliva2014,cluver2014,lange15,moffett16,beeston,bellstedt20}, and in particular the low-$z$ galaxy stellar mass function \citep{baldry2012,moffett16,wright17}: the benchmark for most numerical simulations. 

GAMA extends over 5 regions of sky covering 250\,deg$^2$, and over the past decade we have obtained $\sim 230\,000$ spectroscopic redshift mesurements with a median accuracy of $\pm 35$\,km\,s$^{-1}$ \citep{liske2015}, and complementary imaging, either directly or via collaboration, extending from the UV to the far-IR, i.e., 20-band photometry (see \citeauthor{driver2016} \citeyear{driver2016}) extending from $0.2-500\mu$m. 

To date there have been three GAMA data releases \citep{driver2011,liske2015,baldry2018}, and in this paper we now provide the fourth (GAMA DR4), which includes all redshifts (including those obtained by GAMA or by other surveys), all spectra, and our revised 20-band UV to far-IR photometry for those galaxies with spectroscopic redshifts \citep{bellstedt20b}. In addition, we provide over 30 value added data tables or Data Management Units (DMUs). These consist of many measured (Level 2) and derived quantities (Level 3), created  by the GAMA team, providing quality controlled science-ready products to the global community (see {\url{http://www.gama-survey.org/dr4/})

The paper concludes with a revised measurement of one of our headline goals, the galaxy stellar mass function and its sub-division by morphological type. It extends our previous estimates from $10^8$\,M$_{\odot} h_{70}^{-2}$ to a new lower mass-limit of $10^{6.75}$\,M$_{\odot} h_{70}^{-2}$ at $z<0.1$.

In Section\,2 we incorporate our recent image analysis \citep{bellstedt20b} of the ESO KiDS \citep{kidsdr4} and ESO VIKING \citep{viking} Public Survey data with the GAMA spectroscopic data, and explore our effective redshift completeness for each region and the combined primary regions. In Section\,3 we provide new or revised Data Management Units (DMUs) including photometric redshift estimates for all objects in our revised Input Catalogue, and morphological classifications for all objects with $z<0.08$ and $r_{\rm KiDS DR4} < 19.65$ mag. Section 4 describes the contents of Data Release\,4. In Section\,5 we provide a revised measurement of the Galaxy Stellar Mass Function ($z<0.1$), and its sub-division by morphological type ($z<0.08$). Finally, in Section\,6 we discuss the implication for the cosmic stellar mass density at $z<0.1$, including a re-normalisation from the 230\,deg$^2$ GAMA area to a 5\,012\,deg$^2$ region of the SDSS area, reducing our cosmic variance uncertainty at $z<0.1$ from 12 per cent to 6.5 per cent.

We adopt a concordance `737 cosmology’, with $(H_0,\Omega_M,\Omega_\Lambda)=(70,0.3,0.7)$ throughout, all magnitudes and fluxes are corrected for Galactic extinction, and all magnitudes are reported in the AB system. For all values which are dependent on Hubble's constant, $H_0$ we include this dependency via $h_{70}=H_0/70$\,km/s/Mpc.

\begin{figure*}
	\centering     
	\includegraphics[width=\linewidth]{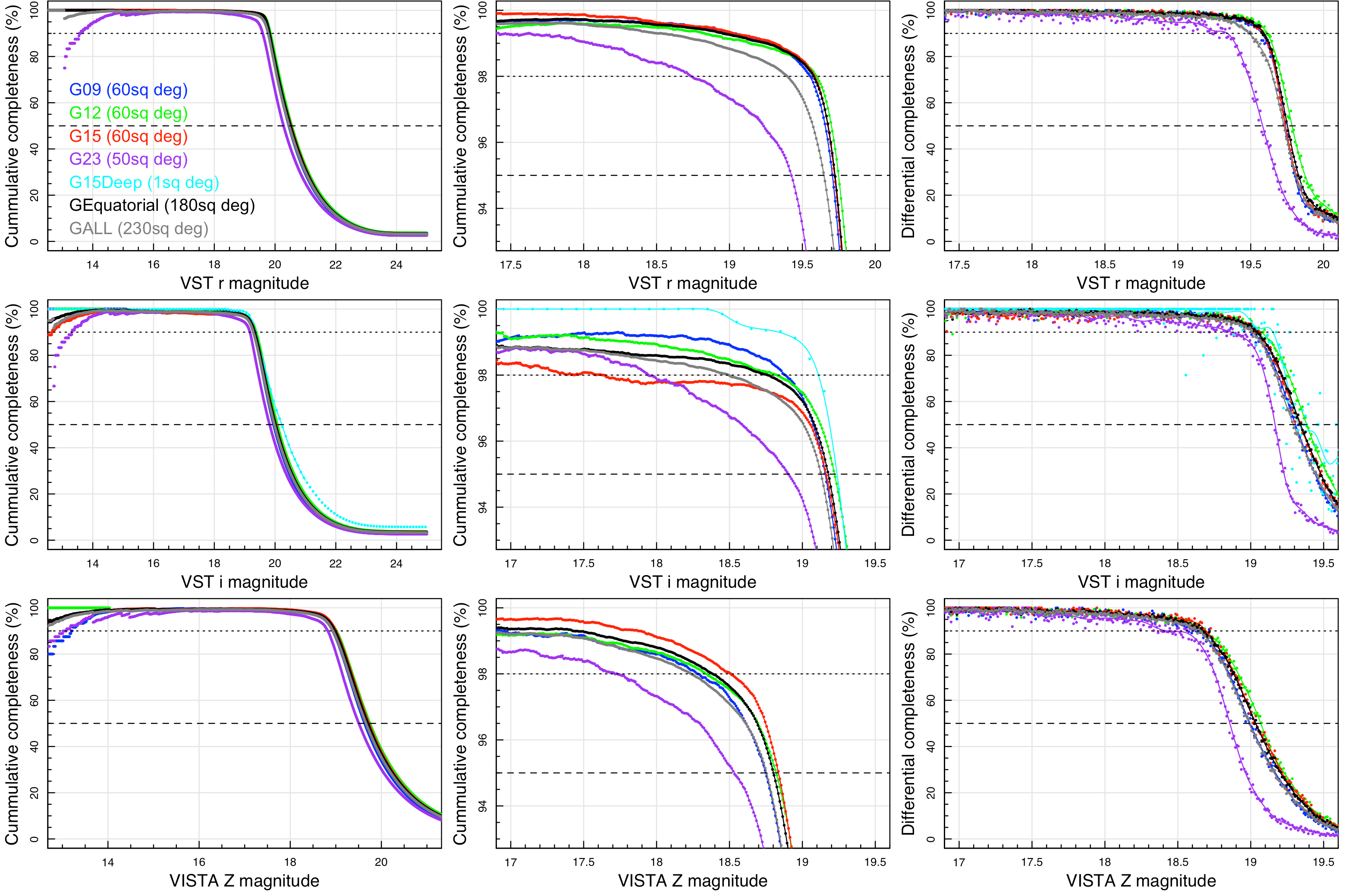}
	\caption{(left panels) The cumulative spectroscopic completeness of the GAMA survey as a function of limiting VST $r$ (upper), VST $i$ (middle) and VISTA $Z$ (lower) wavebands with the 50 per cent (dashed) and 90 percent (dotted) lines shown. (centre panels) A zoom in to the critical turn-over in completeness. The lines now show the 98 and 95 per cent completeness. (right panels) The differential completeness around the cutout flux limits, the lines show the 50 and 90 percent limits. See Table\,\ref{tab:completeness} for more precise cutoff limits. \label{fig:completeness}
}
\end{figure*}

\begin{table*}
\caption{The magnitude limits for individual GAMA fields and combinations, for a range of filters and spectroscopic completeness limits. The number of galaxies with spectroscopic redshifts within the associated magnitude limit are shown in brackets.   
\label{tab:completeness}}
\begin{tabular}{c|c|c|c|c} \hline
Field(s) & \multicolumn{4}{c}{Spectroscopic completeness limits: mag (number with reliable spec-$z$)} \\ \hline
& \multicolumn{4}{c}{$r_{\rm KiDSDR4}$-band magnitude limit to achieve a spec-$z$ completeness of: } \\ \cline{2-5}
& 50\%  & 90\%  & 95\% & 98\%  \\ \cline{2-5}
G09  & $ 20.45 $ ($ 58740 $)  & $ 19.80 $ ($ 53519 $)  & $ 19.71 $ ($ 51084 $)  & $ 19.55 $ ($ 43880 $) \\ 
G12  & $ 20.55 $ ($ 68454 $)  & $ 19.85 $ ($ 61831 $)  & $ 19.75 $ ($ 58798 $)  & $ 19.60 $ ($ 51834 $) \\ 
G15  & $ 20.52 $ ($ 65276 $)  & $ 19.81 $ ($ 58547 $)  & $ 19.72 $ ($ 56254 $)  & $ 19.58 $ ($ 50098 $) \\ 
G23  & $ 20.28 $ ($ 41415 $)  & $ 19.59 $ ($ 37842 $)  & $ 19.42 $ ($ 33388 $)  & $ 18.74 $ ($ 16033 $) \\ 
G09+G12+G15  & $ 20.51 $ ($ 192483 $)  & $ 19.82 $ ($ 173640 $)  & $ 19.72 $ ($ 165294 $)  & $ 19.58 $ ($ 146225 $) \\ 
G09+G12+G15+G23  & $ 20.46 $ ($ 233199 $)  & $ 19.77 $ ($ 210272 $)  & {\bf 19.65} $ (${\bf 195432}$)^\dagger$  & $ 19.40 $ ($ 153601 $) \\ \hline
& \multicolumn{4}{c}{$i_{\rm KiDSDR4}$-band magnitude limit to achieve a spec-$z$ completeness of: } \\ \cline{2-5}
& 50\%  & 90\% & 95\%  & 98\%  \\ \cline{2-5}
G09  & $ 19.95 $ ($ 58694 $)  & $ 19.30 $ ($ 50185 $)  & $ 19.16 $ ($ 44749 $)  & $ 18.89 $ ($ 33053 $) \\ 
G12  & $ 20.07 $ ($ 68343 $)  & $ 19.38 $ ($ 58581 $)  & $ 19.22 $ ($ 51833 $)  & $ 18.82 $ ($ 34129 $) \\ 
G15  & $ 20.02 $ ($ 64954 $)  & $ 19.32 $ ($ 54978 $)  & $ 19.16 $ ($ 48581 $)  & $ 17.62 $ ($ 8298 $) \\ 
G23  & $ 19.82 $ ($ 41320 $)  & $ 19.17 $ ($ 37799 $)  & $ 18.91 $ ($ 29834 $)  & $ 17.95 $ ($ 9801 $) \\ 
G15Deep  & $ 20.22 $ ($ 1826 $)  & $ 19.36 $ ($ 1464 $)  & $ 19.24 $ ($ 1380 $)  & $ 19.11 $ ($ 1248 $) \\ 
G09+G12+G15  & $ 20.02 $ ($ 191996 $)  & $ 19.33 $ ($ 163127 $)  & $ 19.18 $ ($ 145007 $)  & $ 18.76 $ ($ 91966 $) \\ 
G09+G12+G15+G23  & $ 19.98 $ ($ 232967 $)  & $ 19.28 $ ($ 197009 $)  & $ 19.13 $ ($ 175087 $)  & $ 18.49 $ ($ 85776 $) \\ \hline
& \multicolumn{4}{c}{$Z_{\rm VISTA}$-band magnitude limit to achieve a spec-$z$ completeness of: } \\ \cline{2-5}
& 50\%  & 90\%  & 95\%  & 98\%  \\ \cline{2-5}
G09  & $ 19.63 $ ($ 58477 $)  & $ 18.93 $ ($ 45812 $)  & $ 18.75 $ ($ 38268 $)  & $ 18.28 $ ($ 21214 $) \\ 
G12  & $ 19.75 $ ($ 67714 $)  & $ 19.01 $ ($ 54536 $)  & $ 18.82 $ ($ 46162 $)  & $ 18.34 $ ($ 27062 $) \\ 
G15  & $ 19.72 $ ($ 64617 $)  & $ 19.00 $ ($ 52161 $)  & $ 18.84 $ ($ 45712 $)  & $ 18.50 $ ($ 31654 $) \\ 
G23  & $ 19.51 $ ($ 41091 $)  & $ 18.82 $ ($ 35221 $)  & $ 18.53 $ ($ 26449 $)  & $ 17.71 $ ($ 10008 $) \\ 
G09+G12+G15  & $ 19.70 $ ($ 190736 $)  & $ 18.98 $ ($ 152305 $)  & $ 18.80 $ ($ 129629 $)  & $ 18.38 $ ($ 80050 $) \\ 
G09+G12+G15+G23  & $ 19.66 $ ($ 231471 $)  & $ 18.94 $ ($ 185366 $)  & $ 18.75 $ ($ 156176 $)  & $ 18.24 $ ($ 86546 $) \\ \hline
\end{tabular}

\noindent
$^\dagger$ Adopted GAMA Data Release 4 Main Survey (GAMA MS).
\end{table*}

\begin{figure*}
	\centering     
	\includegraphics[width=0.8\textwidth]{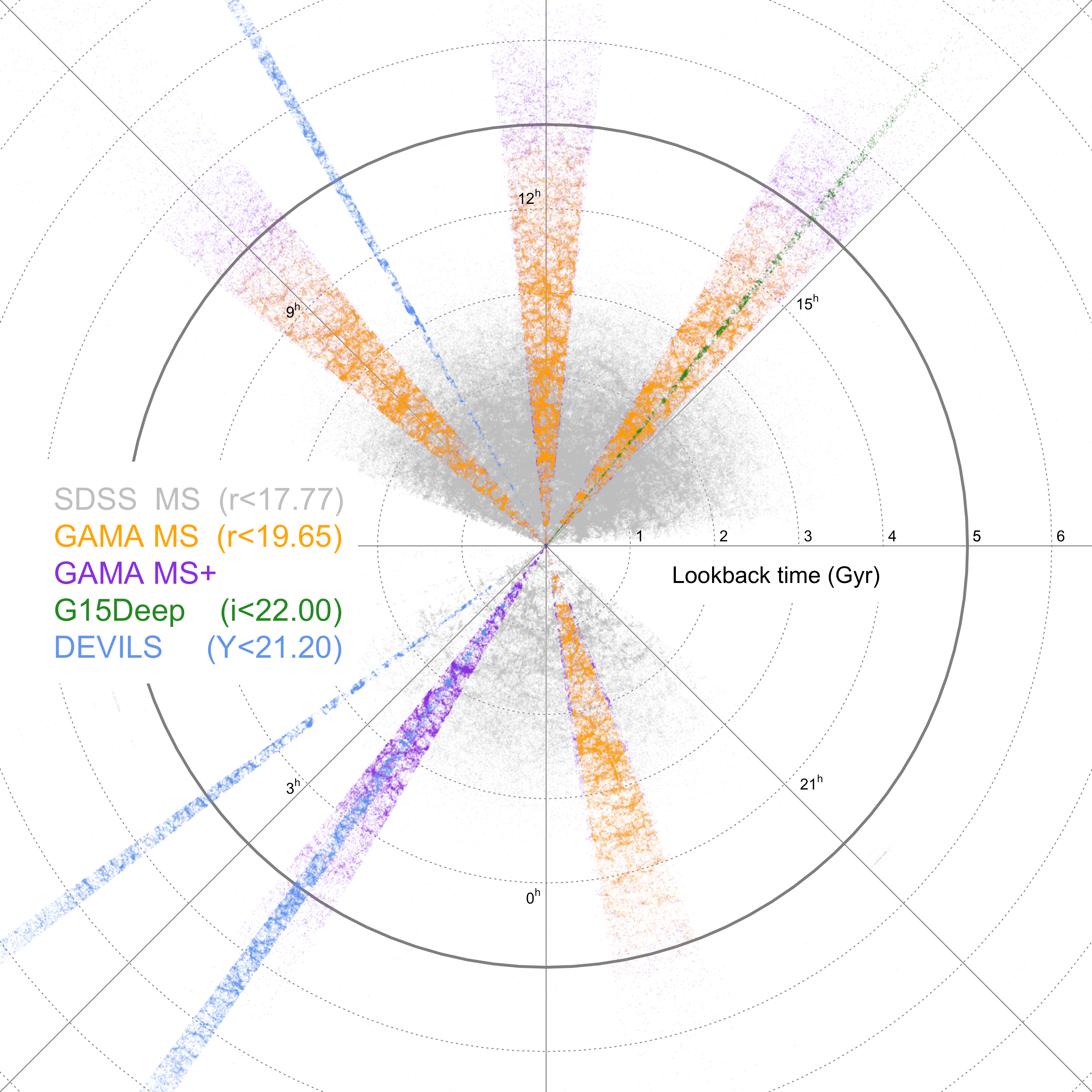}
	
	~
	
	\includegraphics[width=0.9\textwidth]{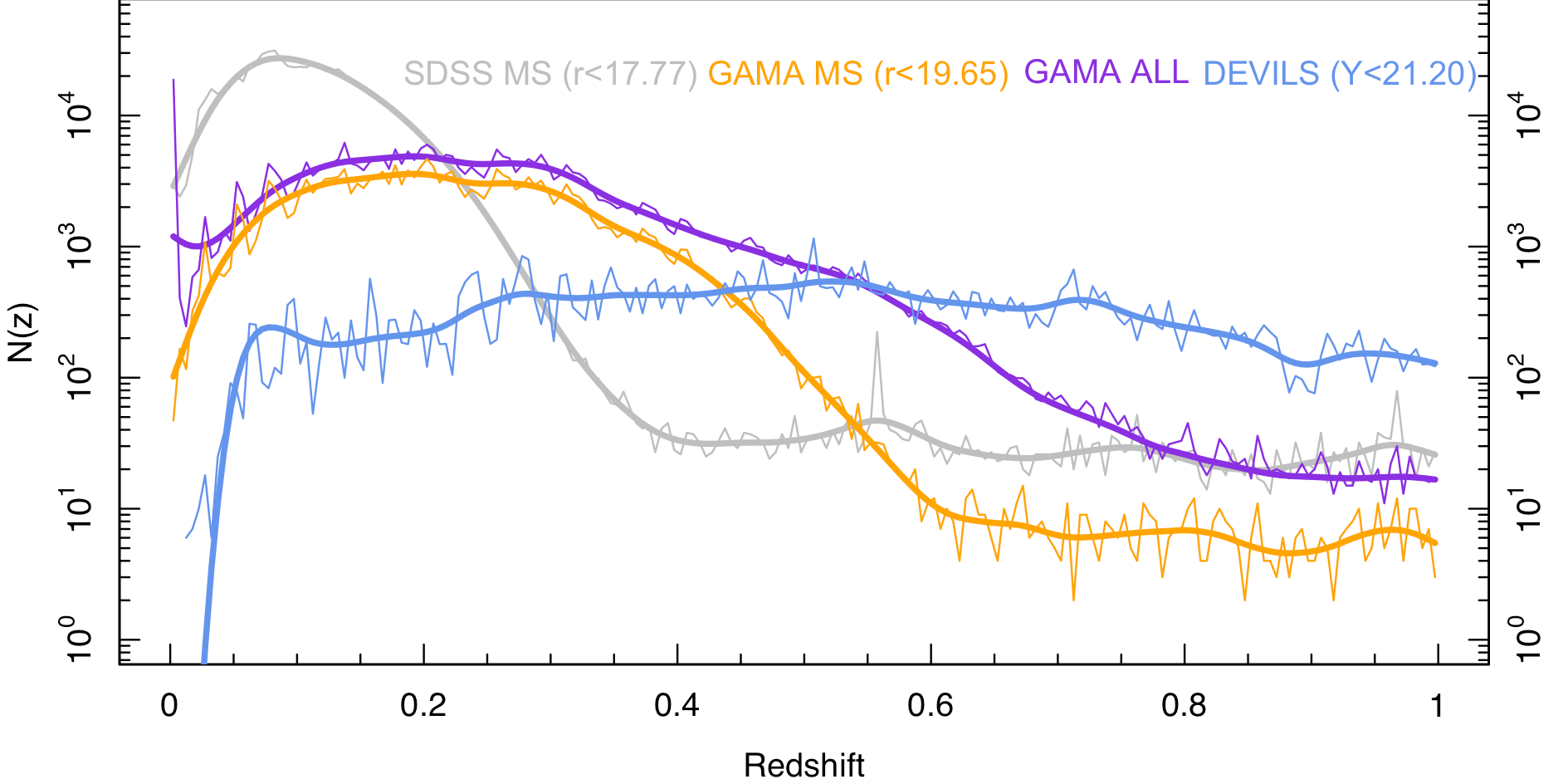}
	\caption{(upper panel) A cone diagram showing the RA and lookback time distribution of the SDSS, GAMA and ongoing DEVILS datasets. These are the currently existing high-completeness ($\sim 90$ per cent) spectroscopic surveys from which reliable merger rates and group catalogues can be constructed. (lower panel) The redshift histogram for SDSS main survey, GAMA main survey, GAMA any and DEVILS (as indicated).}
	\label{fig:cone}
\end{figure*}

\section{Unification of GAMA equatorial (G09, G12 and G15) and 23$^h$ regions (G23)} 
In this data release we include a replacement of the original SDSS (equatorial fields) and ESO VLT Survey Telescope (VST; G23) input catalogues, with deeper homogeneous $ugri$ imaging from the ESO VST Kilo-Degree Survey (KiDS) data release 4 (\citeauthor{kidsdr4} \citeyear{kidsdr4}; \citeauthor{bellstedt20b} \citeyear{bellstedt20b}). In this section we introduce the new data, and quantify the implications of replacing our underlying Input Catalogue in terms of revised magnitude limits for a range of desired spectroscopic completeness limits.

\subsection{Incorporating Kilo-Degree Survey imaging}
The Galaxy And Mass Assembly survey \citep{driver2009,driver2011} conducted its first spectroscopic observations in 2008 (see \citeauthor{liske2015} \citeyear{liske2015}). The spectroscopy was based on an initial target catalogue for the three equatorial regions \citep{baldry2010} drawn from the $6^{th}$ data release of the Sloan Digital Sky Survey \citep{sdssdr6}. Later the survey was extended with an input catalogue for the G02 region created from the CFHT Lensing Survey (CFHTLenS; \citeauthor{erben2013} \citeyear{erben2013}) and for the G23 region from the ESO VST Kilo-Degree Survey Data Release 1 \citep{kidsdr1}. These optical imaging data, along with the UKIDSS Large Area Survey near-IR data \citep{ukidss}, formed the basis of our input catalogues using angular size and concentration, combined with an additional $(J-K_s)$ colour selection to recoup compact galaxy systems (see full details in \citeauthor{baldry2010} \citeyear{baldry2010}). 

The GAMA optical/near-IR input catalogue data were later complemented by UV to far-IR imaging data from GALEX, ESO VISTA VIKING, WISE and the Herschel Space Observatory, as described in \cite{driver2016}. In particular, the large majority of the GALEX data covering the G23 region was acquired in a dedicated observing campaign as part of the All-sky UV Survey Extension, following the hand over of GALEX to Caltech and prior to
decommissioning. Ultimately the G02 spectroscopic survey was not completed to its full extent, and all available G02 data were released as part of GAMA DR3 \citep{baldry2018}. 

Recently, we have updated the original equatorial SDSS imaging data with deeper and higher spatial resolution $ugri$ observations that extend to $r=25$ mag, from the European Southern Observatory's VLT Survey Telescope Kilo-Degree Survey Data Release 4 (KiDS; \citeauthor{kidsdr4} \citeyear{kidsdr4}). The KiDS data complement the existing panchromatic data from UV to far-IR, providing consistent imaging data to unify the equatorial and G23 regions onto a single photometric and astrometric reference frame in $FUV,NUV,ugri,ZYJHK, W1234,P100/160,S250/350/500$ wavebands. The construction of the KiDS catalogue for GAMA (i.e., {\sc gkvInputCatv01}) resulted in the detection and measurement of over 18 million objects extending to $r \sim 24$\,mag and is described in detail in \cite{bellstedt20b}. 

The reanalysis of the FUV-far-IR data used a new source finding algorithm designed for these data, {\sc ProFound} \citep{profound}, and is based on the precepts of dilated isophotal segments and watershed deblending. This reanalysis included improved star-masking based on GAIA DR2 \citep{gaiadr2}, and improved Galactic Extinction corrections based on Planck dust extinction maps \citep{planckdust}. A careful verification and reconstruction of all bright and dense regions with multiple abutting segments was conducted, to ensure the integrity of the bright, large and diffuse galaxies, i.e., those that are well suited to studies with integral field units (e.g. SAMI, \citeauthor{sami} \citeyear{sami}; Hector, \citeauthor{hector} \citeyear{hector}) and/or radio observatories (ASKAP, \citeauthor{askap} \citeyear{askap}; MWA, \citeauthor{mwa} \citeyear{mwa}; etc). 

Most importantly of all, the revised catalogue now allows us to bring together the three equatorial fields and the G23 field with fully consistent and homogeneous photometric measurements from the UV to far-IR using identical facilities to comparable sensitivity limits.

\subsection{Spectroscopic completeness against KiDS \label{sec:kids}}
A key issue raised in``replacing the tablecloth'' (i.e., swapping the SDSS with KiDS photometry), is a change in the spectroscopic completeness profile from one with a sharp spectroscopic selection boundary ($r_{\rm SDSS}^{\rm Petro} <19.8$\,mag in the equatorial fields and $i_{\rm KiDS DR1}<19.2$\,mag in G23), to one with a soft edge. This is because some galaxies with SDSS photometry brighter than our original SDSS flux limit are now found to be fainter in KiDS and vice versa resulting in a less sharp cutoff in spectroscopic completeness. While the KiDS-based catalogues should represent a significant improvement over the original SDSS data, due to the depth of the VST observations, the spectroscopic completeness remains tied to the original SDSS data. 

The simplest way to overcome this is to pull back slightly in terms of the completeness limit, and to attempt to identify a revised shallower limit with a spectroscopic completeness comparable to that of the original spectroscopic survey. The advantage is the ability to use the improved photometry without the need to consider complex selection functions, the disadvantage is the inevitable loss of depth (statistical significance), as some fraction of the spectroscopic redshifts are scattered fainter and some larger fraction of sources for which redshifts were not sought are scattered brightwards. For GAMA the loss of depth is compensated for, if the G23 region can be brought into selection alignment with the equatorial fields, i.e., while the survey depth is slightly diminished (0.15 mag, see Figure\,\ref{fig:completeness} and subsequent discussion), the survey area is increased (by 28 per cent), and the overall cosmic (sample) variance (CV) is reduced by 15 per cent. 

In \cite{liske2015} we reported a combined spectroscopic completeness of 98.48 per cent to $r_{\rm SDSS}=19.8$\,mag in the equatorial fields (G09+G12+G15), 95.5 per cent in the 20 square degree high-completeness portion of the G02 region to $r_{\rm CFHT}=19.8$ mag, and 94.19 per cent to $i_{\rm KiDS DR1} < 19.2$\,mag in the G23 field. Hence we aspire, with the revised photometry, to achieve comparable completeness levels of 95 or 98 per cent. 

All G02 data were released in \cite{baldry2018} and as its panchromatic coverage is quite different we consider it no further. In Figure\,\ref{fig:completeness} we show the revised completeness of the remaining fields (coloured lines), for the equatorial regions combined (black line), and for all four fields combined (grey lines), as a function of KiDS $r$ (top panels), KiDS $i$ (centre panels) and VIKING $Z$ (lower panels) magnitudes. The left-side and centre-column panels show the cumulative distributions with the central panels representing a zoom in of the left-side panels. The right-side panels show a zoom in of the differential distribution. Table\,\ref{tab:completeness} reports the 50, 90, 95 and 98 per cent completeness limits for each of these bands, for each field and for various combinations.

For the remainder of this paper we now consider the GAMA main survey catalogue (GAMA MS) to be defined by $r_{\rm KiDSDR4} < 19.65$\,mag from the combined G09+G12+G15+G23 regions. This contains 205\,540 galaxies for which 195\,432 have reliable (i.e., $NQ>2$) redshifts (i.e., 95.1 per cent). GAMA MS+ comprises a further 135\,110 redshifts which consist of those galaxies in the G02 region, galaxies fainter than our revised limit, and galaxies on the periphery of the four main survey fields. Figure\,\ref{fig:cone} shows a cone-plot of the RA and lookback-time distribution (orange for GAMA MS and purple for GAMA MS+), along with the SDSS Main Survey (grey), and the ongoing DEVILS survey (blue). These surveys (SDSS, GAMA, and DEVILS) highlight our current {\it high-completeness} insight into the $z<1$ Universe. 

\section{New and updated Data Management Units (DMUs)}

\begin{figure*}
	\centering     
	\includegraphics[width=\textwidth]{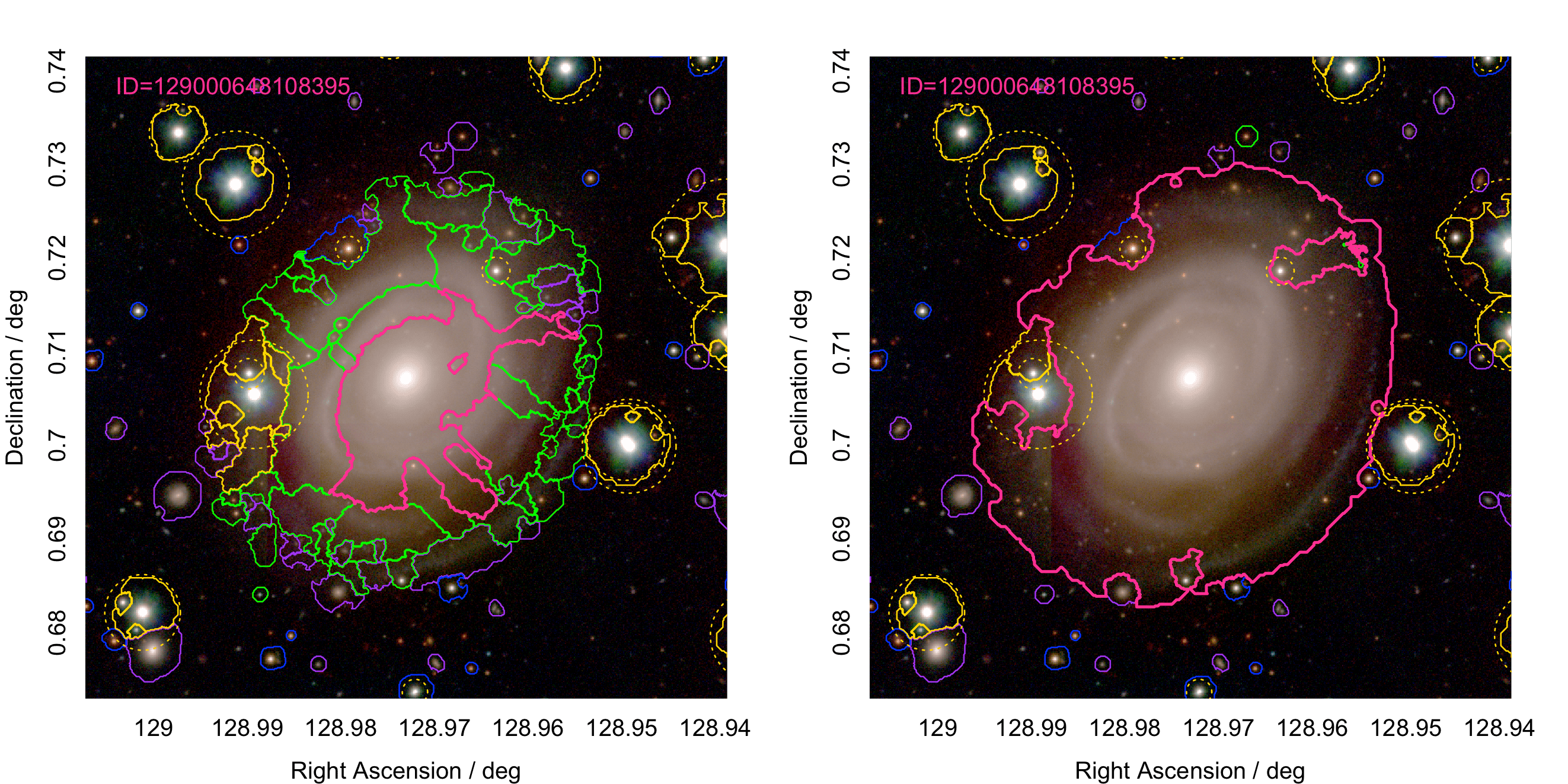}
	\includegraphics[width=\textwidth]{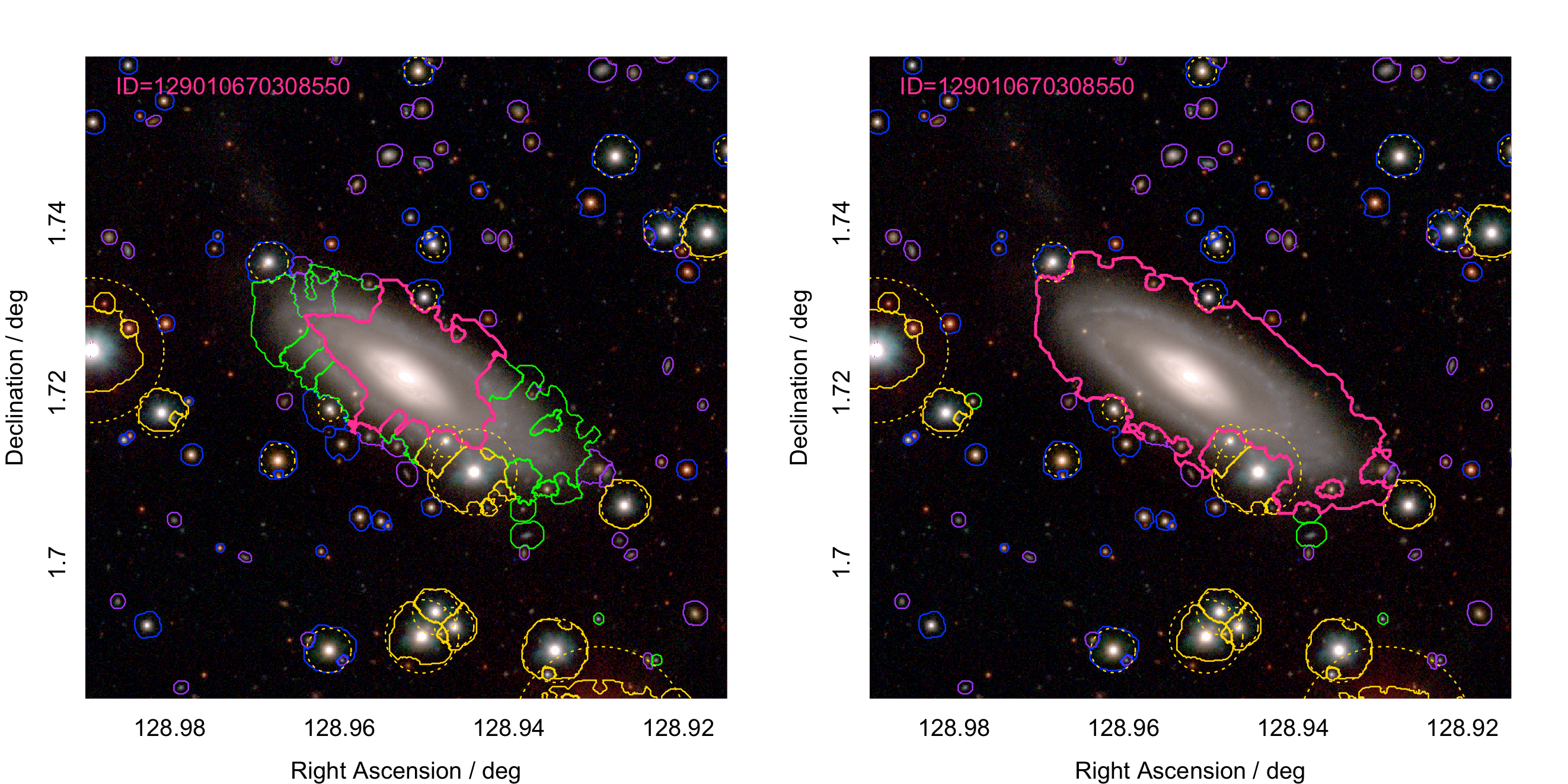}
	\caption{(left side) Two examples of nearby bright galaxies that required manual fixing and (right side) after fixing. The images shows $grZ$ colour composites with the main object shown with a magenta segment outline, ambiguous objects in green, unmasked stars in blue, masked objects in yellow, and galaxies as mauve. The dotted yellow circles show the extend of the starmasks.}
	\label{fig:fixed}
\end{figure*}

\subsection{The GAMA DR4 input catalogue v01 and v02}
In using the {\sc gkvInputCatv01} catalogue \citep{bellstedt20b} we identified a minor flaw in our galaxy rebuilding selection. This resulted in 77 large bright galaxies being heavily fragmented and 1 compact group requiring deblending. For the 77 galaxies these were typically galaxies which intersected with a bright star, and as a consequence were not selected for manual fixing (see \citeauthor{bellstedt20b} \citeyear{bellstedt20b}). While unlikely to impact on any statistical analysis, these very nearby large bright galaxies are of particular interest for a number of nearby low redshift follow-on programmes. Hence we take this opportunity to fix the apertures for these 77 galaxies, and to rerun our measurement and post-processing pipelines for these systems. As part of this process we removed 687 fragments associated with these objects, and replaced their photometry with the 77 revised and rebuilt systems to produce {\sc gkvInputCatv02}.

Figure\,\ref{fig:fixed} shows before and after images for two of these bright galaxies. We note that we also revise our far-IR photometry and our SED analysis to produce {\sc gkvProSpectv02} following the exact processes outlined in \cite{bellstedt20b} and \cite{bellstedt20}. The revised v02 catalogues are made available via GAMA DR4 and the original v01 catalogues are held in the team database (i.e., they are not included in the GAMA DR4 release). Note that the one blended group (uberID=215020829601469) we do not directly fix at this stage. However, in Section\,5 where we calculate the galaxy stellar mass function we replace this system by a bespoke reanalysis, in which we identify six Elliptical components, and reassign its total stellar mass according to their fractional flux (28, 37, 16, 14, 4 and 1 per cent). 

\subsection{GAMA G15-deep}
As part of the GAMA observing programme (July-Sept 2014), we experimented with pushing to a deeper magnitude limit of $i_{\rm SDSS} < 22$\,mag within a 1\,deg$^2$ sub-region of the G15 field ($218.5<$RA$<219.5$, $-1.0<$DEC$<0.0$). Within this region we observed 3\,241 galaxies, and reliable (i.e., $NQ>2$ with $P(z)>0.9$) redshifts were obtained using AUTOZ \citep{autoz} for 736, which includes some duplicates with GAMA MS. These deeper redshifts are potentially useful to constrain photo-$z$ efforts which extend to fainter fluxes than the GAMA MS, and hence we include them in DR4 as {\sc G15DeepSpecCatv01}. We show their location and radial distribution on Figure\,\ref{fig:cone} as the green data points. Further efforts may be made to increase the completeness in this region to complement the DEVILS survey, and further assist in the definition of photometric redshift calibration. Users interested in obtaining access or contributing to this effort should contact the GAMA Exec\footnote{gama@eso.org}.

\begin{figure*}
	\centering     
	\includegraphics[width=\linewidth]{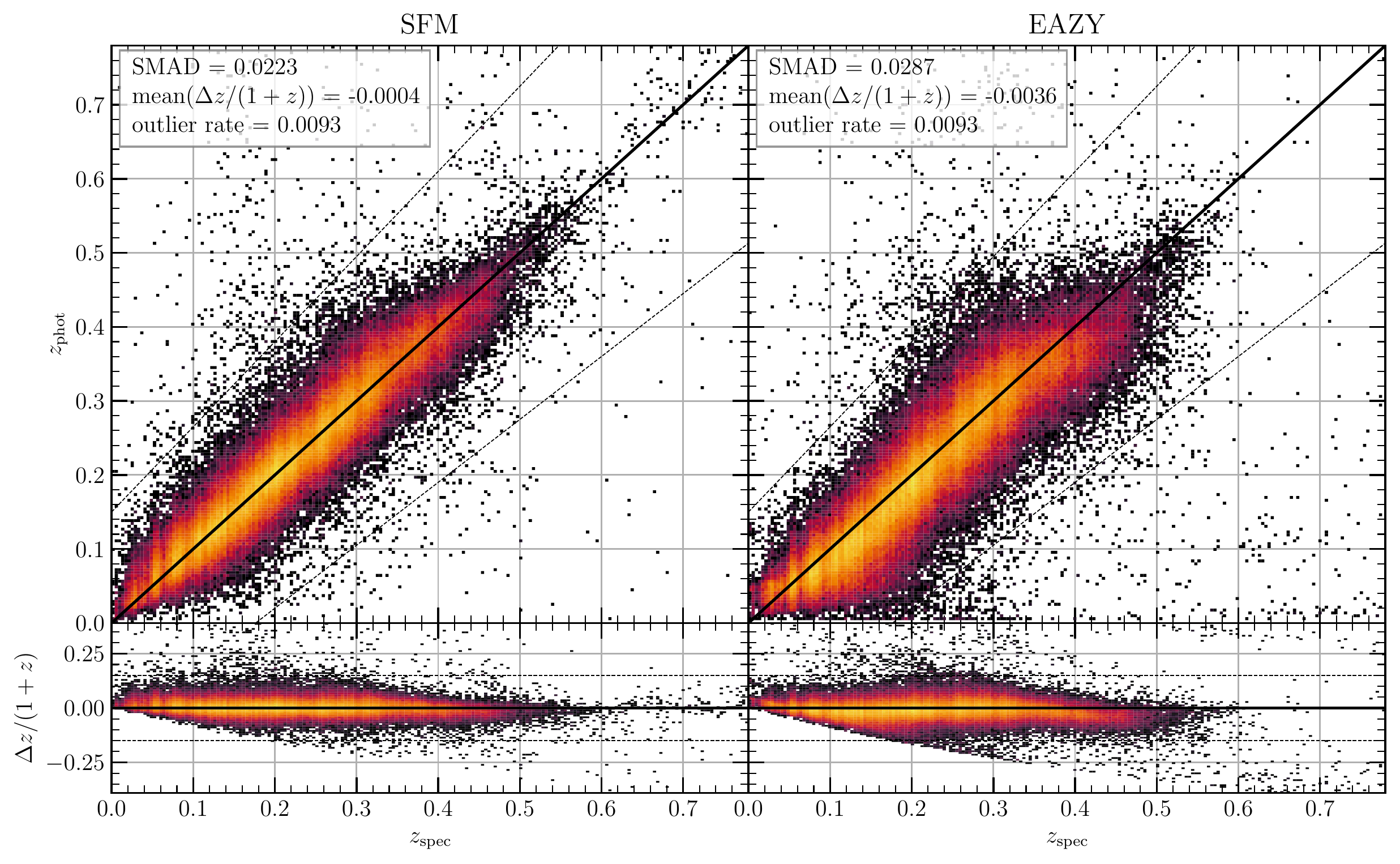}
	\caption{Comparison of spectroscopic and photometric redshifts for the GAMA sample using the scaled flux matching method on the left, and the EAZY template-matching method on the right.}
	\label{fig:photoz_accuracy}
\end{figure*}

\begin{figure*}
	\centering     
	\includegraphics[width=\linewidth]{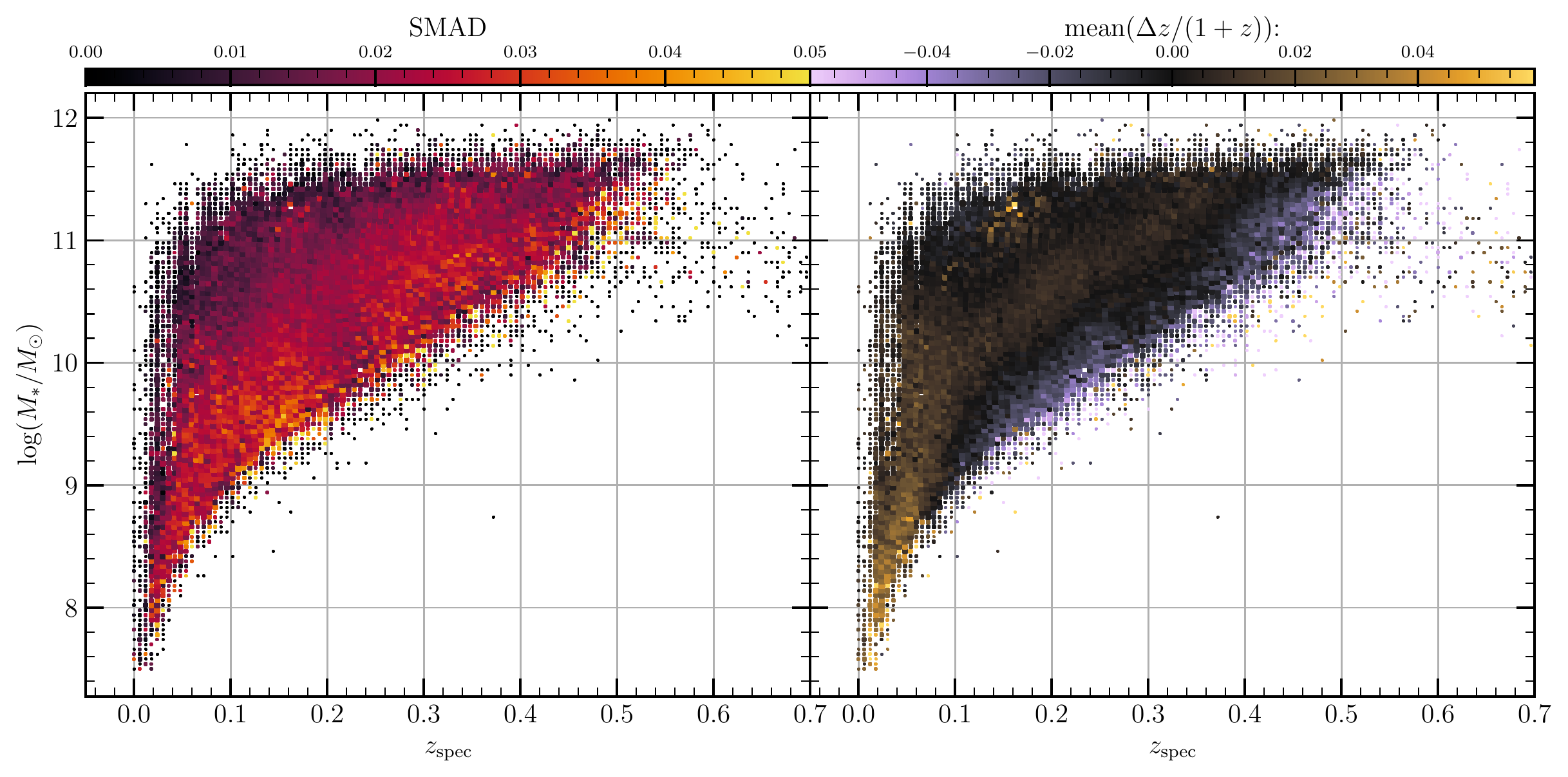}
	\caption{Demonstration of the photometric redshift accuracy for the SFM photo-$z$ sample, as compared with the GAMA spectroscopic redshifts showing the variance (left panel) and bias (right panel) as a function of stellar mass and spectroscopic redshift.
	Note that the minimal colour variation in the right-hand panel is an indication that there is very little bias of the photometric redshifts across the GAMA sample. }
	\label{fig:photoz_accuracy_mass}
\end{figure*}

\subsection{Scaled-flux matched photometric redshifts for main survey sample}
With the redefinition of the GAMA main sample to KiDS+VIKING $r<19.65$\,mag selection, a number of new galaxies are introduced for which redshifts were not sought. 
In total there are now 10\,107 galaxies without spectroscopic redshifts ($z_{\rm spec}$) within our revised magnitude limit. In order to provide an estimate of their likely redshift we employ the empirical method of Scaled Flux Matching (SFM) recently described by \cite{baldry2021} to derive photometric redshifts ($z_{\rm phot}$). 
In this method, we compare the $u$/$g$/$r$/$i$/$Z$/$Y$/$J$/$H$/$K_S$/W1/W2 fluxes of each galaxy with all other galaxies, for which redshift NQ$ > 2$ and the \textsc{ProSpect} fit likelihood is $>-60$,\footnote{This likelihood cut eliminates galaxies with very noisy SEDs from the comparison sample.} and determine match probabilities (with free normalization allowed). This matched sample consists of 222\,991 galaxies. 
Relative band errors are applied in each of the bands in quadrature, consistent with the floor values used in the \textsc{ProSpect} analysis by \citet{bellstedt20b}. 
We thus produce a redshift probability distribution function (PDF) for each object as the smoothed density kernel of all scaled templates, weighted by the data-model likelihood. This allows us to derive the maximum-probability, and also the marginalised redshift value for each object. These values are indicated by the \texttt{zmax} and \texttt{zexp} columns respectively in the relevant DMU.
An uncertainty estimate is made by determining half the 16-84th percentile range of the PDF, which is provided as \texttt{zerr}. 

The accuracy of these photometric redshifts is demonstrated in Figure\,\ref{fig:photoz_accuracy} for the overall sample in terms of the Scaled Median Absolute Deviation (SMAD) given by ${\rm SMAD}(x) = 1.4826\times {\rm median} (|x-{\rm median}(x)|)$, and the mean offset, i.e., mean[$(1+z)^{-1}\Delta z$]. 
The overall values of the SMAD and mean[$(1+z)^{-1}\Delta z$] are $0.0223$ and $-0.0004$ respectively, which represents a significant improvement over more readily used, template-based methods such as EAZY \citep[Easy and Accurate Zphot from Yale, ][see the discussion in Sec. \ref{sec:EAZY}]{eazy+++}.
We note however, that the accuracy of these redshifts is surpassed by those recently presented for the KiDS-bright sample \citep{Bilicki21} using ANNz, where SMAD and mean[$(1+z)^{-1}\Delta z$] values of $0.0180$ and $0.00012$ respectively were achieved.

In Figure \ref{fig:photoz_accuracy_mass}, we show how the SMAD and mean[$(1+z)^{-1}\Delta z$] vary across the sample as a function of both redshift and stellar mass. The SMAD values tend to be similar over the redshift range, displaying a slight trend towards higher values at lower stellar masses within each epoch. 
This highlights that the photo-z values are more precise for high-mass objects. 
The mean[$(1+z)^{-1}\Delta z$] displays more systematic variation across the sample. While the values are overall small (as is evident by the overall dark colour of the right-hand panel of Figure \ref{fig:photoz_accuracy_mass}), it is notable that out to $z\sim0.4$ the $z_{\rm phot}$ values are biased slightly high for galaxies around the median stellar mass, and beyond $z\sim0.1$, low-mass galaxies tend to have their $z_{\rm phot}$ values underestimated. 
Such trends are not apparent when assessing the bias of the sample overall. 
The $z_{\rm photo}$ values for all galaxies in the $r < 19.65$\,mag sample are provided in this release as {\sc gkvSFMPhotozv01}. 
For the sample of 10\,107 galaxies without spectroscopic redshifts, the $z_{\rm photo}$ values and the \textsc{ProSpect}-derived stellar masses, SFRs and gas-phase metallicities \citep[derived in the same manner as described by][]{bellstedt20b}} are released as {\sc gkvSFMPhotozProSpectv01}. 

\subsection{EAZY photometric redshifts for all sources}
\label{sec:EAZY}
Template-fit photometric redshift estimates have been derived for every SED in the \textsc{gkvInputCatv02} DMU using EAZY \citep{eazy+++}, in combination with the \citet{BrownSpectra} atlas of 129 nearby galaxy spectra.  

The main rationale for the choice of templates is as a complement to heavy training; the main value of these estimates lies in the use of the best available empirical templates without prejudice. Overall, we do find that the Brown et al. templates yield slightly better photoz-specz agreement in the $r_{\rm KiDS DR4} < 19.8$\,mag regime than the default EAZY template set. Hence a potential concern is that the fixed template set is not quite flexible enough to fully map the SED-z space.  We therefore experimented with two-template combinations within EAZY, and find only very minor variations in the output photo-zs, suggesting that this is not a leading source of error.  The implicit assumption in using a static empirical template set is that it covers a sufficiently wide range to contain an adequate description of the optical SED for any given target: i.e. not that galaxies do not evolve, but that a low-z analogue can be found for any high-z SED.  This will clearly not be true for rare and/or extreme populations (e.g. extremely metal-poor or sub-mm galaxies, etc.) or for classes that are not represented in the template set (e.g. quasars), but again the primary motivation here is to have a broadly applicable benchmark to complement more sophisticated future approaches.

These photometric redshift values are shown in Figure\,\ref{fig:photoz_accuracy} (right), where they are compared to the GAMA main survey spectroscopic redshifts. As for the SFM analysis the SMAD and mean[$(1+z)^{-1}\Delta z$] are derived, and found to be 0.0287 and -0.0036 respectively, and with a comparably small outlier rate.

These template fit photometric redshifts are intended as a valuable complement to those from machine learning and/or training sets, in several distinct but interrelated ways.
First, these template-fit results are grounded in astrophysics, in the sense that they are based on actual integrated spectra from real galaxies.
Second, because the process involves forward modelling the template spectra over many trial redshifts, it is straightforward to derive the full posterior PDF, $P(z)$.
Third, unlike trained approaches, they can in principle be extrapolated beyond the limit of any representative spectroscopic training/reference set.
In these ways, template fit photometric redshift estimates can be extremely useful as a sanity check for, and especially in probing potential systematic biases in, results derived in other ways.

For the purposes of SED-fitting, only the $u$--$K$ bands are used; the inclusion of the GALEX UV and WISE IR bands does not improve the $\zphot$--$\zspec$ agreement.
We also do not make use of EAZY's facility for template combination, having trialled two-template combination and found no significant improvement.
We adopt the default \texttt{eazy\_v1.0} template error function \citep{eazy+++}, with amplitude 0.5, and a 0.02\,mag `systematic' uncertainty in the photometry to soften template mismatch effects.
The redshift grid spans the range 0.004--4.3 in 209 steps, with grid steps proportional to $\log (1+z)$.
We also include an original $r$-band luminosity prior, which comes from a descriptive model of the GAMA $N(m_z, \zspec)$ distribution, extrapolated down to $r < 28$ mag.
Note that this prior operates mostly to exclude too-low redshift solutions that would lead to implausibly high luminosities.  
It therefore has a relatively large impact on the $\zphot$-$\zspec$ statistics, and plays less of a role for fainter galaxies.

The \textsc{gkvEAZYPhotoz} DMU packages the full EAZY outputs, including both maximum likelihood estimates and minimum variance estimates, evaluated with and without the luminosity prior.
The preferred redshift estimate for any given galaxy is the \texttt{z\_peak} value.
This estimator is not well documented, but is the prior-weighted, minimum variance estimate, evaluated in the vicinity of the maximum likelihood peak.
Note that because this quantity is derived by marginalising over the PDF, it will converge to some central value where there is insufficient information in the SED to properly constrain the redshift.
We have also propagated the best-fit template SED corresponding to the \texttt{z\_peak} solution.
This value is used to compute the Posterior Predictive P-Value (PPP), which is a Bayesian summary statistic that is similar to the frequentist reduced-$\chi^2$, inasmuch as it provides an indication of goodness-of-fit.
Assuming a particular model (in this case, the best-fit template SED at \texttt{z\_peak}), the PPP gives the chances of obtaining data that give a less good fit: thus 0.5 indicates the ideal fit with reduced-$\chi^2 = 1$; 0 would indicate a catastrophically bad fit; a value close to 1 would indicate overfitting.
For each galaxy, a random draw from the posterior PDF is also given as \texttt{z\_mc}; this is appropriate for describing the ensemble with Monte Carlo redshift error propagation.

The \textsc{gkvEAZYPhotoz} DMU comprises of the full photometric catalogue of 18+M sources, including artefacts and stars as well as galaxies, quasars, etc.  
Artefacts and stars can be excluded based on the photometric quality control flags, but it can also be useful to explore how the photometry is mapped to $\zphot$ in these cases.
No attempts have been made to account for SED/spectral types outside the \cite{BrownSpectra} spectral atlas; e.g. rare spectral types, broad-line AGN, or QSOs; any photometric redshift estimates for such objects are likely to be meaningless.
Further, there is the danger of some degree of contamination from such objects in any photometric-redshift-selected galaxy sample.
In addition to the main photometric redshift catalogue, this DMU also includes the full posterior $P(z)$ for every photometric detection.  
We also provide analogues of the \textsc{StellarMasses} DMU (see Sec. \ref{sec:StellarMasses} below) in a separate \textsc{gkvEAZYPhotozStellarMasses} DMU. Values are derived using both the $\texttt{z\_peak}$ and $\texttt{z\_mc}$ values, containing stellar mass estimates, restframe photometry, and ancillary stellar population properties.

\begin{figure*}
	\centering     
	\includegraphics[width=\textwidth]{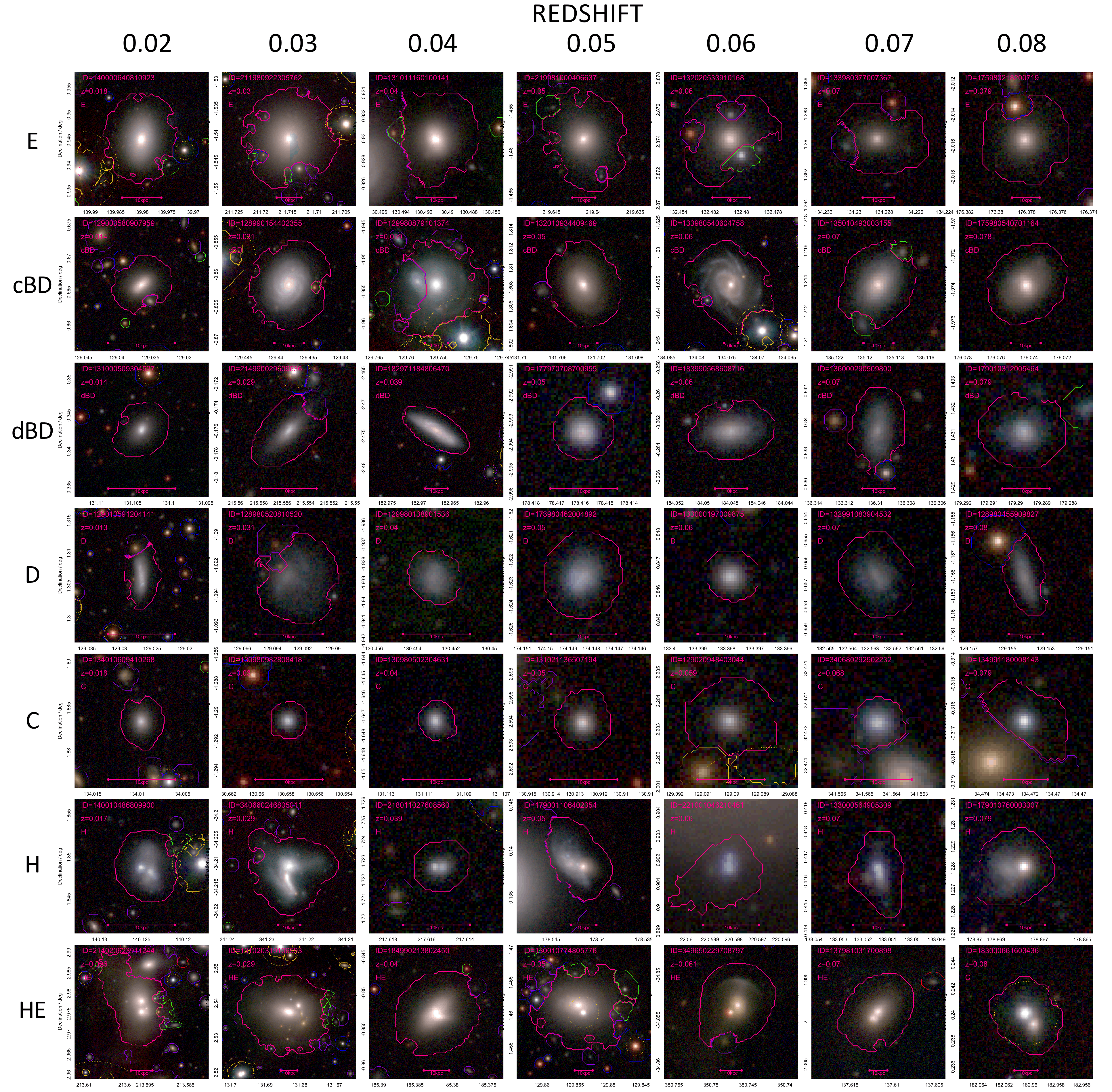}
	\caption{A sample of galaxy postage-stamp images by type and redshift, similar to those used for the morphological classification process. All stamps are displayed in $grZ$ wavebands, at a resolution of $30 h^{-1}_{70}$\,kpc $\times 30 h^{-1}_{70}$\,kpc or larger (as indicated) if the image was likely to extend outside the box. The dynamical range extends from 15 mag/sq arcsec to negative median of the stamp pixel values. This selection was chosen to highlight both high and low level features using an {\rm arcsinh} scaling. The magenta outline shows the segment defining the object from the Bellstedt et al. {\sc ProFound} catalogue. Other coloured lines indicate segments for GAIA stars (yellow), faint stars (blue), GAMA galaxies (mauve), and artefacts (green).  \label{fig:montage}
}
\end{figure*}

\subsection{Morphological classification of the $\mathbf{z<0.08}$ GAMA sample \label{sec:morph}}
The improved resolution of KiDS data (FWHM $\sim 0.7''$), over SDSS (FWHM $\sim 1.5''$), along with the deeper surface brightness limit ($\Delta \mu \sim 2$\,mag per square arcsec), allows us to review our previous morphological classifications. It also allows us to produce new and consistent morphological classifications across the four GAMA primary regions, and to our new nominal completeness limit of $r_{\rm KiDS DR4} < 19.65$\,mag (see Table\,\ref{tab:completeness}). We adopt a redshift limit of $z<0.08$ (at which point $1'' = 1.51$\, $h^{-1}_{70}$\,kpc), which is matched to the redshift selection of our bulge-disc decomposition DMU (Casura et al.~2022). To perform the classifications we create postage stamp images from $grZ$ imaging (i.e., VST \& VISTA). The image stamps are extracted at 30\,$h^{-1}_{70}$\,kpc\,$\times$\,$h^{-1}_{70}$\,$30$\,kpc scales, and with {\rm arcsinh} scaling extending from $\mu_r =15$\,mag per square arcsec to the sky level. For galaxies that overflow the spatial range we increase the stamp size accordingly based on its {\sc R100} value from {\sc gkvInputCatv02}, which represents the approximate elliptical semi-major axis containing 100 per cent of the flux. Figure\,\ref{fig:montage} shows a random selection of images similar to those used for the classification process. Within these limits ($r_{\rm KiDS DR4} < 19.65$\,mag and $z < 0.08$) we have 15\,330 galaxies which we classify into: 

\begin{description}
\item{\hspace{-0.5cm} \bf E:} an early-type system with a single visual component
\item{\hspace{-0.5cm} \bf cBD:} a two-component system with a compact high-surface brightness bulge
\item{\hspace{-0.5cm} \bf dBD} a two-component system with a diffuse or extended bulge (or bulge complex)
\item{\hspace{-0.5cm} \bf D:} a late-type system with a single visual component
\item{\hspace{-0.5cm} \bf C:} a compact system too small to accurately classify
\item{\hspace{-0.5cm} \bf H:} hard to classify due to extreme asymmetry (including merging components)
\item{\hspace{-0.5cm} \bf HE:} hard to classify but the underlying galaxy is an early-type with a single visual component
\item{\hspace{-0.5cm} \bf FRAG}: fragment of a galaxy
\item{\hspace{-0.5cm} \bf STAR:} stellar-like and most likely not a galaxy 
\end{description}

We note that the HE class specifically denotes early-types with what appear to be multiple cores, indicative of late-time major mergers, multiple galaxies within the halo (i.e., a compact group), or possible line-of-sight coincidences. In most of the discussion going forward we combine the E and HE classes, but keep the distinction in the catalogue in case someone is interested in quickly finding multiple-cored early-types.

\begin{figure}
	\centering     
	\includegraphics[width=\columnwidth]{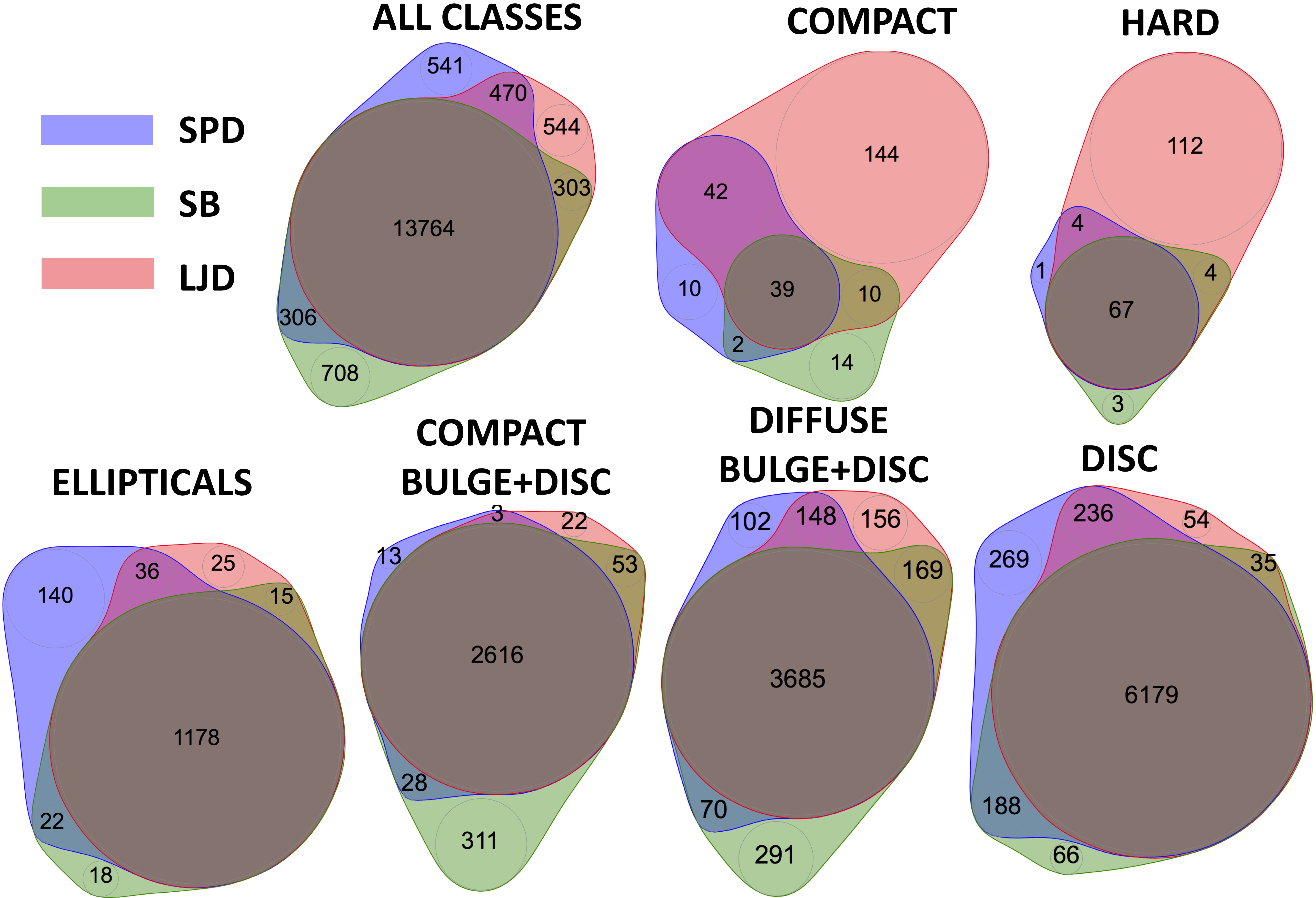}
	\caption{Venn diagrams showing the consistency and biases of the three classifiers, as indicated for the total sample (top left), or sub-classes as labelled (note that the Elliptical includes both the E and HE classes).}
	\label{fig:classes}
\end{figure}

The classification process we follow is similar to that described in \cite{Hashemizadeh2021}. First we distribute the galaxies into classification directories based on criteria such as colour, size, and mass. We then assign a classifier to each directory who extracts objects for which the classification is wrong or uncertain. These are assigned to a temporary classification folder. The custodian of each class views the temporary classification folder for their class and either accepts or rejects the classification into their master set. The process is repeated until all objects are assigned. As in \cite{Hashemizadeh2021} this process was found to be flawed, as ultimately one person is responsible for each class, and their exact definition of where the boundaries lie will vary. There is also no ability to assess the accuracy of the classifications. Hence we implemented a final phase in which {\it all} classifications were reviewed and reassigned independently by SPD, S(abine)B, and LJD. This resulted in three fully independent sets of classifications allowing an assessment of classification accuracy.

\begin{figure}
	\centering     
	\includegraphics[width=\linewidth]{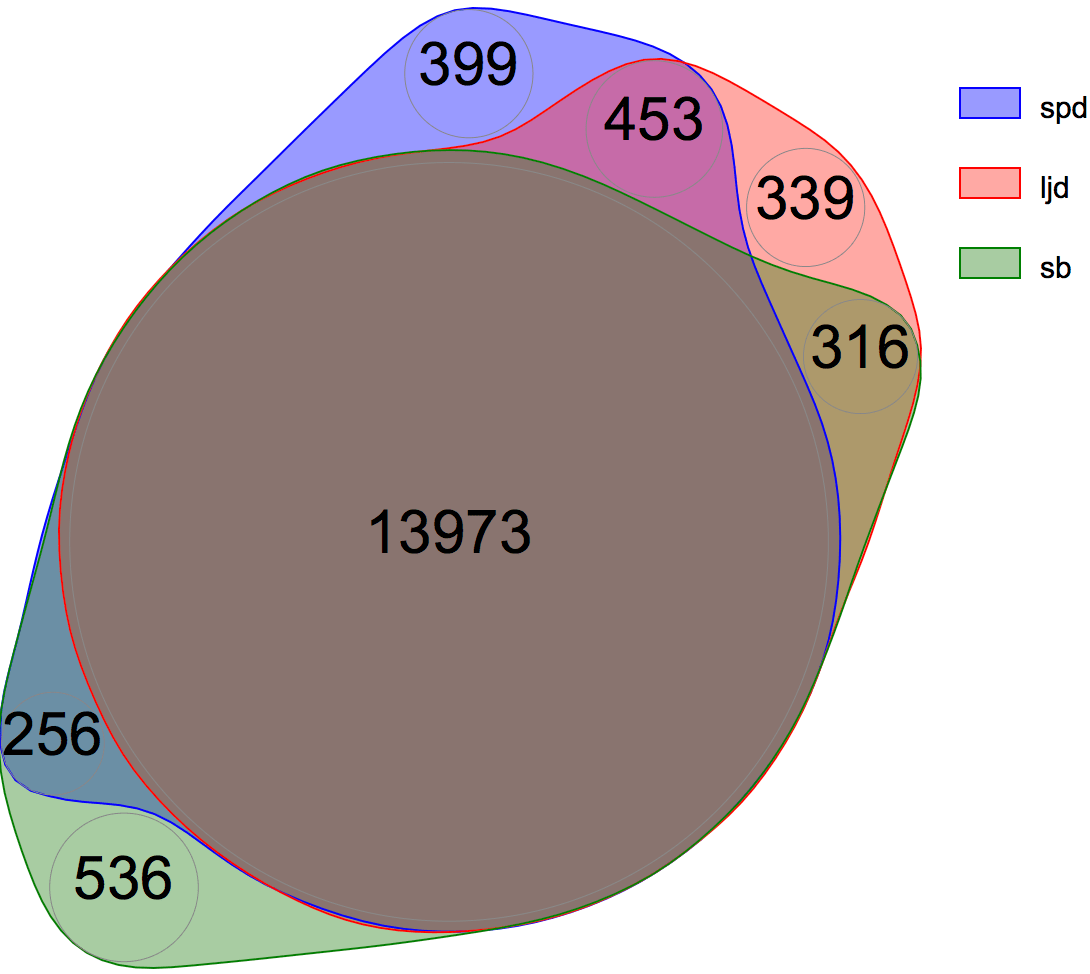}
	\caption{A Venn diagram showing the consistency of the classifications when combining the E+C+HE, D+H and dBD+cBD classes.}
	\label{fig:classes2}
\end{figure}

Figure\,\ref{fig:classes} shows the resulting Venn diagrams for our three classifiers, and the full classification set and for each of the 6 sub-classes (having merged the E+HE classifications and removing the very few objects classified as STAR or FRAG). In general the agreement is at the 90 per cent level throughout. 
From Figure\,\ref{fig:classes} we can see some consistent disparities between the classifiers with SPD having classified more Ellipticals than LJD or SB (denoted by the blue shading in Figure\,\ref{fig:classes} lower left). LJD identified more objects as Compact or Hard (orange shading), and SB perhaps has a slightly different dBD/cBD boundary definition (green shading). For the final classification we take the majority decision, or in the rare case of a three-way disagreement, a final review and decision is made by SPD. The morphology DMU ({\sc gkvMorphologyv02}) contains the starting classifications, classifications after the initial sort, the classifications of SPD, SB, and LJD and the final adopted classification. 

\begin{figure}
	\centering     
	\includegraphics[width=\linewidth]{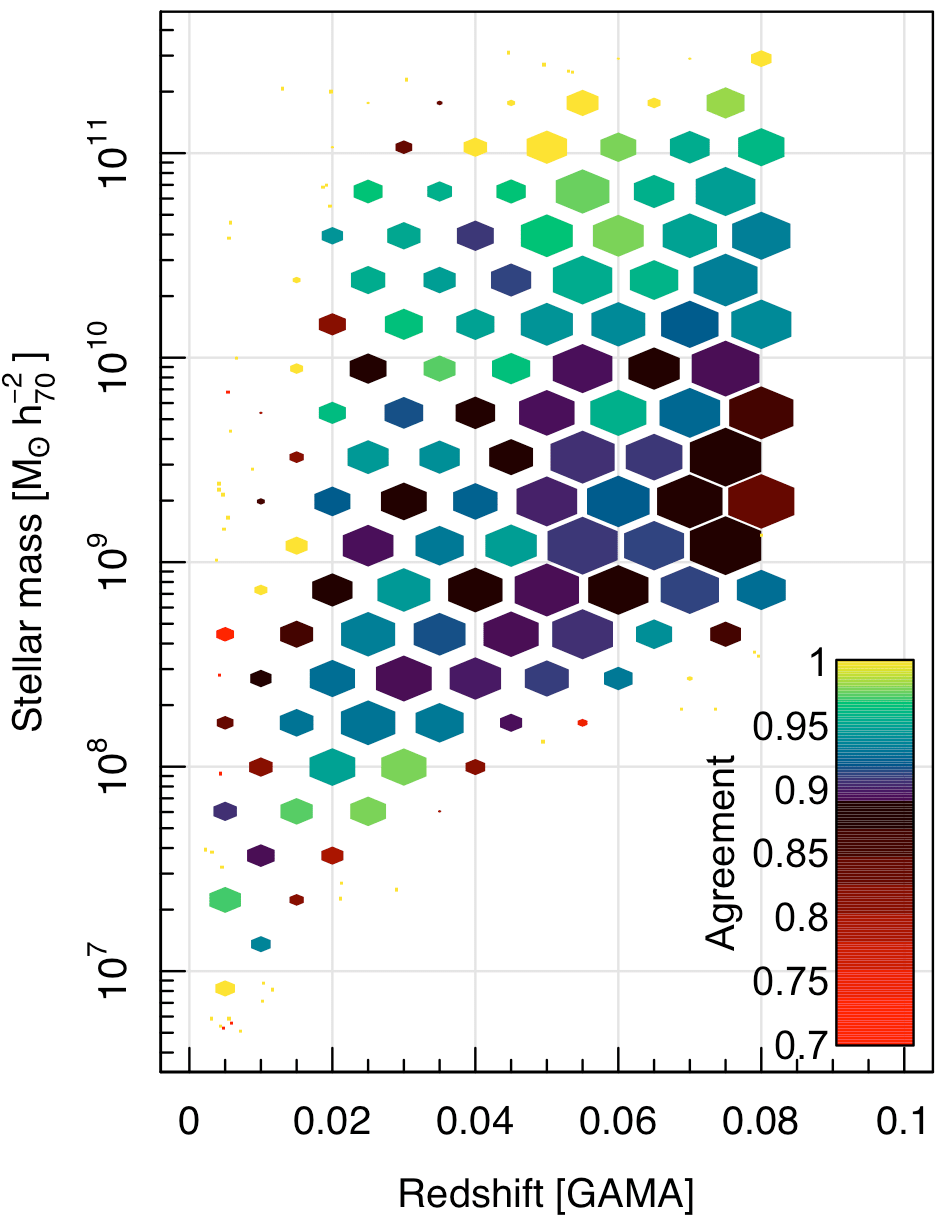}
	\caption{The morphological accuracy as a function of redshift and stellar mass. The colour of the symbol indicates the degree of agreement (see colour scale) and the size of the symbol represents the number of objects in that cell. In general all bins are yellow, green, blue or black indicating better than 87.5 per cent agreement across the full parameter range.}
	\label{fig:classes3}
\end{figure}

Note that as part of this process we have attempted to divide the double component systems into those with a compact-bulge (cBD), or diffuse-bulge (dBD). This classification into cBD and dBD is based solely on the visual appearance of the bulge-component, as either high surface-brightness and point-like (cBD), or low surface-brightness and extended (dBD). In due course comparisons to IFU data such as that drawn from the SAMI survey \citep{sami} can be made to determine the veracity of these sub-classifications.

 If we combine the dBD and cBD classes into a single BD class, that one assumes the Compact classes are predominantly poorly resolved early-types, and the Hard class are predominantly morphologically disturbed late-types (as inspection suggests, see also Figure\,\ref{fig:montage}), then the overall classification consistency changes to that shown in Figure\,\ref{fig:classes2}. This represents a surprisingly modest improvement suggesting that our division into 6 galaxy classes is meaningful. Note that only the spec-$z$ sample is included in this analysis, as the photo-$z$ sample was added later, and LJD did not classify this subset. Hence, the photo-$z$ sample is essentially the classifications of SPD alone (and which will either agree or disagree with those of SB). Nevertheless the overall agreement of the classification process, for the spec-$z$ sample only, is 13\,764/15\,081 (91.3 per cent), or 13\,974/15\,081 (92.7 per cent) if classes are merged as E=E+HE+C, D=D+H, and BD=cBD+dBD.

To explore whether our classification accuracy is biased in stellar mass or redshift, Figure\,\ref{fig:classes3} shows how the agreement varies with these two parameters. The scale is set so that 87.5 per cent agreement is coloured black and anything below varies from dark red to bright red, while higher agreement ranges from mauve to blue to green to yellow (100 per cent agreement). Here agreement is defined for each galaxy as 0.0 if all classifiers disagree, 0.5 if two classifiers agree, and 1 if all three classifiers agree. The final value, for each bin, is then the mean of these agreement values. The size of the symbol for each  bin denotes the number of objects in that bin on a logarithmic scale. Agreement across the M$*$-$z$ plane is generally consistent and above 87.5 per cent in almost all bins. There is a slight bias towards lower agreement at the lower-mass high-$z$ limit, i.e., in the direction of decreasing signal-to-noise, but still well above 87.5 per cent throughout. Hence we conclude that our morphological classifications are robust to $\ge 90$ per cent over the majority of the $M-z$ plane that we are sampling. Nevertheless, we note and acknowledge that morphological classification is an inexact and subjective process, but useful in informing whether the currently available data quality demands a two-component or one-component decomposition. 

\subsection{Stellar mass estimates and stellar populations}
\label{sec:StellarMasses}
Since DR3, the code for stellar mass estimation that was first described in \citet{taylor2011} has been completely refactored.  
Compared to \citet{taylor2011}, the most significant change is to weight the observed SEDs such that the stellar population synthesis (SPS) modelling is done using an approximately fixed wavelength range of 3000--11000${\rm\AA}$. 
The modelling assumes \citet{BC03} stellar evolution models, assuming a \cite{Chabrier2003} stellar initial mass function (IMF) and the \cite{Calzetti2000} dust curve. 
The SPS models used in the fitting are defined via a static grid in four parameters \citep[see Section 3.1 of][]{taylor2011}; namely: time since formation (i.e.\ age; $8 \le \log t_\mathrm{form} \le 10.1$); $e$-folding time for the (exponentially declining) star formation history ($7.5 \le \log \tau < 10$); stellar metallicity ($ -4 \le \log Z \le -1.3$); and dust attenuation ($0 \le A_V \le 2.43$).

The values of all derived parameters given in the DMU, including the formal uncertainties, have been derived in a Bayesian way \citep[Sections 3.2-3.4 of][]{taylor2011}, with flat priors in $t_\mathrm{form}$, $\log \tau$, $\log Z$, and $A_V$.
For DR4, the StellarMasses DMU has been updated to include stellar mass and stellar population parameters based on all the major photometric catalogues included within the GAMA database, including: Source Extractor photometry from the Panchromatic Data Release \citep[PDR][]{driver2016}, matched aperture photometry from LAMBDAR \citep[][]{wright2016}, and the latest ProFound photometry \citep{bellstedt20b} used in this paper, as well as using SDSS- or CFHT-derived photometry in the G02 field.

The differences between these simple estimates and the more sophisticated ones from ProSpect are small: random scatter of 0.13 dex; systematic offset (ProSpect masses being heavier) of 0.06 dex \citep[see Fig.\,34 in][and associated discussions]{prospect}.
Compared to ProSpect, the principal virtue of these stellar mass estimates is their simplicity. Using only optical--NIR photometry and simple star formation histories makes it straightforward for other surveys and teams to derive directly comparable results. 
In other words, they provide a practical basis for robust cross-survey comparisons.
For example, \citet{taylor2011} has shown very good agreement (random scatter of 0.07 dex; systematic offest of 0.01 dex) between our $M/L$s and those used by SDSS. 

\subsection{Velocity dispersions}
With DR4, we fill a long-standing gap in the GAMA dataset with the inclusion of central stellar velocity dispersions as measured from 1D spectra.
As a reflection of the depth of a galaxy's central potential well, modulo structure, velocity dispersion is a valuable complement to stellar mass estimates, and as a tracer of galaxy formation/stellar assembly history \citep[e.g.,][]{sheth2003, bernardi2010, taylor2010, bezanson2011}.
The addition of velocity dispersions into the GAMA database is particularly powerful, as they can be connected to all the many other global galaxy properties GAMA provides, including SED-derived masses, ages, SFRs, etc.;\ spectral absorbtion and emission diagnostics as tracers of age, metallicity, SFR, etc.;\ morphology, sizes, and S\'ersic parameters from optical-NIR imaging; environmental metrics, group associations and masses; and more.

In brief: the velocity dispersion values are derived using pPXF \citep{cappellari2017} with the MILES stellar spectral library \citep{sanchez2006, Falcon2011} as templates.
Following \cite{bezanson2018}, we use both multiplicative and additive Legendre polynomials for broad continuum subtraction of both observed and template spectra, to account for potential errors in spectral background subtraction and flux calibration.
A two-pass scheme is used to identify and account for strong emission lines when fitting to the continuum: in the first pass, we complement the stellar templates with 8 templates for the main emission lines; then in the second and final measurement, we retain only those lines with significant ($> 5 \sigma$) detections. 
Measurements are made and reported for all spectra in the SpecObjAll DMU that have originated from GAMA, SDSS \citep{ahn2014}, 2dFGRS \citep{colless01}, and 6dFGRS \citep{jones2004, 6dfgs} and which have median continuum S/N $> 10$ over 6383---6536${\rm\AA}$, as reported in the SpecLineSFRv05 DMU \citep{hopkins2013, gordon2017}.
The main challenge to overcome has been the need to calibrate/cross-validate measurements based on the heterogenous set of spectra available.
As well as direct comparison to measurements by \cite{said2020}, we have used both intra- and inter-survey comparisons to quantify/calibrate random and systematic errors in the measurements as a function of S/N, velocity dispersion, or survey (see Table~\ref{tab:specres}).
At a median S/N of 10 ${\rm\AA}^{-1}$, typical formal errors are $\sim 0.06$ dex for GAMA spectra, versus $\sim 0.03$ for SDSS spectra, and $\geq 0.1$ for both 6dFGS and 2dFGRS, and scaling approximately inversely for higher S/N.
We caution that measurements from 2dFGRS spectra show greater systematic variations when compared to other data sources, presumably related to its coarser spectral resolution. 
A full description of the new {\sc VelocityDispersions} DMU will be given by Dogruel et al.\ (in prep.).

\begin{table*}
 \begin{center}
 \caption{Summary of the VelocityDispersions DMU, including indicative spectral ranges and resolutions for the four spectral data sources.}\label{tab:specres}
 \begin{tabular}{lcccc}
 \hline
 Data Source & Spec.\ Range & Spec.\ Res.\ & Num.\ Spectra & Num.\ Galaxies \\ 
GAMA & 3730--8850 \AA & 4.4 \AA & 88504 & 85687 \\ 
SDSS & 3600--10300 \AA & 3.5 \AA & 26818 & 23122 \\ 
2dFGRS & 3600--8000 \AA & 9.0 \AA & 14720 & 13782\\ 
6dFGS & 3950--7600 \AA & 6.4 \AA & 974 & 952 \\ 
Total & & & 131016 & 111830 \\ \hline
 \end{tabular}
 \end{center}
\end{table*}

\section{GAMA Data Release 4}
Tables~\ref{tab:dmus}\,\&\,\ref{tab:dmus2} show the DMUs provided as part of GAMA Data Release 4. These are downloadable FITS tables that have been vetted via our internal quality control process and accessible via the GAMA DR4 Schema Browser, along with accompanying documentation. Any dependencies on other DMUs are clearly provided, along with the reference describing the production of the DMU (see Tables~\ref{tab:dmus}\,\&\,\ref{tab:dmus2} Col 4). Note that in most cases the DMUs have been updated from the original versions (see version numbers, Tables~\ref{tab:dmus}\,\&\,\ref{tab:dmus2} Col 2), but the methodologies remain as detailed in the papers listed in the final column of Tables~\ref{tab:dmus}\,\&\,\ref{tab:dmus2}.

\subsection{Data access and good usage policy}
All data are available in the form of downloadable DMUs from the GAMA DR4 website \url{http://www.gama-survey.org/dr4} which also contains a number of basic functions allowing for DMU downloads via the Schema Browser, SQL searches, table merging, image extraction, and a single object viewer. We kindly request that researchers that make use of these data products try to adhere to the following guidelines:

\begin{description}
\item{\hspace{-0.5cm} (1)} List the DMU name and version number of any DMU used, along with the specific column names to ensure reproduceability.
\item{\hspace{-0.5cm} (2)} Consider contacting one of the DMU authors directly, to ensure proper usage of the DMU.
\item{\hspace{-0.5cm} (3)} Include the standard GAMA acknowledgement given at \url{http://www.gama-survey.org/pubs/ack.php}
\item{\hspace{-0.5cm} (4)} Reference the key GAMA survey description papers:
\begin{description}
\item{(i)} GAMA in general: \cite{driver2011}
\item{(ii)} GAMA equatorial input catalogues: \cite{baldry2010}
\item{(iii)} The GAMA spectroscopic pipeline: \cite{hopkins2013}
\item{(iv)} The GAMA redshift measurements: \cite{liske2015}
\item{(v)} The GAMA Data Release 4: Driver et al. (2021), i.e., this paper
\item{(vi)} Any DMU references as indicated in Tables\,\ref{tab:dmus}\,\&\,\ref{tab:dmus2}
\end{description}
\end{description}

\begin{table*}
\caption{Data Management Units (DMUs) specifically built for the GAMA Date Release 4 (and prefixed with gkv for GAMA/KiDS/VIKING), along with version numbers, key individuals responsible for creating the DMU, and the published reference which provides the detailed description of how the DMU was produced. \label{tab:dmus}}
\begin{tabular}{lcp{3cm}p{5.0cm}l} \\ \hline
DMU name & version & creators/contacts & description & Reference \\ \hline
\multicolumn{5}{c}{GAMA/KiDS/VIKING DMUs in the DR4 database} \\ \hline
{\sc gkvInputCat} & v02 & Bellstedt, Driver, Robotham & {\sc ProFound} photometry in FUV, NUV, $ugri$, $ZYJHK$, W1,W2 bands & \cite{bellstedt20b} \\
{\sc gkvSpecCat} & v02 & Liske, Baldry & Spectroscopic redshifts & This paper \\
{\sc gkvScienceCat} & v02 & Driver, Bellstedt, Robotham & Main survey selection including z's & \cite{bellstedt20b} \\
{\sc gkvFarIR} & v02 & Bellstedt, Robotham & {\sc ProFound} fluxes in W3, W4, P150, P180, S250, S350, S500 bands & \cite{bellstedt20b} \\
{\sc gkvSFMPhotoZ} & v02 & Bellstedt, Robotham, Baldry & Probalistic photo-$z$'s for all $r_{\rm KiDSDR4}<19.65$ & \cite{baldry2021} \\ 
{\sc gkvProSpect} & v02 & Bellstedt, Robotham & {\sc ProSpect} derived info (M$*$, SFR etc) for {\sc GAMA MS} & \cite{bellstedt20} \\
{\sc gkvEAZYPhotoZ} & v02 & Taylor & EAZY photo-$z$'s for {\it all} objects in gkvInputCatv02 & This paper \\
{\sc gkvStellarMasses} & v01 & Taylor & Stellar Mass estimates for {\it all} objects in gkvInputCatv02 with reliable spectroscopic redshifts. & This paper \\
{\sc gkvPhotoZStellarMasses} & v01 & Taylor & Stellar Mass estimates for EAZY and SFM photo-$z$'s for {\it all} objects in gkvInputCatv02 & This paper \\
{\sc gkvMorphology} & v02 & Driver, Bellstedt, Davies & Visual morphologies to $z<0.08$ & This paper \\ \hline
\multicolumn{5}{c}{GAMA/KiDS/VIKING DMUs in preparation} \\ \hline
{\sc gkvProFit} & v01 & Casura, Liske & Profit analysis of all main survey galaxies & Casura et al (in prep.) \\
{\sc gkvGroups} & v01 & Bravo, Robotham & F-o-F group catalogue for the revised GAMA main survey & Bravo et al. (in prep.) \\
{\sc gkvFilaments} & v01 & Gurvarinder, Taylor, Cluver & Filament and tendril catalogue for the revised main survey & Gurvarinder et al (in prep.) \\ \hline
\end{tabular}
\end{table*}

\begin{table*}
\caption{{\sc GAMAII} Data Management Units (DMUs) included in the GAMA Date Release 4, along with version numbers, key individuals responsible for creating the DMU, and the published reference which provides the detailed description of how the DMU was produced. \label{tab:dmus2}}
\begin{tabular}{lcp{3cm}p{5.0cm}l} \\ \hline
DMU name & version & creators/contacts & description & Reference \\ \hline
\multicolumn{5}{c}{GAMAII DMUs in the DR4 Database} \\ \hline
{\sc EqInputCat} & v46 & Baldry & Input catalogues for the spectroscopy of the equatorial regions & \cite{baldry2010} \\
{\sc G02InputCat} & v07 & Baldry & Input catalogues for the spectroscopy of the G02 region & \cite{baldry2018} \\
{\sc G23InputCat} & v11 & Moffett, Driver & Input catalogues for the spectroscopy of the G23 region & \cite{liske2015} \\
{\sc SpecCat} & v27 & Liske, Baldry & All redshifts in or near the GAMA regions & \cite{liske2015} \\
{\sc SpecLineSFR} & v05 & Owers & Line flux and equivalent width measurements for selected GAMA II spectra &  \cite{gordon2017} \\
{\sc LocalFlowCorrection} & v14 & Baldry & Redshifts from SpecCat translated into various frames & \cite{baldry2012} \\
{\sc kCorrections} & v05 & Loveday & k-corrections in FUV, NUV, $ugriz, ZYJHK_s$ bands for all galaxies in the equatorial regions & \cite{loveday2012} \\
{\sc FilamentFinding} & v02 & Alpaslan, Robotham & Filament and tendril catalogues & \cite{alpaslan2014} \\
{\sc GALEXPhotometry} & v02 &Seibert, Tuffs & GALEX NUV and FUV photometry for the GAMA II equatorial regions & --- \\
{\sc GeometricEnvironments} & v01 & Eardley, Peacock & Identification of the large scale structure within the GAMA equatorial regions in which each point is classified as a void, sheet, filament or knot & \cite{eardley2015} \\
{\sc GroupFinding} & v10 & Robotham  & GAMA Galaxy Group Catalogue (G3C) for the GAMA II equatorial and G02 fields & \cite{robotham} \\
{\sc WISEPhotometry} & v02 & Cluver. Jarrett & WISE IR photometry for the GAMA equatorial regions & \cite{cluver2014,cluver2020} \\
{\sc HATLASPhotometry} & v03 & Bourne, Liske, Driver & Herschel FIR photometry for Herschel-detected GAMAobjects & \cite{bourne2016} \\
{\sc LambdaPhotometry} & v01 & Wright, Robotham, Driver & 21 band photometry for the GAMA equatorial regions & \cite{wright2016} \\
{\sc PanchromaticPhotom} & v03 & Driver & Combination of various photometry catalogues from GALEX, SDSS, VISTA VIKING, WISE and Herschel-ATLAS & \cite{driver2016} \\
{\sc MagPhys} & v06 & Driver & MAGPHYS analysis of GAMA galaxies using LAMBDAPhotometryv03 & \cite{driver2018} \\
{\sc Randoms} & v02 & Farrow, Norberg & Randomly distributed galaxies with the same selection function as the main spectroscopic survey & \cite{Farrow2015} \\
{\sc SersicPhotometry} & v09 & Kelvin, Driver, Robotham & Serisc fits in u-K bands using GALFIT & \cite{kelvin2012}\\
{\sc StellarMasses} & v24 & Taylor & Stellar Mass measurements for objects with spec-$z$ in SpecObjv27 & \cite{taylor2011} \\
{\sc EnvironmentMeasures} & v05 & Brough & Environmental metrics of the local environment by density and number & \cite{brough2013} \\
{\sc VisualMorphology} & v03 & Driver, Baldry & Visual morphologies based on SDSS images to $z<0.06$ & \cite{kelvin2014} \\
{\sc G15DeepSpecCatv01} & v01 & Davies, Driver & Redshifts obtained in a 1\,deg$^2$ sub-region within the GAMA 15hr region & This paper \\ 
{\sc VelocityDispersions} & v01 & Taylor & Velocity Dispersion measurements for all objects in SpecObjv27 & Dogruel et al. (in prep.) \\ \hline
\end{tabular}
\end{table*}

Please also note that the GAMA DR4 release is intended to be dynamic, and additional catalogues uploaded on an ongoing basis including DMUs submitted for GAMA QC by the community. If you are interested in updates please regularly check the release website and if you are interested in submitting your own DMU to the GAMA QC process please contact gama@eso.org or via the information on the release website.

\begin{figure*}
	\centering     
	\includegraphics[width=\linewidth]{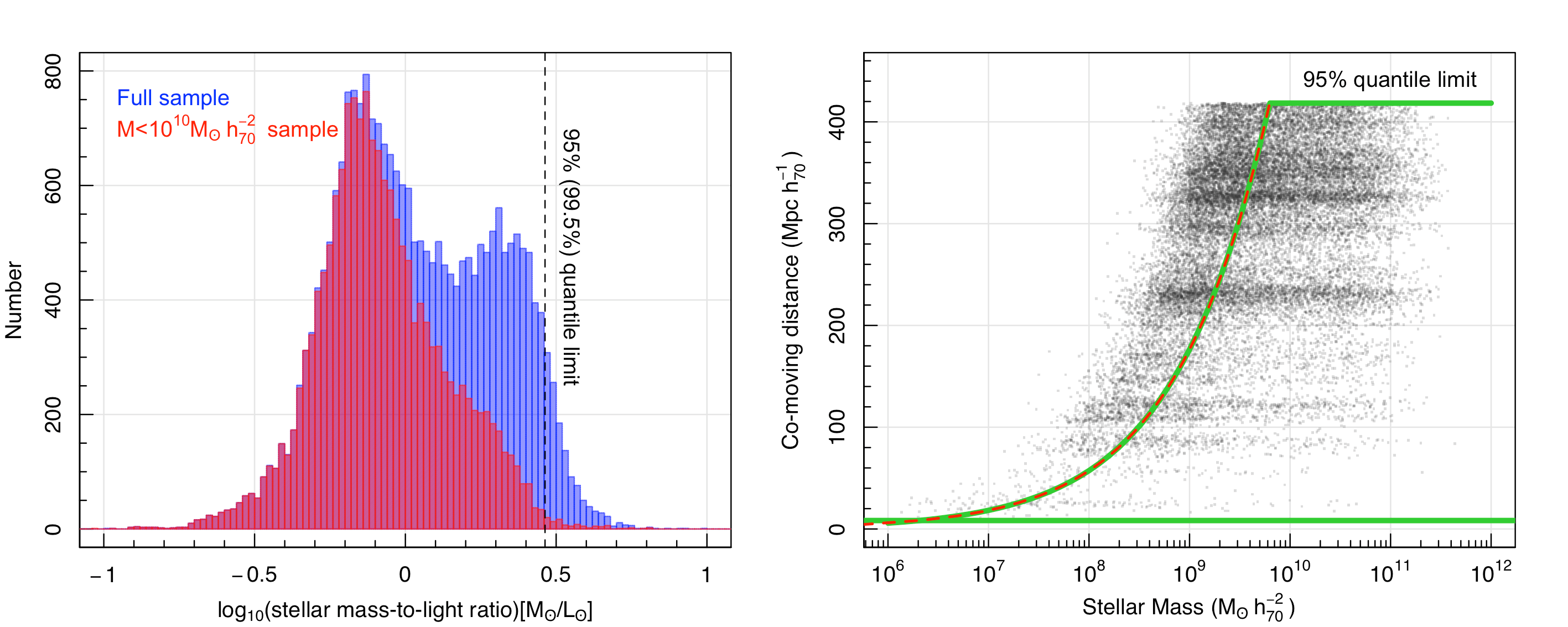}
	\caption{({\it left}) shows the mass-to-light ratio of the galaxies in our sample (blue histogram) along with the dashed line which delineates the 95-percentile value and the lower mass population impacted by this cut (red histogram) and for which our selection encloses 99.45 per cent of the sample. ({\it right}) shows the total sample in the stellar mass distance plane. The red dashed line shows the implied mass completeness limit from a combination of the given flux limit of $r=19.65$\,mag combined with the 95 percentile mass-to-light limit derived from the left panel. The green line shows the final selection function.
	\label{fig:mass2light}}
\end{figure*}

\section{The Galaxy Stellar Mass Function at $\mathbf{\emph{z}<0.1}$}
We conclude this release by providing a revised estimate of the low-$z$ galaxy stellar mass function (GSMF), and its sub-division by morphological type. This builds on earlier GAMA works on these topics from \cite{baldry2012,kelvin2014,moffett16} and \cite{wright17}. Specifically our revised estimate will make use of the following DMUs: {\sc SpecCatv27} for the redshifts, {\sc gkvScienceCatv02} for the photometric measurements, {\sc gkvMorphologyv02} for Hubble Classifications, {\sc gkvProSpectv02} for the Stellar Masses measurements and uncertainties, and {\sc gkvSFMPhotoZv02} for photometric redshifts and stellar masses of Main Survey objects without spectroscopic redshifts. These DMUs are all included in DR4, see Tables~\ref{tab:dmus}\,\&\,\ref{tab:dmus2}, and are available from the release website. The key advances over our previous GSMF estimates are the inclusion of the G23 region, the upgrade to KiDS photometry, revised stellar masses, and the inclusion of photometric data for missing and/or low surface brightness systems.

We note that in this work we do not attempt to identify and remove AGN. In two forthcoming papers Thorne et al. (submitted and in prep.) we explore the AGN contribution and its impact on our {\sc ProSpect} stellar mass estimates in detail.

We now start by adopting the magnitude limit of $r_{\rm KiDSDR4}=19.65$ mag, as discussed in Section\,\ref{sec:kids}, for the four primary GAMA regions (G09, G12, G15, and G23). This sub-sample contains 205\,540 galaxies for which our survey is 95.1 per cent complete in terms of reliable spectroscopic redshifts (see Section\,2.1). For those galaxies without spectroscopic redshifts, we adopt the photometric redshift from {\sc gkvSFMPhotoZv02}, as described in Section\,2.3. Hence, we deem our sample to be 100 per cent redshift complete. We now limit our sample to the nearby Universe by imposing a redshift cutoff of $z<0.1$. Note that no attempt is made to fold in any evolution within the interval $0 < z < 0.1$ (but see later discussion in Section\,6.2).

To reduce the observed sample to an empirical mass function we make use of Modified Maximum Likelihood (MML) estimation, as developed by  \cite{obreschkow2018}. This method avoids binning the data, and is a Bayesian framework for fitting distribution functions to complex multi-dimensional data, developed particularly for galaxy mass functions. By design, the MML framework includes due consideration of the observational measurement errors for each individual object, optimal correction for systemic Eddington bias, the ability to incorporate complex observational selection functions, and the option to correct internally for the underlying large scale structure identified within the survey volume. At its heart the MML approach consists of an iterative fitting algorithm that successively solves a standard maximum likelihood estimation and then updates the data by accounting for the previous fit and the observational uncertainties. The power of this `fit-and-debias' procedure relies on the fact that its solution can be shown to converge towards the exact solution of a much more expensive full Bayesian hierarchical model, in which each observable (e.g., each galaxy mass) is a free parameter with a prior given by the measurement (e.g., flux and redshift).

The MML framework is accessible via {\sc dftools} \citep{obreschkow2018}, an open-source software package for the {\sc R} statistical programming language. The code is fully documented and many examples have been provided by \cite{obreschkow2018}. {\sc dftools} can be used to derive volume-corrected binned mass functions, as well as to fit parametised analytical functions. In both cases, {\sc dftools} can determine the most likely solutions and full co-variance matrices of the relevant model parameters. Here we elect to fit a double Schechter function, able to tackle the characteristic upturn seen at intermediate stellar mass by \cite{baldry2012} and in subsequent studies. This function is defined in \cite{baldry2012} as
\begin{equation}
\phi_MdM=e^{-M/M^*}\left(\phi^*_1(\frac{M}{M_*})^{\alpha_1}+\phi^*_2(\frac{M}{M_*})^{\alpha_2}\right)\frac{dM}{M^*},
\end{equation}
where $\phi_M dM$ is the number density of galaxies in the mass interval $dM$ and $\phi^*_1$, $\phi^*_2$ and $\alpha_1$, $\alpha_2$ describe the normalisation and slope parameters respectively for the two components. Without loss of generality, we can always choose $\alpha_1 > \alpha_2$ such that the second term in Equation~1 dominates at lower masses.

In our usage of {\sc dftools}, we adopt the stellar mass errors identified for each galaxy from the {\sc ProSpect} analysis of \cite{bellstedt20}. The median error for our $z<0.1$ sample is $\Delta \log_{10} M = 0.043$. In those cases, where photometric redshifts are used, the errors also include the uncertainty in the redshift (see Section\,2.3), although we note that in our final analysis, only 98 galaxies with photometric redshifts, (i.e., 0.7 per cent) survive through to the final sample.

The observational selection function is the key component to deriving the appropriate volume correction. Given the data are flux limited in the observed $r$-band, yet we are looking to recover stellar mass functions, this is non-trivial. To overcome this issue we explore the $r$-band mass-to-light distribution as shown in the left panel of Figure\,\ref{fig:mass2light}. The blue distribution shows the spread and the dashed line shows the mass-to-light ratio that encloses 95 per cent of the full distribution, i.e., $\log_{10}(M/L_r) = 0.463$. Note that if we only consider galaxies with stellar masses below $10^{10}$M$_{\odot} h_{70}^{-2}$ (i.e., those not seen over the entire volume), we find only 0.55 per cent with mass-to-light ratios above 0.463 (see the red shaded histogram in Figure\,\ref{fig:mass2light}). Hence, in effect, our cut is valid for 99.45 per cent of the population impacted by the selection boundary. Figure\,\ref{fig:mass2light} (right panel) now shows our full sample in terms of their stellar mass and co-moving distance as grey dots, and where the large scale structure (horizontal banding) is clearly visible. 
Using our $r_{\rm KiDS DR4}$-band limit of 19.65 mag, combined with our 95 per cent mass-to-light limit (dashed line of Figure\,\ref{fig:mass2light} left), we can now define the red dashed line. This denotes the distance limit at each redshift for which our sample will be 99.5 per cent complete for M$<10^{10}$M$_{\odot} h_{70}^{-2}$. 
We fit the dashed red curve on Figure\,\ref{fig:mass2light} (right) with a generalised logistics function, also known as a Richards curve, defined as follows:
\begin{equation}
y(x)=A+\frac{K-A}{(C+e^{-B (x-M)})^{1/\nu}}.
\label{eqn:richards}
\end{equation}
Here $y$ represents the co-moving distance limit, and $x$ the mass-limit, while the fitted parameters: $A, K, C, B, M$ and $\nu$ define the upper and lower asymptotes and the shape of the transition curve. Table\,\ref{tab:richards} shows the fitted Richards curve parameters determined for the full sample (All), or independently for each morphological class. The fit is shown as the green curve on Figure\,\ref{fig:mass2light} (right) and follows the red dashed line extremely closely. Finally, we truncate the green line at our minimum redshift ($z>0.0013$) to avoid stellar contamination, and our maximum redshift ($z<0.1$), to define our final sample selection boundary. Although we have essentially rejected 50 per cent of our original $z<0.1$ sample, we have now transformed the sample from an $r$-band selected one to a near (99.5 per cent) mass-limited one with a very precisely defined selection function.

\begin{figure*}
	\centering     
	\includegraphics[width=\columnwidth]{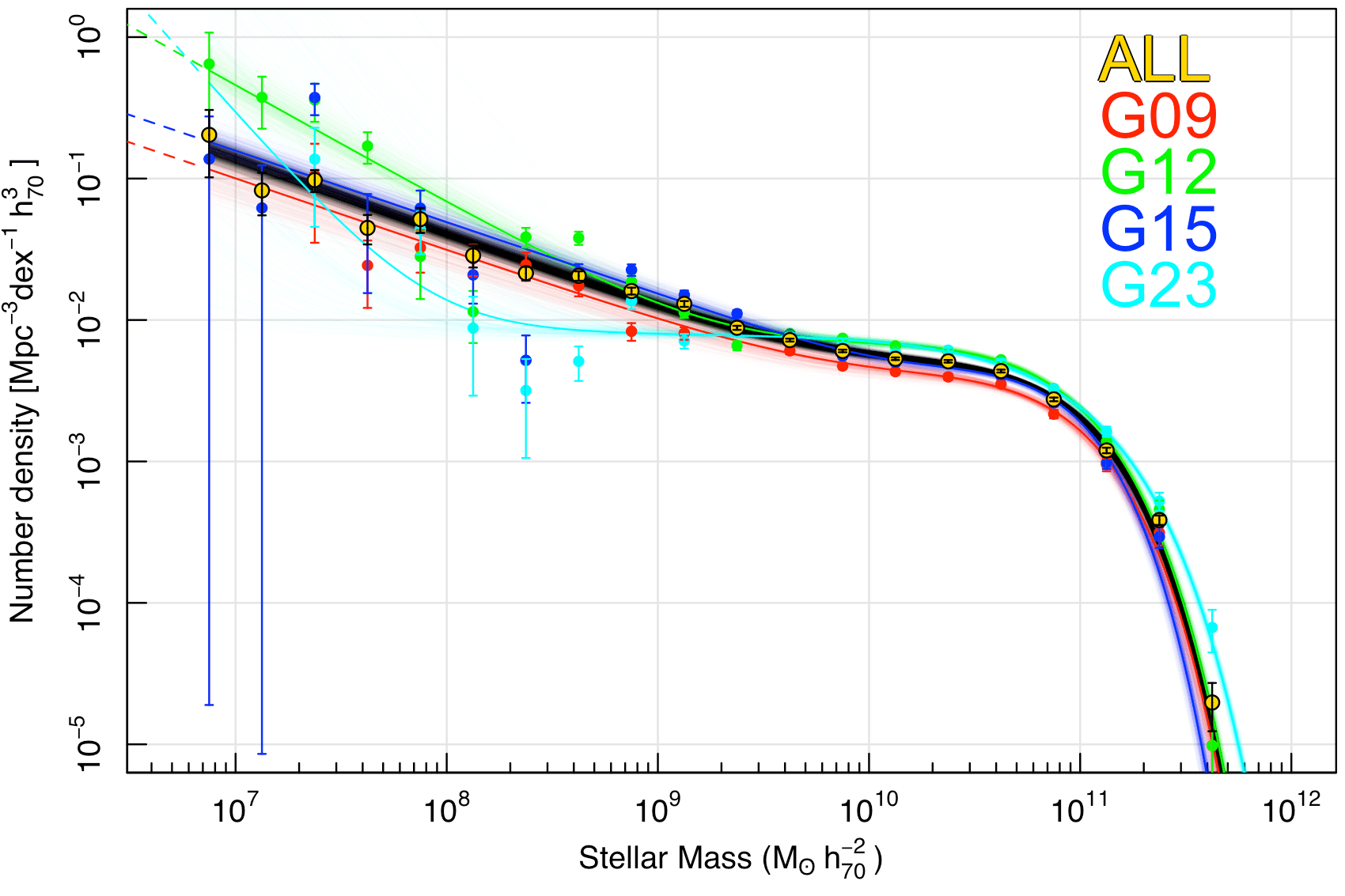}
	\includegraphics[width=\columnwidth]{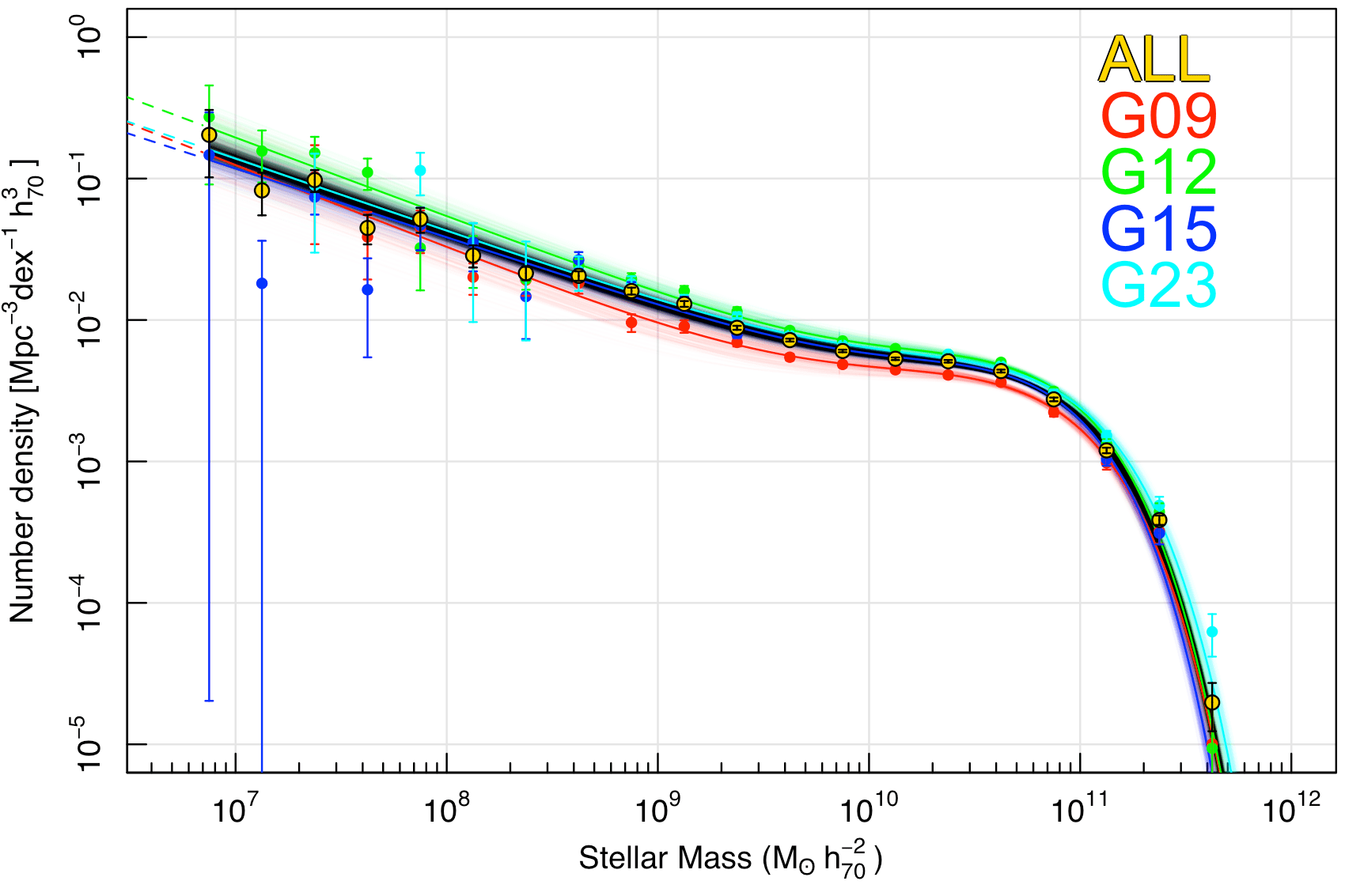}
	\caption{The recovered galaxy stellar mass function for the combined data (shown as the black circles with yellow infill and black lines), and for each of the four regions shown individually with the colours as indicated in the figure key. 
	The left panel shows the results without implementing the {\sc dffit} LSS correction, and the right panel with. 
	Generally the overall GSMF is unchanged, but the GSMF for each region is brought into much greater alignment.
	}
	\label{fig:plot4}
\end{figure*}

\begin{table*}
\caption{Key values defining the Richards curve selection function parameters for various samples (as indicated). \label{tab:richards}}
\begin{tabular}{c|c|c|c|c|c|c|c|c|c} \hline \hline
Galaxy & Redshift & Number of galaxies & $\log_{10}\left[\frac{M}{L_r}\right]_{95\% {\rm limit}}$ & $A$ & $K$ & $C$ & $B$ & $M$ & $\nu$ \\ 
sample &  $z$   &  remaining (starting) & $(M_{\odot}/L_{\odot})$ & & & & & &   \\    \hline
Total & $z<0.10$ & 13\,957 (24,082) & 0.463 & $-0.0160$ & 2742.0 & 0.9412 & 1.1483 & 11.815 & 1.691 \\
E+HE &  $z<0.08$ & 1\,272 (1\,355) & 0.506 & $-0.1232$ & 2512.6 & 0.1411 & 1.1460 & 10.233 & 0.285 \\
cBD & $z<0.08$ & 2\,638 (2\,713) & 0.497 & $-0.0605$ & 2517.7 & 0.4758 & 1.1469 & 11.274 & 0.948 \\
dBD & $z<0.08$ & 2\,799 (4\,138) & 0.405 & $-0.0124$ & 2521.6 & 1.5262 & 1.1465 & 12.205 & 3.042 \\
D & $z<0.08$ & 3\,116 (6\,896) & 0.299 & $-0.0146$ & 2157.0 & 0.5619 & 1.1477 & 11.212 & 1.296 \\ 
C & $z<0.08$ &  38 (112) & 0.396 & $-0.1197$ & 2399.7 &   0.0933 & 1.1460 & 9.761 & 0.197 \\
H & $z<0.08$ & 40 (79) & 0.443 & $-0.1064$ & 3138.3  & 0.4670 & 1.1462 &  11.212 & 0.753\\ \hline
\end{tabular}
\end{table*}

The sample is now restricted to those galaxies that lie within the limits defined by our selection function. We input the selected galaxies' co-moving distances, stellar masses, stellar mass errors, selection function and desired functional form to fit into the {\sc dftools} routine {\sc dffit}. This code returns the binned co-moving space density distribution (see Table\,\ref{tab:massdistributions}), the functional fit (see Table\,\ref{tab:massfunctions}), and the full co-variance matrix for the fitted parameters. Figure\,\ref{fig:plot4} shows the combined galaxy stellar mass function data (black circles with yellow fill), along with the results for each of the four GAMA regions separately (coloured discs). 

Given a survey selection function and a specific model for the MF (e.g. a parametric Schechter function or a binned step-wise MF), the most likely cosmic density {\it variations} as a function of mass caused by cosmic LSS can be determined simultaneously while fitting the free parameters of the MF model. Intuitively speaking, this is possible, because a smooth selection function and a smooth MF normally predict a smooth source count as a function of mass. A comparison to the actual source count, then allows us to infer the LSS-driven fluctuations. The mathematics and explicit form of the likelihood function can be found in \citeauthor{obreschkow2018} (\citeyear{obreschkow2018}, section 2.3). The only unknown in this automatic large scale structure (LSS) correction is the {\it overall} density normalization of the survey volume. This we fix manually by imposing the condition that the total mass within the whole survey volume is unbiased.

In the left panel the {\sc dftools} inbuilt LSS correction is not implemented, and in the right panel it is bringing the fields into good alignment. In general the four regions agree well, showing a variation consistent with what one might expect from cosmic variance considerations of 25 per cent per GAMA region. We note that the binned data for the total distribution (black circles with yellow fill) appear to exhibit a smooth distribution and extent, with reasonable statistics, over the stellar mass range of $10^{6.75}$\,M$_{\odot} h_{70}^{-2}$ to $10^{11.5}$\,M$_{\odot} h_{70}^{-2}$. 

Comparing the GSMF derived for the four individual fields without the LSS correction (i.e., Figure\,\ref{fig:plot4} left), we see that the G23 region has both the highest mass density as well as the steepest low mass upturn. This steep upturn is reminiscent of the mass function seen in the Virgo cluster. There is therefore some possibility that the G23 region may intersect with an as yet undefined very nearby loosely bound group. This is not explored here, but will be considered as we obtain 21cm radio observations in this region. We also note that G12 exhibits a slightly steeper low-mass end, and lies a few degrees offset from the Virgo Southern spur (see \citeauthor{virgo} \citeyear{virgo} their Figure\,1). With the LSS correction implemented (i.e., Figure\,\ref{fig:plot4} right), we see the four fields brought into closer alignment. It is reassuring that both with and without the LSS correction the overall GSMF is identical suggesting that the combination of the four distinct fields goes a long way towards ironing out the LSS.

Finally, we note that both with and without the LSS correction the G09 region (red data points) appears under-dense, this has been a feature noted and highlighted in earlier GAMA papers and arguably sets the G09 field slightly apart. In Section\,6 we will explore GAMA's overall over/under density relative to a 5\,012\,deg$^2$ SDSS selected region. In due course two imminent surveys will improve upon our measurements, namely the Wide Area VISTA Extragalactic Survey \citep{driver2019} which will survey 1\,150\,deg$^2$ in two distinct regions to $m_{Z} \sim 21$\,mag, and the recently commenced DESI Bright Galaxy Survey \citep{desibgs} which will survey 14\,000\,deg$^2$ to a comparable depth as GAMA.

\begin{table*}
\centering
\caption{The galaxy stellar mass number-density distributions for the full sample and for each GAMA region at $z<0.1$. No detections in a bin are indicated by a blank entry. Note that the tabulated values {\it DO NOT} include the re-normalisation to SDSS, or correction to redshift zero (see Section\,6). To apply these corrections add $0.0807$ dex to all number-density values.\label{tab:massdistributions}}
\begin{tabular}{c|c|c|c|c|c} \hline
$\log_{10}(M)$ & \multicolumn{5}{c}{$\log_{\rm 10}$ number-density of galaxies per dex per Mpc$^{3}h_{70}^{-3}$} \\ \cline{2-6}
$(M_{\odot} h_{70}^{-2})$ &  All & G09 & G12 & G15 & G23 \\ \hline
$ 11.875 $ &  - &  - & - & - & - \\
$ 11.625 $ & $ -4.704 \pm 0.138 $ & $ -4.997 \pm 0.301 $ & $ -5.028 \pm 0.301 $ & - & $ -4.204 \pm 0.125 $ \\
$ 11.375 $ & $ -3.414 \pm 0.032 $ & $ -3.492 \pm 0.075 $ & $ -3.356 \pm 0.060 $ & $ -3.507 \pm 0.067 $ & $ -3.310 \pm 0.06 $ \\
$ 11.125 $ & $ -2.922 \pm 0.019 $ & $ -3.010 \pm 0.043 $ & $ -2.878 \pm 0.035 $ & $ -2.985 \pm 0.041 $ & $ -2.818 \pm 0.034 $ \\
$ 10.875 $ & $ -2.561 \pm 0.013 $ & $ -2.652 \pm 0.029 $ & $ -2.504 \pm 0.023 $ & $ -2.561 \pm 0.025 $ & $ -2.517 \pm 0.025 $ \\
$ 10.625 $ & $ -2.361 \pm 0.010 $ & $ -2.440 \pm 0.022 $ & $ -2.299 \pm 0.018 $ & $ -2.371 \pm 0.021 $ & $ -2.324 \pm 0.020 $ \\
$ 10.375 $ & $ -2.292 \pm 0.009 $ & $ -2.389 \pm 0.021 $ & $ -2.244 \pm 0.017 $ & $ -2.284 \pm 0.019 $ & $ -2.242 \pm 0.018 $ \\
$ 10.125 $ & $ -2.274 \pm 0.009 $ & $ -2.352 \pm 0.020 $ & $ -2.202 \pm 0.017 $ & $ -2.277 \pm 0.018 $ & $ -2.256 \pm 0.018 $ \\
$ 9.875 $ & $ -2.219 \pm 0.009 $ & $ -2.314 \pm 0.020 $ & $ -2.148 \pm 0.015 $ & $ -2.232 \pm 0.018 $ & $ -2.178 \pm 0.017 $ \\
$ 9.625 $ & $ -2.142 \pm 0.010 $ & $ -2.263 \pm 0.023 $ & $ -2.074 \pm 0.019 $ & $ -2.117 \pm 0.019 $ & $ -2.109 \pm 0.021 $ \\
$ 9.375 $ & $ -2.055 \pm 0.014 $ & $ -2.158 \pm 0.027 $ & $ -1.940 \pm 0.031 $ & $ -2.101 \pm 0.025 $ & $ -1.976 \pm 0.030 $ \\
$ 9.125 $ & $ -1.886 \pm 0.020 $ & $ -2.044 \pm 0.042 $ & $ -1.794 \pm 0.035 $ & $ -1.866 \pm 0.031 $ & $ -1.871 \pm 0.048 $ \\
$ 8.875 $ & $ -1.795 \pm 0.024 $ & $ -2.018 \pm 0.058 $ & $ -1.710 \pm 0.041 $ & $ -1.771 \pm 0.039 $ & $ -1.744 \pm 0.053 $ \\
$ 8.625 $ & $ -1.688 \pm 0.032 $ & $ -1.739 \pm 0.064 $ & $ -1.613 \pm 0.044 $ & $ -1.575 \pm 0.055 $ & $ -1.655 \pm 0.105 $ \\
$ 8.375 $ & $ -1.669 \pm 0.045 $ & $ -1.717 \pm 0.089 $ & $ -1.711 \pm 0.064 $ & $ -1.834 \pm 0.176 $ & $ -1.667 \pm 0.222 $ \\
$ 8.125 $ & $ -1.544 \pm 0.071 $ & $ -1.697 \pm 0.097 $ & $ -1.552 \pm 0.146 $ & $ -1.453 \pm 0.138 $ & $ -1.537 \pm 0.222 $ \\
$ 7.875 $ & $ -1.287 \pm 0.079 $ & $ -1.351 \pm 0.125 $ & $ -1.490 \pm 0.176 $ & $ -1.331 \pm 0.125 $ & $ -0.943 \pm 0.125 $ \\
$ 7.625 $ & $ -1.349 \pm 0.092 $ & $ -1.413 \pm 0.176 $ & $ -0.956 \pm 0.097 $ & $ -1.786 \pm 0.222 $ & - \\
$ 7.375 $ & $ -1.011 \pm 0.071 $ & $ -0.986 \pm 0.222 $ & $ -0.819 \pm 0.114 $ & $ -1.130 \pm 0.097 $ & $ -1.046 \pm 0.222 $ \\
$ 7.125 $ & $ -1.084 \pm 0.125 $ & - & $ -0.806 \pm 0.146 $ & $ -1.741 \pm 0.301 $ & - \\
$ 6.875 $ & $ -0.691 \pm 0.176 $ & - & $ -0.564 \pm 0.222 $ & $ -0.833 \pm 0.301 $ & - \\
$ 6.625 $ & - & - & - & - & - \\
$ 6.375 $ & $ 0.202 \pm 0.176 $ & - & $ 0.173 \pm 0.301 $ & $ -0.098 \pm 0.301 $ & - \\
$ 6.125 $ & - & - & - & - & - \\ \hline
\end{tabular}
\end{table*}

\begin{table*}
\caption{GAMA galaxy stellar mass functions for various regions ($z<0.1)$. Note that the $\log_{10}(\phi*)$ values {\it DO NOT} include the renormalisation to SDSS, or the correction to redshift zero (see Section\,6). To apply these corrections add $0.0807$ dex to both $\log_{10}(\phi*)$ values.\label{tab:massfunctions}}
\begin{tabular}{c|c|c|c|c|c|c} \hline
Dataset & $\log_{10}(M*)$ & $\log_{10}(\phi^*_1)$ & $\log_{10}(\phi^*_2)$ & $\alpha_1$ & $\alpha_2$ & $\log_{10}(\rho_*)$ \\
        &  $(M_{\odot} h_{70}^{-2})$    &  (Mpc$^{-3} h_{70}^3$)   & (Mpc$^{-3} h_{70}^3$) &        &  & (M$_{\odot}$ Mpc$^{-3} h_{70}$) \\ \hline
All & $10.745 \pm 0.020$ & $-2.437 \pm 0.016$ & $-3.201 \pm 0.064$ & $-0.466 \pm 0.069$ & $-1.530 \pm 0.027$ & $8.392\pm 0.006$ \\
G09 & $10.764 \pm 0.044$ & $-2.513 \pm 0.035$ & $-3.462 \pm 0.204$ & $-0.577 \pm 0.144$ & $-1.583 \pm 0.084$ & $8.310\pm 0.017$ \\
G12 & $10.737 \pm 0.037$ & $-2.373 \pm 0.029$ & $-3.155 \pm 0.114$ & $-0.469 \pm 0.126$ & $-1.557 \pm 0.045$ & $8.453\pm 0.012$ \\
G15 & $10.666 \pm 0.038$ & $-2.411 \pm 0.030$ & $-3.048 \pm 0.088$ & $-0.265 \pm 0.141$ & $-1.474 \pm 0.041$ & $8.374\pm 0.012$ \\
G23 & $10.800 \pm 0.043$ & $-2.438 \pm 0.037$ & $-3.167 \pm 0.166$ & $-0.515 \pm 0.157$ & $-1.512 \pm 0.075$ & $8.456\pm 0.017$ \\ \hline
\end{tabular}

$^\dagger$ Cosmic (sample) variance error.
\end{table*}

\begin{figure*}
	\centering     
	\includegraphics[width=\textwidth]{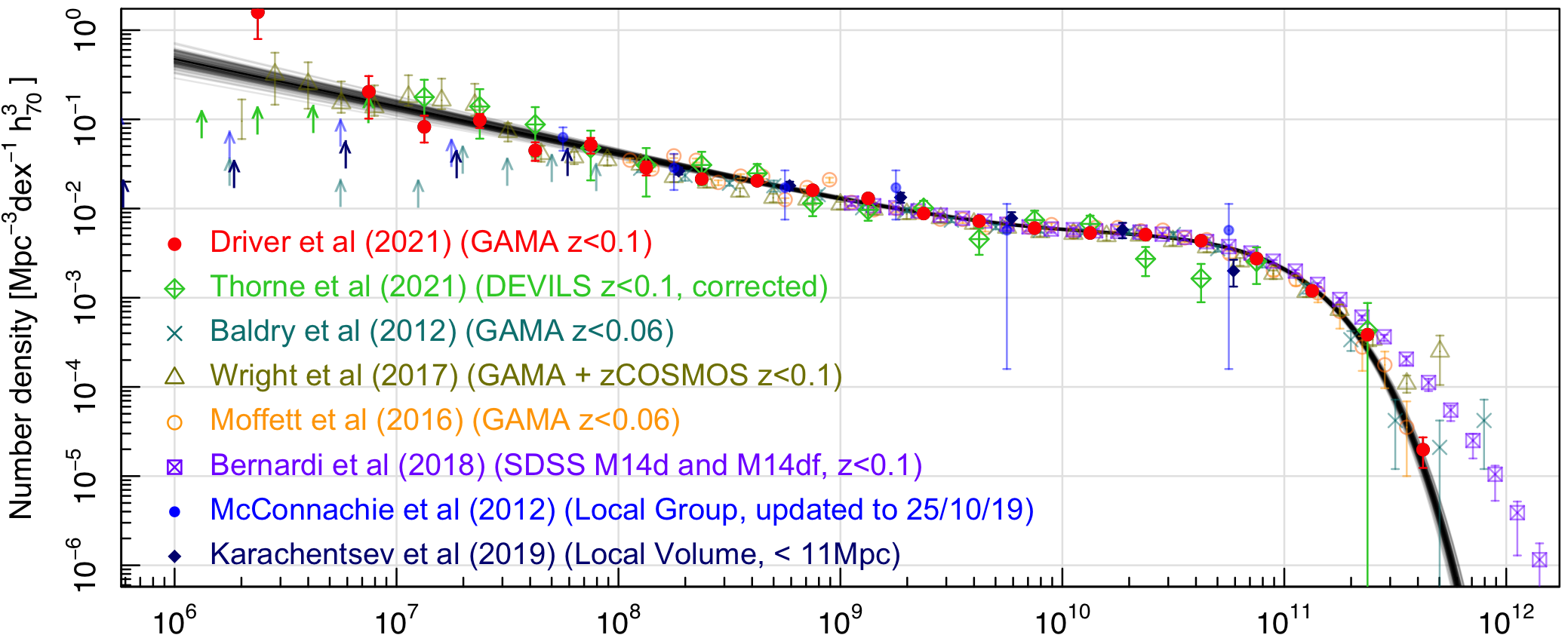}
	\includegraphics[width=\textwidth]{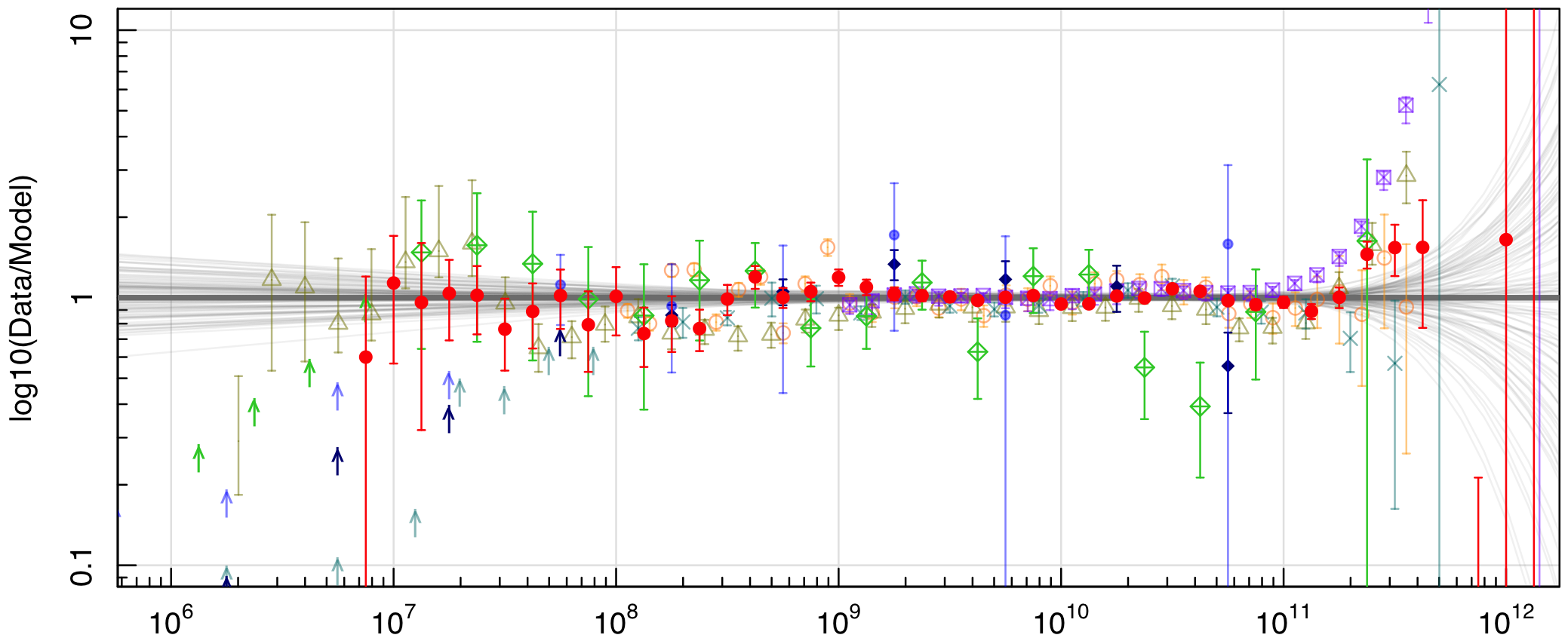}
	\includegraphics[width=\textwidth]{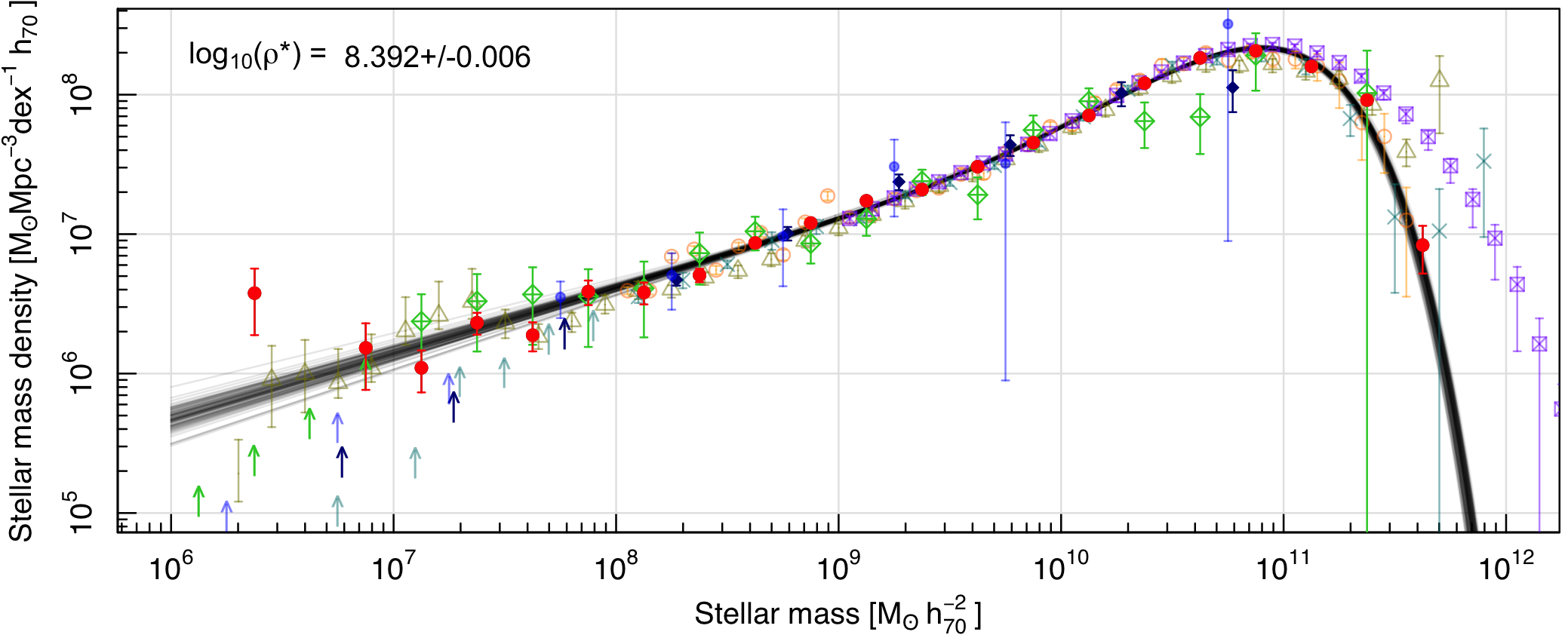}
	\caption{({\it upper}) The nearby galaxy stellar mass function recovered in this study and compared to recent measurements as indicated. ({\it centre}) the same data but with the functional fit to the red data points (this study) removed, to highlight discrepancies relative to the quoted errorbars. The grey lines show fits to the data points when randomly perturbed by their individual errors. ({\it lower}) The contribution of each stellar mass interval to the total stellar mass density. The distribution appears well bounded at high and low stellar masses suggesting the majority of the stellar mass has been identified.}
	\label{fig:gsmf}
\end{figure*}

\subsection{Comparison to previous measurements}
Figure\,\ref{fig:gsmf} (upper) shows our derived galaxy stellar mass function (with the LSS correction implemented), for the combined GAMA sample (red dots), and compared to our earlier GAMA measurements as well as notable literature values (as indicated). Also shown is the local group compendium of \cite{mcconnachie2012} (updated to October 2019 via \url{http://www.astro.uvic.ca/~alan/Nearby_Dwarf_Database.html}), and the local sphere compendium of \cite{karachentsev2019}. Note that many of the surveys shown, start to suffer from incompleteness at masses below $10^8$\,M$_{\odot}$, and these are indicated by lower limits (shown as arrows). In general the plot shows reasonable consistency across the datasets with the exception of the very high mass and very low mass ends.

On Figure\,\ref{fig:gsmf} (centre), we replot the same base data but with the fitted double Schechter function now divided out and the GAMA data points shown in 0.125dex mass bins (rather than 0.25dex mass bins in the upper and lower panels). Note that the grey lines show a 1$\sigma$ sampling of the co-variance matrix of the fit (in all three panels), to highlight the fit uncertainty. Not surprisingly the red data points, to which the double Schechter function has been fitted, scatter around the flat line. Most of the other surveys show overlap within their quoted errors. The one obvious discrepancy is with the SDSS data of \cite{bernardi2018}, where we see what looks like a systematic offset at the very high mass end. 
\cite{bernardi2018} explore in detail the difficulties of estimating the very high-mass end correctly, and provide a range of possible mass estimations, highlighting that the uncertainty is due to many potential factors related to: significant photometric corrections applied to the high-mass SDSS data; the importance of the stellar population assumptions; and the role of dust. At this stage we are not overly concerned, but note that where both GAMA and SDSS statistics are good the surveys appear to agree well within the errors, with perhaps some indication of incompleteness at the SDSS low mass end at $\sim 10^9$M$_{\odot} h_{70}^{-2}$ (as one might expect from its shallower surface brightness limit of 23\,mag per arcsec$^{2}$). We revisit this topic later in Section 6 following renormalisation of the GAMA data to the SDSS volume.

Included in the data shown on Figure\,\ref{fig:gsmf} are the narrow-band photometric redshift data from the COSMOS15 release \cite{cosmos15} as used by \cite{wright17} to determine a GSMF to very low stellar masses (gold triangles). The \cite{wright17} result used the combination of GAMA and COSMOS15 data, extending the GSMF down to $10^7$\,M$_{\odot} h_{70}^{-2}$ but with the caveat of increased errors, and the potential for a systematic bias in the very low photometric redshift estimates (and reflected in the errors). Despite these caveats the agreement is good, although see further discussion in Section~\ref{sec:knee}. In the end our deeper analysis suggests these corrections may have been over-estimated. We extend this work by now showing the $z<0.1$ GSMF from the DEVILS survey \citep{devils}. 

The DEVILS data are a combination of many contributing surveys that provide narrow-band photometric, spectroscopic, or grism data, as detailed in \cite{thorne2021} (see their Appendix C). Stellar masses for these data were determined by \cite{thorne2021} using the same {\sc ProSpect} code as used for the GAMA data (i.e., \citeauthor{bellstedt20} \citeyear{bellstedt20}). In plotting the \citeauthor{thorne2021} data we need to incorporate an Eddington bias correction. This is due to the combination of the photometric redshift error with the solid angle on the sky, resulting in more high-$z$ galaxies being scattered to $z<0.1$, than low-$z$ galaxies scattered above. We estimate the scale of this effect by running a set of Monte Carlo simulations in which we peturb the redshift values by their quoted errors, and recompute the DEVILS GSMF after scaling the masses for the redshift change. We estimate the Eddington bias from the change in the GSMF measurements, and correct the original GSMF by this amount. In effect we are introducing an additional Eddington bias to determine its approximate impact, and then removing this from the original distribution. The data points move {\it systematically} downwards but within their original errors. Following this correction we see that the DEVILS $z<0.1$ data (green diamonds) agree well with the GAMA $z<0.1$ data to the GAMA mass limit. 

While the errors on the DEVILS data are large, this agreement is encouraging as DEVILS imaging is based on much deeper Subaru data than the ESO KiDS data. This agreement would be unlikely to occur if DEVILS was identifying a significant additional low-$z$ population not seen in the ESO KiDS data, e.g., due to surface brightness considerations. While these agreements are tentative --- because one cannot rule out a bias in the photometric estimation that acts to emulate a surface brightness bias in the GAMA data --- the consistency between these results, \cite{wright17} and \cite{thorne2021} is reassuring.

As noticed in previous papers, and in particular \cite{baldry2012}, the galaxy stellar mass function appears to exhibit a plateau around $10^{9.5}$ to $10^{10.5}$\,M$_{\odot} h_{70}^{-2}$. This feature {\it may} be due to a higher late-time merger rate at the high mass end, as argued by \cite{robotham14}. This distinctive feature has been shown to emerge at lower redshift ($z<1$) by \cite{wright2018} in their compendium of GSMF's from $z=5$ to $z=0$. At lower masses the galaxy stellar mass function turns up at $10^{9.5}$\,M$_{\odot} h_{70}^{-2}$ and exhibits a linear slope (in $\log(\phi)-\log($M$_*$) space) to the completeness limit. This trend is now extended by the new data to $10^{6.75}$\,M$_{\odot} h_{70}^{-2}$ with no obvious sign of any significant downturn or flattening. Hence the most numerous type of galaxy in the nearby Universe must have a mass at, or more likely below, $10^{6.75}$\,M$_{\odot} h_{70}^{-2}$. This is consistent with basic Jeans' mass arguments that, in a Cold Dark Matter dominated Universe, the lowest mass system able to collapse rather than dissipate, should be around $10^{4.5}$\,M$_{\odot} h_{70}^{-2}$ (assuming the baryonic mass is fully converted to stars). Hence we are gradually encroaching upon this limit, but are still just over two orders of magnitude away. 

It is worth noting that the mass function is significantly less steep ($\alpha_2=-1.53 \pm 0.03$, see Table\,\ref{tab:massfunctions}) than the theoretical halo mass function ($\alpha=-1.8$), and various studies have argued that this decrease in the stellar mass to dark matter ratio, may be due to the role of supernova feedback or stellar winds ejecting baryonic mass as well as shutting down star-formation in a mass dependent manner (see review of this topic by \citeauthor{wechsler2018} \citeyear{wechsler2018}). An interesting aside is that for this mechanism to work, star-formation must occur in order to generate the SN and AGB winds. This mechanism would therefore suggest that {\it every} dark matter halo should contain some residual stellar mass from this initial burst of star-formation, albeit potentially extended and diffuse. 

This may eventually create a significant problem, as for stellar feedback (SN and Winds) to be the sole mechanisms responsible for the discrepant slopes (HMF v GSMF), we eventually require an extreme number of very low stellar mass galaxies residing in intermediate mass haloes, i.e., with exceptionally high dark matter to stellar mass ratios. While some very low mass systems do contain very high mass-to-light ratios (e.g., \citeauthor{battaglia2013} \citeyear{battaglia2013}), it is unclear whether these are fully representative of all low mass haloes. The obvious solutions are that these extreme systems are of exceptionally low surface brightness and rendered undetectable, that some other process prevents stars from ever forming in lower mass haloes (i.e., a failure to spark, \citeauthor{bullock2000} \citeyear{bullock2000}), or that some external process prevents the gas collapsing \citep{darkhaloes}. A viable example might be the ambient radiation field that a higher mass galaxy exerts on the surrounding environment to prevent the cooling of gas in nearby less massive haloes. At present while we are finding relatively small numbers of very dark matter dominated systems, these are typically restricted to rich cluster environments, thus far. Similarly 21cm studies have yet to identify strong cases of neutral gas only systems which cannot be explained as ejected mass from a nearby companion.

From integrating our GSMF to $10^{6.75}$M$_{\odot} h_{70}^{-2}$ we find a space density of $0.24 \pm 0.04$ galaxies per Mpc$^3h_{70}^{-3}$. This same density of dark matter haloes is reached for a standard Halo Mass Function (HMF, see \citeauthor{murray2013} \citeyear{murray2013}) at a dark matter integration limit of $\sim 10^{9.6}$M$_{\odot} h_{70}^{-2}$. This implies that our lowest mass systems must have dark matter to stellar mass ratios above 700, to reconcile our GSMF with a standard $\Lambda$CDM HMF without recourse to fully dark haloes above $\sim 10^{9.6}$M$_{\odot} h_{70}^{-2}$. While high, this is not entirely inconsistent with measurements of some nearby, albeit lower mass, dwarf systems, e.g., Segue 1 with 99.9 per cent dark matter, \cite{segue1}, and consistent with the conclusions of the simulations community summarised in \cite{wechsler2018} (see their Figure\,2).

As well as exploring the stellar mass distribution it is also worth reviewing the total stellar mass density derived from integrating our stellar mass density function (i.e., Figure\,\ref{fig:gsmf} upper). Figure\,\ref{fig:gsmf} (lower) shows the same data but multiplied through by the abscissa to make clear the contribution of each stellar mass interval to the total stellar mass density. Here we see that the peak is relatively narrow and centred around $10^{10.8}$\,M$_{\odot} h_{70}^{-2}$ highlighting how $M_*$-galaxies dominate the contribution to the stellar mass density. We find that 90 per cent of the stellar mass lies in the range $10^{9.6}$\,M$_{\odot} h_{70}^{-2}$ to $10^{11.6}$\,M$_{\odot} h_{70}^{-2}$. The distribution drops more steeply towards higher-masses, indicating a minimal ($<1$ per cent) stellar mass contribution from super-massive galaxies ($>10^{11.6}$\,M$_{\odot} h_{70}^{-2}$), and drops more gradually towards lower stellar masses. Nevertheless the contribution of each mass interval to the total stellar mass density has dropped by a factor of 100 from the peak to our limiting stellar mass of $10^{6.75}$\,M$_{\odot} h_{70}^{-2}$. This informs us that while the most numerous galaxy per decade of stellar mass has a stellar mass below $10^{6.75}$\,M$_{\odot} h_{70}^{-2}$, its contribution to the total stellar mass density is likely to be minimal ($\sim 1$ per cent from a simple extrapolation from $10^{6.5}$\,M$_{\odot} h_{70}^{-2}$ to $0$\,M$_{\odot} h_{70}^{-2}$. 

Here we can report that by integrating the contribution to the stellar mass density we recover a value of $\rho_* = (2.47 \pm 0.04) \times 10^8$\,M$_{\odot}$ Mpc$^{-3} h_{70}$ for a 737 cosmology. This equates to $\Omega_* = (1.81 \pm 0.03) \times 10^{-3} h_{70}^{-1}$ (also for a 737 cosmology). We note that these values are as yet uncorrected for any over/under density in the overall GAMA footprint, and are effectively a measurement at the median redshift of $z=0.079$ and both of these issues will be addressed in Section\,6.

Putting aside the issue of stripped stellar mass not accounted for in the integral of the galaxy stellar mass function,
there are three caveats to our $\Omega_*$ measurement worth considering. The first is that one can never rule out a dramatic upturn to the distribution below our mass-limit ($10^{6.75}$\,M$_{\odot} h_{70}^{-2}$), e.g., a high space density of free-floating globular cluster-like mass systems at $10^6$\,M$_{\odot} h_{70}^{-2}$. While this would appear unphysical, it is not impossible, but not supported by any observation or simulation. The second is whether our sample is missing low surface brightness systems. Without deeper data this is always hard to assess, but we do note that the extensive search for low surface brightness galaxies in 200\,deg$^2$ of the Hyper Suprime-Cam Survey by \cite{greco2018}, identified no low surface brightness galaxies above the GAMA magnitude limit ($r_{\rm KiDSDR4}<19.65$\,mag). The third is a fairly subtle effect related to the first caveat. Its basis is that the galaxy population shows a significant diversity of form (faint-end slopes). Hence, a valid question is whether the total stellar mass density should be derived from the extrapolation of the total distribution, or the extrapolations of the distinct morphological types. For example, in earlier studies of the GAMA morphological mass functions, extending down to $10^9$\,M$_{\odot} h_{70}^{-2}$, \cite{moffett16} identified divergent slopes to some particular types, the extrapolation of which would lead to an infinite stellar mass density. We now reconsider this notion by deriving the galaxy stellar mass functions for each morphological class. 

\begin{figure*}
	\centering     
	\includegraphics[width=\linewidth]{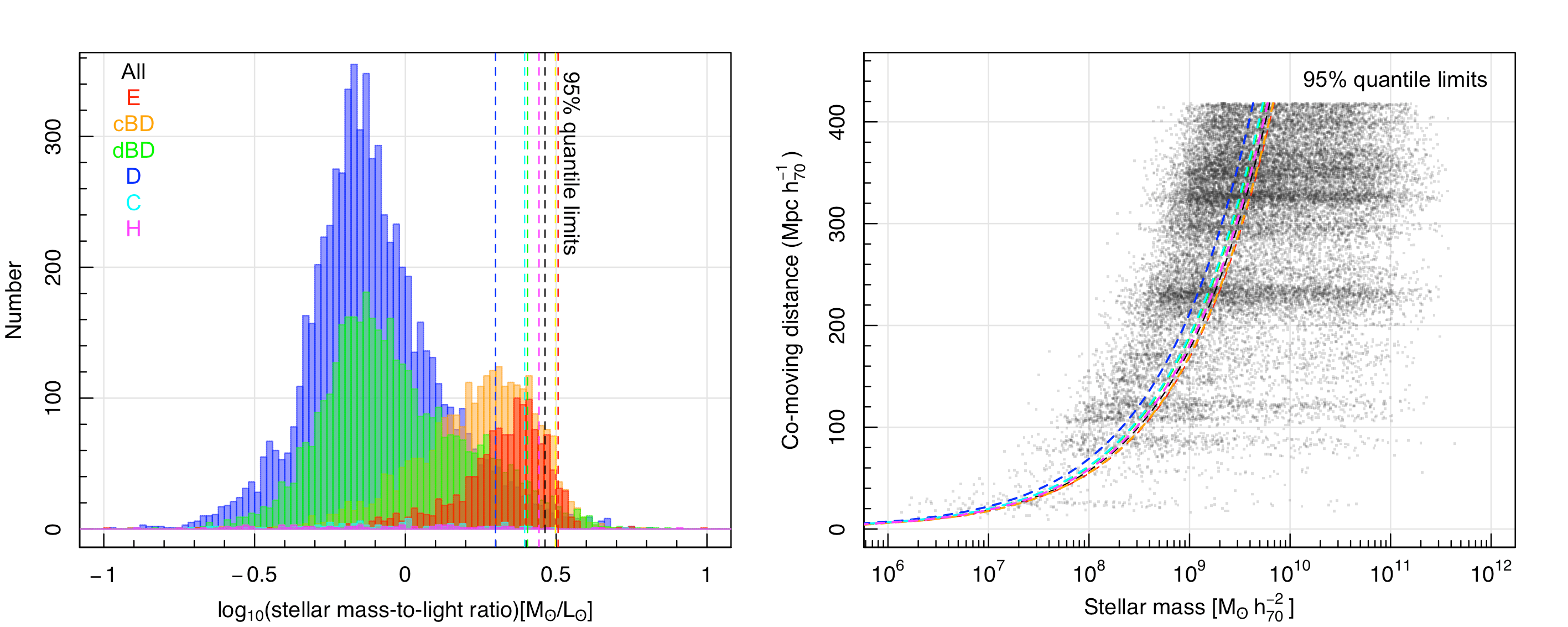}
	\caption{As for Figure\,\ref{fig:mass2light} except now showing the mass-to-light ratio and selection function for each morphological type as indicated.}
	\label{fig:mass2light_morph}
\end{figure*}

\begin{figure*}
	\centering     
	\includegraphics[width=\textwidth]{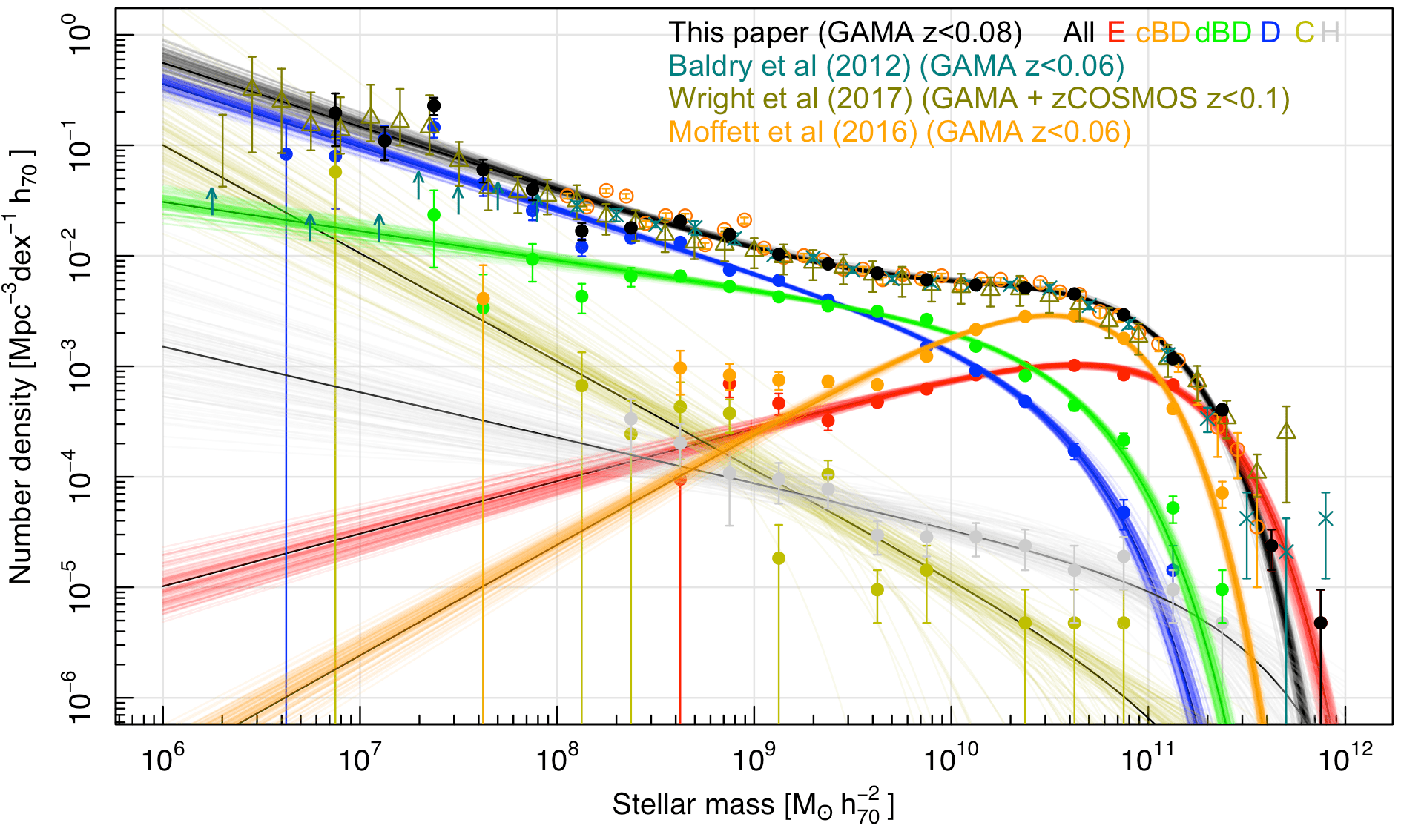}
	\includegraphics[width=\textwidth]{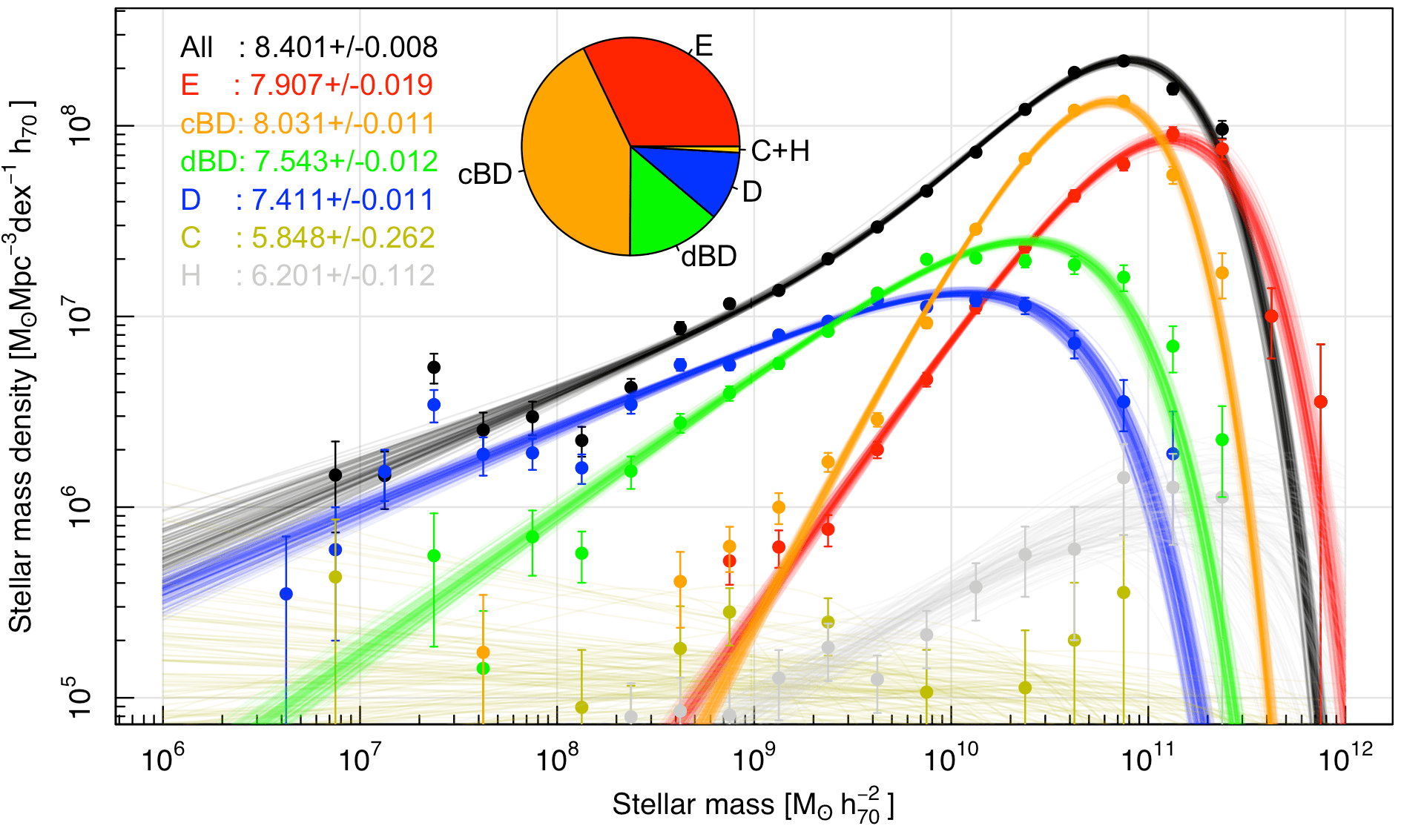}
	\caption{(upper) The GSMF for various morphological types (E, red; cBD, orange; dBD, green; D, blue; C, gold; and H, grey) and compared to previous measurements. (lower) the contribution of each log stellar mass interval to the overall stellar mass density. Shown as an insert is a pie diagram indicating the relative contribution of each class to the overall stellar mass density. Next to the inset are the stellar mass densities in logarithmic units from integrating the fitted mass functions (see Table~\ref{tab:massdistributionsmorph}).}
	\label{fig:gsdmf_morph}
\end{figure*}

\begin{table*}
\centering
\caption{The galaxy stellar mass number-density distributions for various morphological types for $z < 0.08$. No detections in a bin are indicated by a blank entry. Note that the tabulated values {\it DO NOT} include the re-normalisation to SDSS, or correction to redshift zero (see Section\,6). To apply these corrections add $0.0866$ dex to all number-density values.\label{tab:massdistributionsmorph}}
\begin{tabular}{c|c|c|c|c|c|c} \hline
$\log_{10}(M)$ & \multicolumn{6}{c}{Number-density of galaxies per dex per Mpc$^{3}h_{70}^{-3}$} \\ \cline{2-7}
$(M_{\odot} h_{70}^{-2})$ & E+HE & cBD & dBD & D & H & C \\  \hline
$ 11.875 $ & - & - & - & - & - & - \\
$ 11.625 $ & $ -4.623 \pm 0.146 $ & - & - & - & - & - \\
$ 11.375 $ & $ -3.483 \pm 0.048 $ & $ -4.146 \pm 0.103 $ & $ -5.021 \pm 0.176 $ & - & $ -5.322 \pm 0.301 $ & - \\
$ 11.125 $ & $ -3.164 \pm 0.035 $ & $ -3.383 \pm 0.043 $ & $ -4.281 \pm 0.105 $ & $ -4.845 \pm 0.222 $ & $ -5.021 \pm 0.176 $ & - \\
$ 10.875 $ & $ -3.074 \pm 0.031 $ & $ -2.746 \pm 0.021 $ & $ -3.669 \pm 0.063 $ & $ -4.322 \pm 0.114 $ & $ -4.720 \pm 0.176 $ & $ -5.322 \pm 0.301 $ \\
$ 10.625 $ & $ -2.992 \pm 0.029 $ & $ -2.543 \pm 0.017 $ & $ -3.354 \pm 0.044 $ & $ -3.766 \pm 0.067 $ & $ -4.845 \pm 0.222 $ & $ -5.322 \pm 0.301 $ \\
$ 10.375 $ & $ -3.008 \pm 0.029 $ & $ -2.549 \pm 0.017 $ & $ -3.084 \pm 0.031 $ & $ -3.318 \pm 0.041 $ & $ -4.623 \pm 0.146 $ & $ -5.322 \pm 0.301 $ \\
$ 10.125 $ & $ -3.077 \pm 0.031 $ & $ -2.667 \pm 0.020 $ & $ -2.819 \pm 0.024 $ & $ -3.039 \pm 0.031 $ & $ -4.544 \pm 0.125 $ & - \\
$ 9.875 $ & $ -3.202 \pm 0.035 $ & $ -2.909 \pm 0.026 $ & $ -2.576 \pm 0.018 $ & $ -2.824 \pm 0.024 $ & $ -4.544 \pm 0.125 $ & $ -4.845 \pm 0.222 $ \\
$ 9.625 $ & $ -3.323 \pm 0.041 $ & $ -3.166 \pm 0.035 $ & $ -2.502 \pm 0.016 $ & $ -2.537 \pm 0.017 $ & $ -4.528 \pm 0.125 $ & $ -5.021 \pm 0.176 $ \\
$ 9.375 $ & $ -3.491 \pm 0.075 $ & $ -3.138 \pm 0.047 $ & $ -2.453 \pm 0.020 $ & $ -2.401 \pm 0.016 $ & $ -4.110 \pm 0.125 $ & $ -3.978 \pm 0.125 $ \\
$ 9.125 $ & $ -3.334 \pm 0.087 $ & $ -3.125 \pm 0.074 $ & $ -2.372 \pm 0.027 $ & $ -2.222 \pm 0.019 $ & $ -4.021 \pm 0.146 $ & $ -4.736 \pm 0.301 $ \\
$ 8.875 $ & $ -3.157 \pm 0.097 $ & $ -3.081 \pm 0.103 $ & $ -2.278 \pm 0.037 $ & $ -2.129 \pm 0.026 $ & $ -3.965 \pm 0.222 $ & $ -3.425 \pm 0.125 $ \\
$ 8.625 $ & $ -4.022 \pm 0.301 $ & $ -3.015 \pm 0.155 $ & $ -2.183 \pm 0.047 $ & $ -1.878 \pm 0.030 $ & $ -3.695 \pm 0.176 $ & $ -3.366 \pm 0.222 $ \\
$ 8.375 $ & - & - & $ -2.186 \pm 0.077 $ & $ -1.837 \pm 0.043 $ & $ -3.475 \pm 0.176 $ & $ -3.613 \pm 0.301 $ \\
$ 8.125 $ & - & - & $ -2.367 \pm 0.114 $ & $ -1.919 \pm 0.071 $ & - & $ -3.175 \pm 0.301 $ \\
$ 7.875 $ & - & - & $ -2.031 \pm 0.138 $ & $ -1.591 \pm 0.074 $ & - & - \\
$ 7.625 $ & - & $ -2.386 \pm 0.301 $ & $ -2.471 \pm 0.301 $ & $ -1.348 \pm 0.089 $ & - & - \\
$ 7.375 $ & - & - & $ -1.629 \pm 0.222 $ & $ -0.838 \pm 0.077 $ & - & - \\
$ 7.125 $ & - & - & - & $ -0.938 \pm 0.114 $ & - & - \\
$ 6.875 $ & - & - & - & $ -1.098 \pm 0.222 $ & - & $ -1.241 \pm 0.301 $ \\
$ 6.625 $ & - & - & - & $ -1.080 \pm 0.301 $ & - & - \\ \hline
\end{tabular}
\end{table*}

\begin{table*}
\caption{Single Schechter function fits to the galaxy stellar mass functions for various morphological classes at $z<0.08$, and for the full sample with redshift cuts as indicated. Note that the $\log_{10}(\phi*)$ values {\it DO NOT} include the renormalisation to SDSS, or the correction to redshift zero (see Section\,6). To apply these corrections add $0.0866$ dex to both $\log_{10}(\phi*)$ values.\label{tab:massfunctionsmorph}}
\begin{tabular}{c|c|c|c|c} \hline
Dataset & $\log_{10}($M$*)$ & $\log_{10}(\phi^*)$ & $\alpha$ &  $\log_{10}(\rho_*)$ \\
        &  $($M$_{\odot} h_{70}^{-2})$    &  $($Mpc$^{-3} h_{70}^{3})$   &         &  $($M$_{\odot}$Mpc$^{-3} h_{70}$) \\ \hline
E+HE & $10.954 \pm 0.028$ & $-2.994 \pm 0.025$ & $-0.524 \pm 0.037$ & $7.906\pm 0.018$ \\
cBD & $10.499 \pm 0.016$ & $-2.469 \pm 0.011$ & $+0.003 \pm 0.039$ & $8.030\pm 0.011$ \\
dBD & $10.513 \pm 0.031$ & $-3.065 \pm 0.035$ & $-1.264 \pm 0.023$ & $7.543\pm 0.013$ \\
D & $10.436 \pm 0.038$ & $-3.332 \pm 0.044$ & $-1.569 \pm 0.018$ & $7.411\pm 0.011$ \\
H & $11.435 \pm 0.354$ & $-5.423 \pm 0.324$ & $-1.412 \pm 0.11$ & $6.211\pm 0.111$ \\
C & $11.170 \pm 0.970$ & $-6.419 \pm 1.123$ & $-1.978 \pm 0.118$ & $5.814\pm 0.176$ \\ 
SUM & - & - & - & $8.400 \pm 0.015$ \\ \hline
All($z<0.10$) & $10.745 \pm 0.020$ & $-2.437 \pm 0.016$ & $-0.465 \pm 0.069$ & $8.392\pm 0.006$ \\
All($z<0.08$) & $10.774 \pm 0.026$ & $-2.424 \pm 0.022$ & $-0.601 \pm 0.078$ & $8.401\pm 0.008$ \\\hline
\end{tabular}
\end{table*}

\subsection{The morphological galaxy stellar mass functions at $\mathbf{\emph{z}<0.0}8$}
We repeat the process of the previous section except for two changes. First, we extract morphological sub-samples as either: (E+HE), cBD, dBD, D, C or H, see Section\,\ref{sec:morph}, and secondly we elect to fit a simpler single Schechter function given by:
\begin{equation}
\phi(M)dM=\phi^* e^{-M/M^*}\left(\frac{M}{M^*}\right)^\alpha dM.
\end{equation}
As before the three fitted parameters are defined by a characteristic mass, M$^*$, a characteristic normalisation, $\phi^*$, and the faint-end slope, $\alpha$.

To obtain the fitted parameters we again make use of {\sc dftools} and provide the selected data, errors and selection function. To obtain the selection function
we examine the mass-to-light ratio distributions for each sub-population (see Figure\,\ref{fig:mass2light_morph} left) and determine the point at which the  mass-to-light ratio encloses 95 per cent of that population. These mass-to-light values are then combined with the $r$-band flux limit ($r_{\rm KiDSDR4}<19.65$\,mag), to derive the selection functions shown on Figure\,\ref{fig:mass2light_morph} (right). These curves are fitted with a Richards curve (see Eqn~\ref{eqn:richards}) and the values are reported in Table\,\ref{tab:richards}. These selection functions are used to select the relevant morphological sample, i.e., only those objects with masses higher than that indicated by the Richards curve are fed into the {\sc dffit} routine to derive the morphological mass function for that class. Tables\,\ref{tab:massdistributionsmorph} \& \ref{tab:massfunctionsmorph} shows our results. Note that in the fitting process we need to incorporate one subtlety, which was to ensure that the fits for each of the morphological classes adopted the same underlying LSS correction. To do this we stored the LSS correction identified from the full sample fit to $z<0.08$ (i.e., the black curve from Figure\,\ref{fig:gsdmf_morph}), and forced {\sc dffit} to use this LSS solution for the subsequent fits for each morphological class. 

The resulting distributions and fits are shown in Figure\,\ref{fig:gsdmf_morph} (upper panel), and the contribution to the stellar mass density from each morphological class is shown in the lower panel along with a pie chart representation (inset).
As can be seen in Figure\,\ref{fig:gsdmf_morph} the summed morphological data agrees reasonably well with our earlier total estimate, i.e., $(2.51 \pm 0.05) \times10^8$\,M$_{\odot}$ Mpc$^{-3} h_{70}$ now versus $(2.47 \pm 0.04) \times10^8$\,M$_{\odot}$ Mpc$^{-3} h_{70}$ previously.

Ellipticals (red) dominate at the highest masses, followed by the cBD systems (orange), the dBD systems (green), and finally the disc only systems (blue) from $10^{9.25}$\,M$_{\odot} h_{70}^{-2}$ and below. The compact (gold) and hard (grey) classes contribute a small amount to the total mass density. Hence the most frequent galaxy type is a disc, and the most massive galaxy type an Elliptical. The C and H classes start to become more significant in number at low mass, and as the C class is nearly divergent there is significant implied mass in its extrapolation and hence its value should be taken with great caution (as reflected in the error). The rise of the C and H classes at lower-masses, also highlights the increasing difficulty in making clear classifications at our lowest mass-limit out to $z=0.08$. Where statistics are good the fits appear to pass through the data points well, and the single Schechter function appears to be a reasonable fit to the data.

Figure\,\ref{fig:gsdmf_morph} (lower) shows the same data but now with the mass multiplied ordinate to reflect the contribution of each type and each mass interval to the total stellar mass density. The dominant contributions come from the cBD and E classes of ($1.07 \pm 0.04$) $\times 10^8$\,M$_{\odot}$ Mpc$^{-3} h_{70}$ and ($0.81 \pm 0.05$) $\times 10^8$\,M$_{\odot}$ Mpc$^{-3} h_{70}$ respectively. Together these classes make up 75 per cent of the total stellar mass density (if one ignores the mainly extrapolated C extrapolated masses).

As noted we now obtain a total stellar mass of $(2.51 \pm 0.05) \times 10^8$\,M$_{\odot}$ Mpc$^{-3} h_{70}$ rather than $(2.47 \pm 0.04) \times 10^8$\,M$_{\odot}$ Mpc$^{-3} h_{70}$ from the double Schechter function fit to the full dataset which is consistent within the errors.

\section{The cosmic stellar mass density}
A single GAMA region, at $z<0.1$, covers $\sim 400\,000$\,Mpc$^{3}$ with a CV of $\sim \pm 25$ per cent (see \citeauthor{driver2010} \citeyear{driver2010} Eqn.\,3). With four fully independent sight-lines the overall CV reduces to $\pm 12.5$ per cent for the combined GAMA fields (for $z < 0.1$ measurements). This error dominates over random or formal fitting errors but can be reduced further by determining any over/under density relative to a wider area similar depth spectroscopic survey, with uniform characteristics, and which encompass the GAMA regions. The only suitable survey, at the present time, is the Sloan Digital Sky Survey (SDSS) which covers almost $50\times$ the area of GAMA and fully contains the G12 and G15 regions.

\subsection{Normalisation of GAMA to SDSS}
\begin{figure*}
	\centering     
	\includegraphics[width=\textwidth]{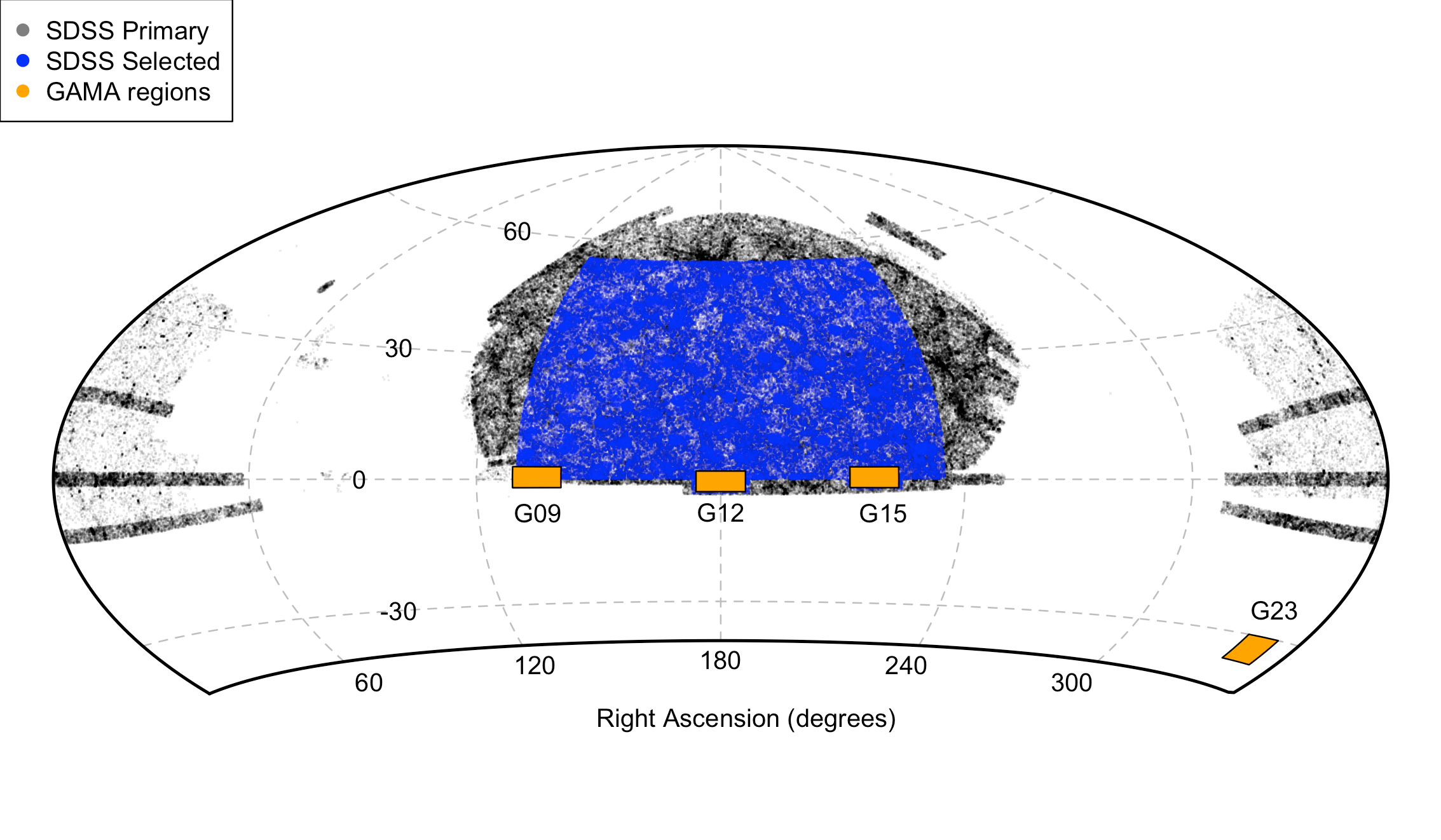}
	\caption{An Aitoff projection showing the data used for the renormalisation of GAMA to SDSS. This includes the SDSS Main Survey Primary footprint (grey), the region selected for our normalisation check covering 5\,012.134\,deg$^{2}$ (blue), and the location of the GAMA fields as labelled (orange boxes). }
	\label{fig:renorm}
\end{figure*}

Figure\,\ref{fig:renorm} shows the full SDSS DR16 spectroscopic survey in grey, and the region we have selected to normalise to in blue. The four GAMA regions are shown by the four orange boxes and only G12 and G15 are fully subsumed within the SDSS spectroscopic survey footprint. Our selected SDSS region is given by $130-235$ in Right Ascension (deg) and $0-55$ in Declination (deg), plus two small extensions to cover the G12 and G15 regions fully given by: $173-187$ in R.A.\,and $-3.5-0.0$ in Dec.\,(G12 extension); and $210.5-224.5$ in R.A.\,and $-2.5-0.0$ in Dec.\,(G15 extension). This equates to a contiguous area of 5\,012.134\,deg$^2$ with an estimated CV \citep{driver2010} of $\pm 6.5$ per cent. Hence by bootstrapping to SDSS we can theoretically rescale our results, and reduce our CV error by $\times 2$.

As only two of our regions lie within the SDSS footprint, we need to implement a double bootstrap. We first compare the density of the $z<0.1$ (or $z<0.08$) G12 and G15 regions to the SDSS selected area, using an SDSS tracer. We then compare the G12 and G15 regions to the full GAMA area, using a GAMA tracer. By combining the two density ratios we can determine a normalisation correction for the combined GAMA volume to the effective SDSS selected volume. 

There are a number of subtleties worth highlighting. This includes an underlying implicit assumption that the SDSS completeness and masking within the G12 and G15 regions are broadly consistent with the overall SDSS selected region. From Figure\,2 of \cite{driver2010} we see that this is a reasonable assumption. We also need to identify an appropriate density tracer. Previously we and other groups have adopted an intrinsic luminosity range, and quantified the {\it number}-density of this tracer population between the various regions. Here we adopt a slightly more sophisticated and stable measure. This is to use the full $r$-band luminosity density measurement by summing the intrinsic luminosities of all galaxies with $r_{\rm petro} < 17.77$\,mag and $z<0.1$ within the desired volumes. 

We find that G12 and G15 combined have a density $0.8781\times$ that of our SDSS selected region i.e., these GAMA regions are $\sim 12$ per cent under-dense. We also find, by summing the $r$-band flux of all galaxies in the GAMA regions with $r_{\rm KiDSDR4} < 19.65$\,mag and $z<0.1$ (spec-$z$ or photo-$z$), that G12 and G15 are over-dense compared to the combined GAMA regions by $\times1.0178$. Note that this estimate included consideration of the star-mask reported in \citeauthor{bellstedt20b} \citeyear{bellstedt20b}. 

Hence the overall renormalisation from GAMA to SDSS selected at $z<0.1$ is $\times1.159$ (and $\times 1.185$ at $z<0.08$). That is, we need to multiply our total GAMA density measurements by a factor of $1.159$ to renormalise to an effective SDSS selected area of 5012\,deg$^2$. 

This factor is fully consistent with the estimate reported in \cite{driver2011} (see their figure 20), and most GAMA low-$z$ published measurements have incorporated an appropriate upward scaling. We can confirm that with the inclusion of G23 an upward scaling is still required and that overall GAMA represents a slightly under-dense region of the Universe at $z<0.1$. It is worth noting that this under-density mainly stems from  single region, G09. Hence G12, G15 and G23 can be considered reasonably representative.

Finally, we note that if we did determine the correction based on number-density rather than luminosity-density we find a re-normalisation value of $\times 1.195$ ($\times1.220$) at $z<0.1$ ($z<0.08$), providing some indication of the uncertainty of the correction based on the choice of tracer.

\begin{figure}
	\includegraphics[width=\columnwidth]{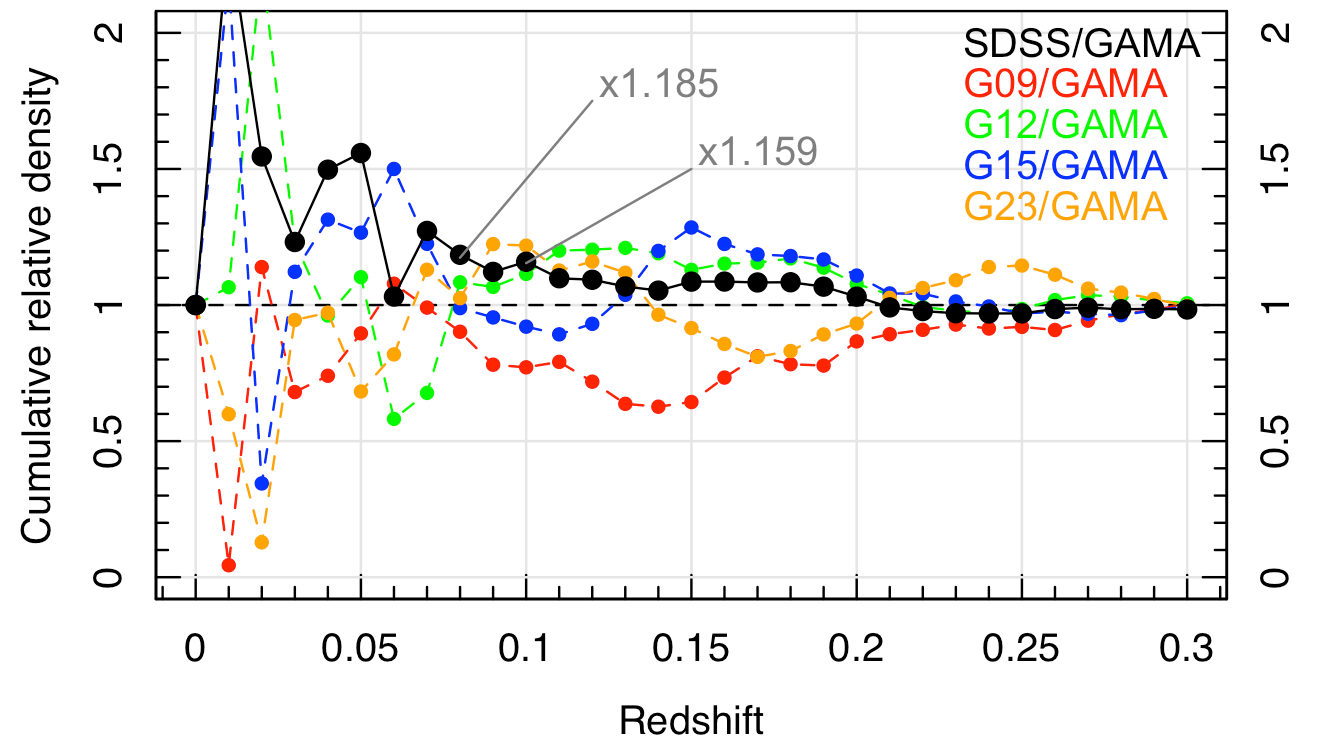}
	\caption{The cumulative density (by intrinsic flux) for GAMA compared to SDSS (black) and for each GAMA region relative to the combined GAMA area. Eventually all converge at $z=0.3$ but at $z=0.08$ and $z=0.1$ we find significant multipliers are required of $\times 1.185$ and $\times 1.159$ respectively.}
	\label{fig:renorm2}
\end{figure}

Figure\,\ref{fig:renorm2} shows the same calculation as described above, but now as a function of redshift (black dots and line). The figure also shows the relative density of each GAMA field compared to the combined GAMA density (coloured dots and lines). At very low redshift the volumes are very small and the local supercluster structure dominates. The variations then damp as the volume grows and converge to within a few per cent by $z \sim 0.3$. Noticeably at $z \approx0.1$ we see a range in luminosity densities of $\pm 14.4$ per cent about the mean, which is less than the \cite{driver2010} $1-\sigma$ CV estimate of $\pm 25$ per cent. 

We have also verified that the Virgo cluster, located in the central region of the SDSS selected area, has no impact by redetermining the correction with a minimum redshift that includes or excludes the Virgo cluster (the impact is in the fifth significant figure). We note that all tables reported earlier {\it DO NOT} include the renormalisation correction to the SDSS selected region. Hence one can choose to adopt the results of Section\,4 as GAMA unnormalised with a predicted CV error of $\pm 12.5$ per cent, or re-scale by the factors shown in Figure\,\ref{fig:renorm2} to generate the GAMA normalised to SDSS values with a CV error of $\pm 6.5$ per cent.

\subsection{Evolution to a $\mathbf{z=0.0}$ CSMD measurement}
One final correction we consider is the potential for some small evolution in the stellar masses over the redshift range explored ($0<z<0.1$). Typically our measurement at intermediate to high mass is weighted towards the median redshift over our full volume ($z \sim 0.079$), however, at lower masses, where volume corrections are required, the density measurement is for a progressively lower redshift range. As most galaxies are still forming stars and have existing stellar populations which are still evolving, there is the possibility of some small mass-dependent bias in our measured GSMF below $10^9$M$_{\odot} h_{70}^{-2}$ as well as a small overall offset from $z \approx 0.079$ to $z=0$. 

We can estimate this effect from the recent DEVILS study by \cite{thorne2021} who determine GSMF's from $z\approx0.8$ to $z=0$ and who fitted a smooth function to the total galaxy stellar mass density. This shows a net overall evolution in the cosmic stellar mass density from $z=0.079$ to $z=0.0$ of $\sim 3.9$ per cent. Hence we advocate for a further correction of $\times 1.039$ to correct our final cosmic stellar mass density measurement to a $z=0$ value of $\rho_* = (2.47 \pm 0.04) \times 10^{8} \times 1.159_{\rm SDSS renorm} \times 1.039_{\rm z=0 correction}$ M$_{\odot} $ Mpc$^{-3} h_{70}$ or $\rho_* = (2.97 \pm 0.04) \times 10^{8}$ M$_{\odot} $ Mpc$^{-3} h_{70}$.

\subsection{Comparison to earlier GAMA and SDSS DR7 GSMFs \label{sec:knee}}
\begin{figure}
	\centering     
	\includegraphics[width=\linewidth]{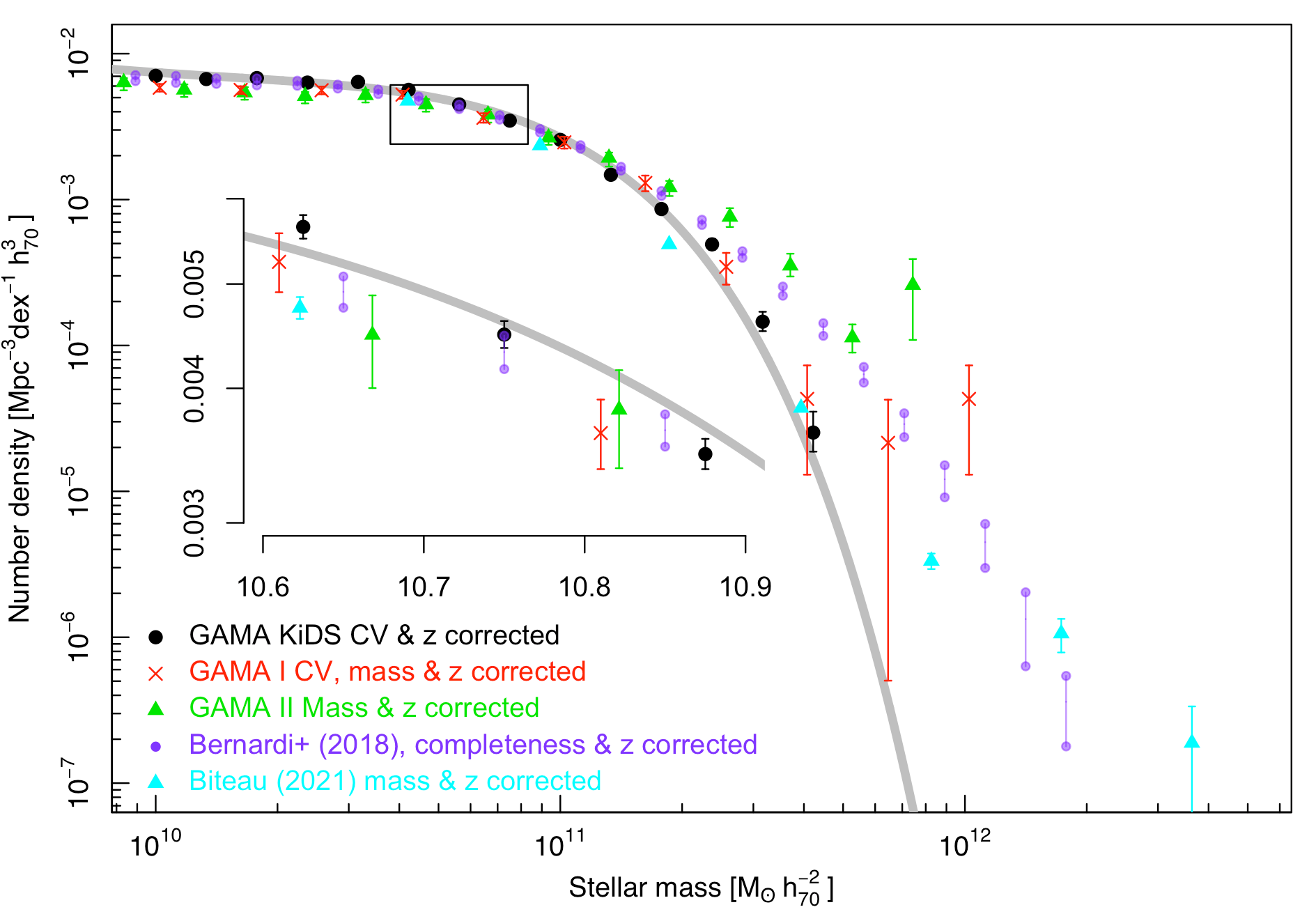}
	\caption{A comparison of recent GAMA and SDSS GSMFs around the normalisation or "knee" region, and often used to calibrate numerical simulations. The inset shows a zoom into the $M*$ region at $10^{10.75}$M$_{\odot} h_{70}^{-2}$, highlighting consistency at the 5-10\% percent level around the "knee", but some more significant variations at the highest masses where high S\'ersic indices and dense cluster cores can make photometric measurements more problematic. \label{fig:knee}}
\end{figure}

Figure.~\ref{fig:knee} shows the ``knee'' region of the GSMF for GAMA I \citep{baldry2012}, GAMA II \citep{wright17}, and GAMA KiDS (this work), as well as for SDSS DR7 \citep{bernardi2018}, and a recent local all sky compendium by \cite{biteau2021}. A number of corrections have been made to make each dataset fully consistent, and these include: mass corrections to GAMA I (+0.11\,dex), GAMA II (+0.17\,dex) and \cite{biteau2021} (+0.04\,dex, priv. comm.), to the ProSpect masses used in GAMA KiDS (see \citeauthor{prospect} \citeyear{prospect} Figure\,34), the cosmic variance correction of $\times 1.159$ for GAMA KiDS (as described above). Note that at $z<0.06$ the CV correction for GAMA I is negligible (see Figure\,\ref{fig:renorm}), and for GAMA II the methodology used a density-defining population from $0.07 < z < 0.19$ within which the CV is expected to be small. 

For \cite{bernardi2018} we implement an 11 per cent spectroscopic completeness correction. This arises from consideration of the SDSS DR7 and DR16 completeness within the G12 and G15 regions, and which suggests 87 per cent total completeness to $r_{\rm PETRO} < 17.77$\,mag, and representing all types of incompleteness including failed redshifts, fibre collisions, and the SDSS mask. As \cite{bernardi2018} incorporated a 2 per cent correction for fibre collision incompleteness, we implement a further 1.13/1.02 per cent upward correction. Finally, all datasets are scaled to redshift zero using the value from DEVILS (see Section\,6.2), of $\times 1.039$ for the $z<0.1$ datasets (GAMA II, GAMA KiDS, SDSS DR7), $\times 1.024$ for the $z<0.06$ dataset (GAMA I and \citeauthor{biteau2021} \citeyear{biteau2021}), and $\times 1.030$ for the $z<0.08$ morphological sample. 

The various mass functions on Figure\,\ref{fig:knee}, show good consistency around the ``knee'' to within 5-10 per cent, but some significant variations at the very high mass end. This is where photometric measurements become problematic for two reasons: extended high-S\'ersic index outer profiles, and blending as massive systems are often in the centres of highly crowded regions. In particular, we note that in some of our earlier works \citep[e.g.,]{wright17} we used extrapolations of our S\'ersic profiles to infer hidden flux, and hence mass, lurking below the isophote and outside the photometric aperture. From our deeper KiDS analysis we can see that these corrections were in some cases excessive leading to an Eddington-like bias at the high-mass end. To ensure we have captured all of the flux for these massive systems, we visually inspected all 20 GAMA KiDS $z<0.1$ objects above $3 \times 10^{11}$M$_{\odot} h_{70}^{-2}$ and concluded that neither photometric errors nor over-blending are likely to be the case for the GAMA KiDS GSMF.

\subsection{Comparison to recent measurements of $\Omega_*$}

\begin{figure*}
	\centering     
	\includegraphics[width=\textwidth]{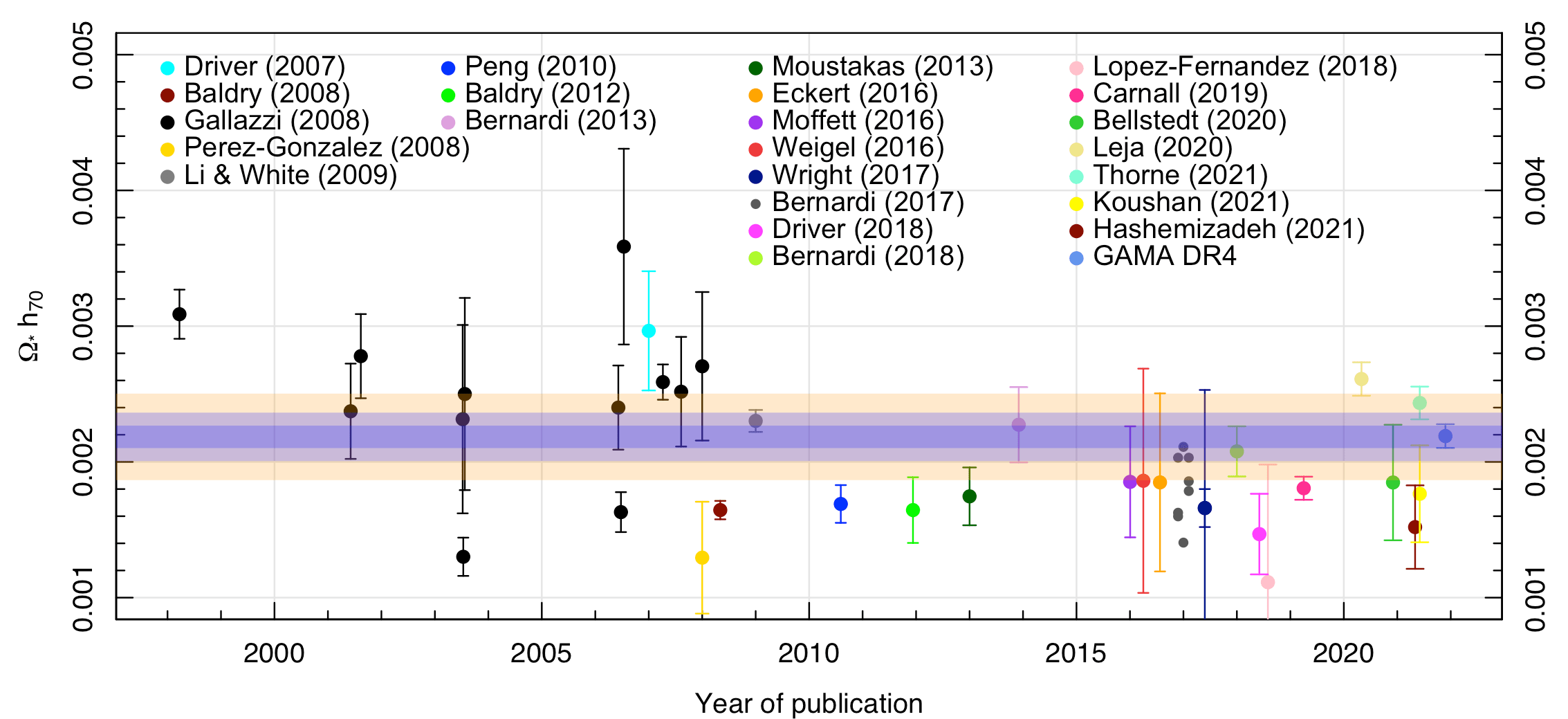}
	\caption{The galaxy stellar mass density at $z<0.1$ as estimated by various studies (as indicated) and including the total measurement derived here (rightmost blue data point and horizontal band). The dark blue band shows the formal statistical fitting error, the light blue band shows the error including cosmic variance, and the orange band incorporating systematic uncertainties in the mass estimation and choice of IMF. All data points have been corrected to a Chabrier IMF hence the appropriate error for comparison is indicated by the fainter blue band. \label{fig:smd}}
\end{figure*}

Figure\,\ref{fig:smd} shows published estimates of the local cosmic stellar mass density \citep{driver2007,baldry2008,perez2008,li2009,peng2010,baldry2012,bernardi2013,moustakas2013,eckert2016,moffett16,weigel2016,wright17,driver2018,lopez2018,carnall2019,bellstedt20,leja2020,thorne2021,koushan2021,Hashemizadeh2021} over the past few decades, along with our new value derived from the total GSMF, renormalised to SDSS and rescaled for evolution to $z=0$ (rightmost blue data-point). The horizontal bands highlight the error-range with the darkest blue band representing the formal fitting error, and the lighter-blue band the error including the CV uncertainty following renormalisation to SDSS (i.e., $\pm 6.5$ per cent). We note that the increase between this and earlier GAMA-based measurements can be traced back to a $\sim 0.06$dex or $\sim 0.14$dex increase in our stellar mass measurements from our recent {\sc ProSpect} analysis \citep{bellstedt20} compared to earlier estimates reported in \cite{taylor2011}, or via {\sc MAGPHYS} by \cite{driver2018}, respectively (with all measurements using a Chabrier IMF, see \citeauthor{prospect} \citeyear{prospect} for details).  

We conclude that the dominant error in the measurement of the cosmic stellar mass density, is now stemming from remaining systematic uncertainties in our stellar mass estimates, see for example the spread of the \cite{bernardi2017} estimates on Figure\,\ref{fig:smd} who discussed mass uncertainties due to flux, mass-to-light and dust attenuation issues (see also \citeauthor{sahu2019} \citeyear{sahu2019}); and \cite{hopkins2018} who discuss mass uncertainties due to issues related to IMF uncertainty/variation. Hence, limiting depth, incompleteness, errors in redshift measurements, or statistical size (as opposed to survey volume) are no longer dominating. We also note the critical importance of the adopted underlying initial mass function (IMF). The values shown on Figure\,\ref{fig:smd} are all for a Chabrier IMF. Note that a range of IMF conversion factors are shown in Table\,1 of \cite{yu2016} with a spread of $\pm 8$ per cent if one excludes a pure Salpeter IMF etc (see also \citeauthor{md14} \citeyear{md14}). 

The conclusion is that the uncertainty in our measurement of the stellar mass density remains more significant than our formal errors from this analysis suggest, with random and fitting uncertainties of $\sim 1$ per cent, CV of $\pm 6.5$ per cent, stellar mass estimates for a fixed IMF of $\pm 18$ per cent, and the plausible IMF range of $\pm 8$ per cent. In quadrature this puts the combined error at $\pm 21$ per cent, this uncertainty is shown as the faint orange band in Figure\,\ref{fig:smd} and indeed encompasses most previous measurements of the past decade (although we have attempted to correct all measurements plotted to a Chabrier IMF throughout).

\section{Summary}
Through this data release we provide full access to the Galaxy And Mass Assembly spectra, redshifts, and Data Management Units assembled over the past 12 years by the GAMA Team. This release now includes over 230\,000 new redshift measurements obtained using AAOmega on the Anglo-Australian Telescope across five regions to $r_{\rm SDSS} < 19.8$\,mag, plus a small 1\,deg$^2$ region in which we obtain 736 redshifts (with $P(z) > 0.9$) to a fainter flux limit of $r_{\rm SDSS}=21.6$\,mag. 

As reported in \cite{bellstedt20b} we have now replaced the original SDSS DR6/7 photometry with new photometry based on the significantly deeper and higher-resolution ESO VST KiDS data \citep{kidsdr4}. As this process shuffles many galaxies faintwards and some brightwards, as well as gaining and losing some galaxies, this requires us to re-evaluate our completeness limits. We now define the GAMA Main Survey to be $r_{\rm KiDS DR4} < 19.65$\,mag across the four primary regions (G09, G12, G15, G23). 

The four primary regions (G09, G12, G15 \& G23) cover $\sim230$\,deg$^2$, contain 205\,540 galaxies with $r_{\rm KiDS DR4} < 19.65$\,mag, for which we have reliable redshift measurements for 195\,432, i.e., 95.1 per cent complete. For the remaining galaxies for which we lack spectroscopic measurements we obtain photometric redshift estimates via a scaled-flux matching method identical to that described in \cite{baldry2021}. We also include {\sc EAZY} photometric redshift estimates for all 18\,million objects in {\sc gkvInputCatv02}.

We have morphologically classified all galaxies with a redshift below 0.08 as either Elliptical (E+HE), compact-bulge plus disc (cBD), diffuse-bulge plus disc (dBD), disc-only (D), compact (C) or hard (H). The process is conducted blindly by three classifiers and we demonstrate better than 90 per cent consistency in almost all mass and redshift bins.

In the final stages of the paper we construct the galaxy stellar mass function for each region, the combined dataset, and sub-divided into each morphological type. We use the Maximum Likelihood method from the {\sc dftools} package and provide a selection function based on the $95^{\rm th}$-percentile mass-to-light distribution combined with our new nominal flux limit of $r_{\rm KiDSDR4}=19.65$\,mag. The final galaxy stellar mass function now reaches down to $ M \sim 10^{6.75}$\,M$_{\odot} h_{70}^{-2}$ extending over an order of magnitude beyond our previous estimates provided by \cite{baldry2012} and \cite{wright17}. The extension contains no surprises, and is fully consistent with the earlier extrapolation and hence leads to no significant change in the overall stellar mass density.

Hence, we conclude that the $z=0$ Universe contains:
$(2.97 \pm 0.04_{\rm random} \pm 0.58_{\rm systematic} \pm 0.20_{\rm CV}) \times 10^8$\,M$_{\odot}$ Mpc$^{-3} h_{70}$ 
of stellar mass bound within the galaxy population. This value {\it includes} renormalisation to a 5\,012\,deg$^2$ region of SDSS ($\times 1.159$), and a correction to $z=0$ ($\times1.039$). However, we do not address how much stellar mass may reside in stripped mass in the intra-halo light, but note that arguments based on the extragalactic background light would suggest this must be fairly minimal ($<5$ per cent, see \citeauthor{driver2018} \citeyear{driver2018} and \citeauthor{koushan2021} \citeyear{koushan2021}).

Further exploration of the galaxy stellar mass function to lower limits will continue, with the forthcoming 4MOST Wide Area VISTA Extragalactic Survey (WAVES; \citeauthor{driver2019} \citeyear{driver2019}) scheduled to commence in 2023, and which should extend GSMF measurements to below $10^6$\,M$_{\odot} h_{70}^{-2}$. 
As yet the galaxy stellar mass function continues to rise to our detection limits and as such the most numerous galaxy type (per decade of log mass) remains to be discovered, and its properties and space density to be quantified.

\section*{Acknowledgements}
We acknowledge the early support of the University of St Andrews, and ongoing support from the International Centre for Radio Astronomy Research, the European Southern Observatory, Australian Astronomical Optics, and Universit\"at Hamburg, in hosting the leadership team (SPD,\,AMH,\,JL), and the GAMA database.

We also acknowledge funding from many sources, but especially that of the Australian Research Council and the UK Science and Technology Facilities Council.

JTAdJ is supported by the Netherlands Organisation for Scientific Research (NWO) through grant 621.016.402

AD acknowledges European Research Council Consolidator Grant (No. 770935).

CH acknowledges support from the European Research Council under grant number 647112, and support from the Max Planck Society and the Alexander von Humboldt Foundation in the framework of the Max Planck-Humboldt Research Award endowed by the Federal Ministry of Education and Research.

H.Hildebrandt is supported by the Heisenberg grant of the Deutsche Forschungsgemeinschaft (Hi 1495/5-1) as well as a European Research Council Grant (No. 770935).

CSF acknowledges support from the European Research Council
through ERC Advanced Investigator grant, DMIDAS [GA 786910].

FDE and AvdW acknowledge funding through the H2020 ERC Consolidator Grant 683184, the ERC Advanced grant 695671
“QUENCH” and support by the Science and Technology Facilities Council (STFC). Velocity dispersion measurements were made possible by the extraordinary work and dedication of Brian Van Den Noortgaete.

M. Bilicki is supported by the Polish National Science Center through grants no. 2020/38/E/ST9/00395, 2018/30/E/ST9/00698 and 2018/31/G/ST9/03388, and by the Polish Ministry of Science and Higher Education through grant DIR/WK/2018/12.

KS acknowledges support from the Australian Government through the Australian Research Council’s Laureate Fellowship funding scheme (project FL180100168)

This work used the DiRAC Data Centric system at Durham University, 
operated by the Institute for Computational Cosmology on behalf of the 
STFC DiRAC HPC Facility (\url{https://dirac.ac.uk/}). This equipment was funded by 
BIS National E-infrastructure capital grants ST/P002293/1, ST/R002371/1 
and ST/S002502/1, Durham University and STFC operations grant 
ST/R000832/1.

GAMA is a joint European-Australasian project based around a
spectroscopic campaign using the Anglo-Australian Telescope. The GAMA
input catalogue is based on data taken from the Sloan Digital Sky
Survey and the UKIRT Infrared Deep Sky Survey. Complementary imaging
of the GAMA regions is being obtained by a number of independent
survey programmes including GALEX MIS, VST KiDS, VISTA VIKING, WISE,
Herschel-ATLAS, GMRT and ASKAP providing UV to radio coverage. GAMA is
funded by the STFC (UK), the ARC (Australia), the AAO, and the
participating institutions. The GAMA website is
\url{http://www.gama-survey.org/}. Based on observations made with ESO
Telescopes at the La Silla Paranal Observatory under programme IDs
177.A-3016, 177.A-3017, 177.A-3018 and 179.A-2004. 

This work was supported by resources provided by the \textit{Pawsey Supercomputing Centre} with funding from the Australian Government and the Government of Western Australia.
We gratefully acknowledge \textit{DUG Technology} for their support and HPC services.

We thank the institutions of the
GAMA/Herschel-ATLAS/DINGO Consortium for making
possible the completion of the coverage of the GAMA survey
area by the GALEX satellite at MIS depth, as part of the
All-sky UV Survey Extension of the GALEX mission organised by
Caltech.

\section{Data Availability}

The data presented in this paper are all publicly available as part of Data Release 4 of the GAMA survey. Each of the Data Management Units outlined in Tables \ref{tab:dmus} and \ref{tab:dmus2} can be accessed at \url{http://www.gama-survey.org/dr4/}.

\bibliographystyle{mnras}
\bibliography{paperlib}

~

\noindent
{$^1$ International Centre for Radio Astronomy Research (ICRAR), University of Western Australia, Crawley, WA 6009, Australia \\
$^2$ Astrophysics Research Institute, Liverpool John Moores University, 146 Brownlow Hill, Liverpool, L3 5RF, UK \\
$^3$ Hamburger Sternwarte, Universit\"at Hamburg, Gojenbergsweg 112, 21029 Hamburg, Germany \\
$^4$ Centre for Astrophysics and Supercomputing, Swinburne University of Technology, Hawthorn, VIC 3122, Australia \\
$^5$ Faculty of Physics and Astronomy, Astronomical Institute (AIRUB), Ruhr University Bochum, D-44780 Bochum, Germany \\
$^6$ Imagen Technologies, 151 W 26th St, New York, NY 10001, USA\\
$^7$ School of Physics \& Astronomy, University of Nottingham, University Park, Nottingham, NG7 2RD, UK \\ 
$^8$ Vera C. Rubin Observatory 950 N Cherry Ave, Tucson AZ 85719 USA\\ 
$^9$ Sydney Astrophotonic Instrumentation Labs, School of Physics, the University of Sydney, Sydney NSW 2006, Australia \\ 
$^{10}$ Centre for Theoretical Physics, Polish Academy of Sciences, al. Lotnikow 32/46, 02-668 Warsaw, Poland \\
$^{11}$ School of Physics, University of New South Wales, NSW 2052, Australia \\
$^{12}$ Department of Physics and Astronomy, University of the Western Cape, Robert Sobukwe Road, Bellville, South Africa \\
$^{13}$ Research School of Astronomy and Astrophysics, Australian National University, Canberra, ACT 2611, Australia \\
$^{14}$ Jodrell Bank Centre for Astrophysics, University of Manchester, Oxford Road, Manchester UK \\
$^{15}$ Leiden Observatory, Leiden University, PO Box 9513, 2300 RA Leiden, The Netherlands \\
$^{16}$ Kapetyn Astronomical Institute, University of Groningen, PO Box 800, 9700 AV Groningen, The Netherlands \\
$^{17}$ Cavendish Laboratory and Kavli Institute for Cosmology, University of Cambridge, Madingley Rise, Cambridge, CB3 0HA, UK \\
$^{18}$ Sterrenkundig Observatorium, Universiteit Gent, Krijgslaan 281 S9, B-9000 Gent, Belgium \\
$^{19}$ FINCA, University of Turku, Vesilinnantie 5, Turku, 20014, Finland \\
$^{20}$ School of Mathematics and Physics, The University of Queensland, Brisbane, QLD 4072, Australia \\
$^{21}$ Max-Planck-Institut f\"ur extraterrestrische Physik, Giessenbachstrasse, 85748 Garching, Germany \\
$^{22}$ Institute for Computational Cosmology, Physics Department, University of Durham, Durham DH1 3LE, UK \\
$^{23}$ Institute for Astronomy, University of Edinburgh, Royal Observatory, Edinburgh EH9 3HJ, UK \\
$^{24}$ Netherlands eScience Centre, Science Park 140, 1098 XG Amsterdam, The Netherlands \\
$^{25}$ European Southern Observatory, Alonso de Cordova 3107, Casilla 19001, Santiago, Chile \\
$^{26}$ Department of Physics and Astronomy, 102 Natural Science Building, University of Louisville, Louisville KY 40292, USA \\
$^{27}$ Australian Astronomical Optics, Macquarie University, 105 Delhi Rd, North Ryde, NSW 2113, Australia \\
$^{28}$ Department of Astronomy, University of Cape Town, Rondebosch, Cape Town, 7700, South Africa \\
$^{29}$ Pathways and Academic Learning Support Centre, University of Newcastle, Callaghan, NSW 2308, Australia \\
$^{30}$ Department of Astrophysical Sciences, Princeton University, 4 Ivy Lane, Princeton, NJ 08544, USA \\
$^{31}$ Armagh Observatory and Planetarium, College Hill, Armagh, BT61 DG, UK \\
$^{32}$ Curtin Institute for Computation, Curtin University, PO Box U 1987, Perth, WA6845, Australia \\
$^{33}$ Astronomy Centre, University of Sussex, Falmer, Brighton BN1 9QH, UK\\
$^{34}$ Indian Institute of Science Education and Research Mohali- IISERM, Knowledge City, Manauli, 140306 Punjab, India \\
$^{35}$ Department of Physics and Astronomy, University of North Georgia, 3820 Mundy Mill Rd., Oakwood GA 30566, USA \\
$^{36}$ School of Physics and Astronomy, Sun Yat-sen University, Guangzhou 519082, Zhuhai Campus, P.R. China \\
$^{37}$ Department of Physics and Astronomy, Macquarie University, NSW 2109, Australia \\
$^{38}$ Astronomy, Astrophysics and Astrophotonics Research Centre, Macquarie University, Sydney, NSW 2109, Australia \\
$^{39}$ INAF - Osservatorio Astronomico di Padova, via dell'Osservatorio 5, 35122 Padova, Italy \\
$^{40}$ Astrophysics Group, School of Physics, Tyndall Avenue, University of Bristol, Bristol, BS8 1TL, UK \\
$^{41}$ E.A.Milne Centre for Astrophysics, University of Hull, Cottingham Road, Kingston-upon-Hull, HU6 7RX, UK \\
$^{42}$ Jeremiah Horrocks Institute, School of Natural Sciences, University of Central Lancashire, Preston, PR1 2HE, UK \\
$^{43}$ The Observatories of the Carnegie Institute for Science, 813 Santa Barbara Street, Pasadena, CA 91101, USA\\
$^{44}$ School of Physics and Astronomy, Queen Mary University of London, Mile End Road, London E1 4NS, UK \\
$^{45}$ Mas-Planck-Institue fuer Kernphysik, Saupfercheckweg 1, D-69117 Heidelberg, Germany}
$^{46}$ ESO, Karl-Schwarzschild-Str. 2, 85748 Garching bei München, Germany \\
\onecolumn

\end{document}